%% file: main.tex
\documentclass[aps,prb,reprint,twocolumn,superscriptaddress,nofootinbib,longbibliography]{revtex4-2}

\usepackage[ruled]{algorithm2e}
\usepackage{amsmath}
\usepackage{amssymb}
\usepackage{amsthm}
\usepackage[australian]{babel}
\usepackage{braket}
\usepackage{hyperref}
\usepackage{orcidlink}
\usepackage{pgfplots}
\usepackage{tikz}

\pgfplotsset{compat=1.18}
\usepgfplotslibrary{colorbrewer}
\usepgfplotslibrary{fillbetween}
\pgfplotsset{every tick label/.append style={font=\footnotesize}}

\newcommand{\drm}{\mathrm{d}}
\newcommand{\erm}{\mathrm{e}}
\newcommand{\irm}{\mathrm{i}}

\newcommand\norm[1]{\left\lVert#1\right\rVert}

\def\pbra#1{\mathinner{({#1}|}}
\def\pket#1{\mathinner{|{#1})}}
\def\pbraket#1{\mathinner{({#1})}}

\newcommand{\defaulttensorsize}{10pt}
\newcommand{\tensorsize}{\defaulttensorsize}

\usetikzlibrary{patterns,patterns.meta}
\usetikzlibrary{shapes.geometric}
\usetikzlibrary{shapes.misc}
\tikzstyle{tensor}=[draw, inner sep=0, outer sep=0, minimum size=\tensorsize]
\tikzstyle{notensor}=[inner sep=0, outer xsep=2pt, outer ysep=0, minimum size=\tensorsize]
\tikzstyle{atensor}=[tensor, circle]
\tikzstyle{ctensor}=[tensor, circle]
\tikzstyle{dtensor}=[tensor, diamond]
\tikzstyle{wtensor}=[tensor]
\tikzstyle{ltensor}=[tensor, rounded rectangle, rounded rectangle left arc=none]
\tikzstyle{rtensor}=[tensor, rounded rectangle, rounded rectangle right arc=none]
\tikzstyle{etensor}=[tensor, minimum height=(1cm/\defaulttensorsize*0.5*2+1)*\tensorsize]
\tikzstyle{widetensor}[2]=[tensor, minimum width=(1cm/\defaulttensorsize*0.75*(#1-1)+1)*\tensorsize]

\tikzstyle{tensornetwork}=[baseline=-0.25em, xscale=0.75, yscale=0.5,
                           every node/.style={inner sep=0, outer xsep=2pt, outer ysep=5pt, text depth=0pt},
                           execute at end picture={\path (current bounding box.south west) +(-2pt, 0) (current bounding box.north east) +(2pt, 0);}]

\newcommand{\stripesize}{4pt}
\pgfdeclarepatternformonly{stripes}{\pgfpointorigin}{\pgfpoint{\stripesize}{\stripesize}}{\pgfpoint{\stripesize}{\stripesize}}
{
    \pgfpathmoveto{\pgfpoint{0}{0}}
    \pgfpathlineto{\pgfpoint{\stripesize}{\stripesize}}
    \pgfpathlineto{\pgfpoint{\stripesize}{0.5*\stripesize}}
    \pgfpathlineto{\pgfpoint{0.5*\stripesize}{0}}
    \pgfpathclose%
    \pgfusepath{fill}
    \pgfpathmoveto{\pgfpoint{0}{0.5*\stripesize}}
    \pgfpathlineto{\pgfpoint{0}{\stripesize}}
    \pgfpathlineto{\pgfpoint{0.5*\stripesize}{\stripesize}}
    \pgfpathclose%
    \pgfusepath{fill}
}
\tikzstyle{striped}=[pattern=stripes, pattern color=lightgray]

\begin{document}
\title{Efficient and systematic calculation of arbitrary observables for the matrix product state excitation ansatz}

\author{Jesse~J. Osborne${}^{\orcidlink{0000-0003-0415-0690}}$}
\affiliation{School of Mathematics and Physics, The University of Queensland, St Lucia, QLD 4072, Australia}
\email{j.osborne@student.uq.edu.au}

\author{Ian~P. McCulloch${}^{\orcidlink{0000-0002-8983-6327}}$}
\affiliation{Department of Physics, National Tsing Hua University, Hsinchu 30013, Taiwan}
\affiliation{Frontier Center for Theory and Computation, National Tsing Hua University, Hsinchu 30013, Taiwan}
\email{ian@phys.nthu.edu.tw}

\begin{abstract}
Numerical methods based on matrix product states (MPSs) are currently the \textit{de facto} standard for calculating the ground-state properties of (quasi-)one-dimensional quantum many-body systems.
While the properties of the low-lying excitations in such systems are often studied in this MPS framework through \emph{dynamics} by means of time-evolution simulations, we can also look at their \emph{statics} by directly calculating eigenstates corresponding to these excitations.
The so-called MPS excitation ansatz is a powerful method for finding such eigenstates with a single-particle character in the thermodynamic limit.
Although this excitation ansatz has been used quite extensively, a general method for calculating expectation values for these states is lacking in the literature: we aim to fill this gap by presenting a recursive algorithm to calculate arbitrary observables expressed as matrix product operators.
This method concisely encapsulates existing methods for—as well as extensions to—the excitation ansatz, such as excitations with a larger spatial support and multi-particle excitations, and is robust enough to handle further innovations.
We demonstrate the versatility of our method by studying the low-lying excitations in the spin-1 Heisenberg chain and the one-dimensional Hubbard model, looking at how the excitations converge in the former, while in the latter, we present a refined method of targeting single-particle excitations inside a continuum by minimizing the energy \emph{variance} rather than the energy itself.
We hope that this technique will foster further advancements with the excitation ansatz.
\end{abstract}

\date{\today}

\maketitle

\section{Introduction}
Over the past two decades, algorithms based on the formalism of matrix product states (MPSs) have emerged as a major powerhouse in computational physics~\cite{white1992,white1993,schollwoeck2011,paeckel2019,verstraete2023}.
This is because, by their nature, MPSs are highly efficient at describing locally entangled states.
This includes not only ground states of gapped local Hamiltonians in one spatial dimension, which obey an entanglement area law, but even critical states, excited states, finite-temperature states, far-from-equilibrium states, as well as two-dimensional states, with a great degree of success.
Expanding beyond the original applications, MPSs, and tensor networks in general~\cite{banuls2023}, have proved to be a valuable tool in computational science, with applications in quantum chemistry~\cite{white1999,chan2011}, numerical analysis~\cite{oseledets2011}, and machine learning~\cite{sengupta2022}.
Recent developments include a renewed interest in the quantics tensor train~\cite{latorre2005,oseledets2009,khoromskij2011,chen2023,ritter2024} and the key role of tensor networks in challenging claims of quantum supremacy~\cite{zhou2020,gray2021,pan2021,liu2021,pan2022,ayral2023,tindall2024}.

This interest in MPSs was initially sparked by the density matrix renormalization group (DMRG) algorithm for finding the ground state of a one-dimensional Hamiltonian~\cite{white1992,white1993,schollwoeck2011,verstraete2023}.
While we can obtain much useful information from the ground state, to fully understand such models, we need to go beyond the ground state and analyse their excitations.
This is relevant, for instance, for the calculation of spectral functions, which can be measured experimentally through inelastic neutron scattering, angle-resolved photoemission spectroscopy~\cite{damascelli2003}, and resonant inelastic x-ray scattering~\cite{ament2011}, among other methods.
One popular avenue for such analysis is through the time-evolution simulation of local or global quenches~\cite{paeckel2019}: while effective, this is limited by the inherent buildup of entanglement during the time evolution (although recent efforts have been made to mitigate this buildup, such as the use of complex time evolution~\cite{grundner2024,cao2024}, time evolution in momentum space~\cite{van-damme2022}, temporal MPSs~\cite{banuls2009,hastings2015,lerose2021,frias-perez2022,lerose2023,carignano2024}, or by converting the entangled time-evolved states into less-entangled mixed states~\cite{surace2019,frias-perez2024,taylor2023}).

In contrast to the calculation of \emph{dynamics} via time evolution, an alternative and complementary method for investigating excitations is to calculate \emph{static} low-lying energy eigenstates.
While we could calculate the low-lying states together with the ground state when performing DMRG for finite-size systems~\cite{mcculloch2007,li2024}, in the thermodynamic limit we can use the excitation ansatz (EA)~\cite{oestlund1995,haegeman2012,vanderstraeten2019a}, which builds up excited states as local disturbances on top of an infinite MPS representation of the ground state with a well-defined momentum.
This EA form is economical, as we reuse the already-known ground-state wave function as a starting point for finding the excited states, which can then be systematically refined.
The EA is highly efficient at describing states with a single-particle character~\cite{haegeman2013b}, and have also been generalized to describe two-particle scattering states~\cite{vanderstraeten2014,vanderstraeten2015a}.
Many other advances have been made to expand the versatility of this method, such as handling long-range interactions~\cite{vanderstraeten2018}, representing fractionalized excitations using conserved quantum numbers~\cite{zauner-stauber2018b}, calculating excitations for 2D systems on a cylinder with well-defined transverse momentum~\cite{van-damme2021b,zhang2024}.
It has also been generalized to other types of tensor networks, including projected-pair entangled states (PEPS)~\cite{vanderstraeten2015b,vanderstraeten2019b,ponsioen2020}.
In some recent works, the excitation ansatz has been used to directly complement time evolution, such as by using EA wave functions to generate real-space wave packets for a time-evolution simulation~\cite{van-damme2021a}, or as basis states to measure the occupation of different particle sectors following a scattering simulation~\cite{milsted2022,belyansky2024}.
The excitation ansatz has been successfully used for investigating key phenomenological issues, such as the confinement of excitations~\cite{bera2017,halimeh2020,vanderstraeten2020}, and many-body scarring by representing scarred eigenstates for frustration-free Hamiltonians~\cite{moudgalya2018a,moudgalya2018b,moudgalya2020,zhang2023}.

Although the MPS excitation ansatz has been extensively studied over the past decade, there is no systematic method in the literature for calculating arbitrary observables for an EA wave function.
For instance, many authors assume the Hamiltonian to have a two-site gate form~\cite{haegeman2012,vanderstraeten2015a,vanderstraeten2019a} instead of a more general matrix product operator (MPO), and each time an extension to the excitation ansatz is proposed, the formulas have to be rederived from scratch.
In this paper, we fill this gap by presenting a recursive algorithm for calculating expectation values of an MPO for any EA wave function.
This is an extension of the algorithm proposed in Ref.~\cite{michel2010} for ordinary infinite MPSs, and results from the interpretation of an EA wave function as a block-upper-triangular MPS, which has not been explored much in the literature.
Indeed, this block-upper-triangular form encapsulates existing extensions to the excitation ansatz, such as extending the spatial support of the excitation with multi-site ‘windows’~\cite{vanderstraeten2020}, and multi-particle wave functions~\cite{vanderstraeten2014,vanderstraeten2015a}, and also allows us to express new extensions in a natural way, such as derivatives of EA wave functions with respect to momentum (discussed in Sec.~\ref{sec:momentum-derivatives}).
This formalism allows us to concisely express objects such as the effective Hamiltonian, and since we are working with MPOs, we can efficiently handle general operators, such as the energy variance.
We demonstrate the usefulness of this by calculating an otherwise inaccessible dispersion relation within a multi-particle continuum by minimizing the energy variance rather than the energy itself (Sec.~\ref{sec:variance-minimization}).

The layout of this paper is as follows.
In Sec.~\ref{sec:ea}, we give a brief overview of the MPS excitation ansatz.
In Sec.~\ref{sec:mpo}, we describe the general algorithm for calculating the fixed points of an MPO, explain how existing methods, such as the EA algorithm~\cite{haegeman2012}, can be expressed in this form, and discuss extensions such as variance minimization (Sec.~\ref{sec:variance-minimization}) and momentum derivatives (Sec.~\ref{sec:momentum-derivatives}).
In Sec.~\ref{sec:results}, we present calculations for the spin-1 Heisenberg model and one-dimensional Hubbard model demonstrating the utility of this general method, and present concluding remarks in Sec.~\ref{sec:conclusion}.

\section{The excitation ansatz}\label{sec:ea}
We begin with some preliminary definitions of basic MPS concepts.
For a system described by a tensor product of \(L\) \(d\)-dimensional local Hilbert spaces, taking the thermodynamic limit \(L \rightarrow \infty\), a generic infinite MPS (iMPS) has the translation-invariant form
\begin{equation}\label{eq:mps}
    \ket{\Psi[A^s]} = \sum_{\mathbf{s}} \mathbf{v}_L \left[ \cdots A^{s_{-1}} A^{s_0} A^{s_1} A^{s_2} \cdots \right] \mathbf{v}_R^\dagger \ket{\mathbf{s}},
\end{equation}
where \(\ket{\mathbf{s}} = \cdots \otimes \ket{s_0} \otimes \ket{s_1} \otimes \ket{s_2} \otimes \cdots\) is the tensor product of local basis states, and the same tensor~\(A^s\) with dimension \(d\times D\times D\) (where \(D\) is the so-called bond dimension) is repeated for each site in the system.
We will generally work with normalized, injective states (for which the transfer operator \(T(E) = \sum_s A^{s\dagger} E A^s\) has a largest eigenvalue of 1 that is nondegenerate): in this case, the boundary vectors \(\mathbf{v}_L\) and \(\mathbf{v}_R^\dagger\) are irrelevant and so we will omit writing them, as well as the basis~\(\ket{\mathbf{s}}\), for brevity.
The \(A^s\)~tensor can be optimized to represent the ground state of some Hamiltonian by means of an algorithm such as infinite DMRG (iDMRG)~\cite{mcculloch2008,schollwoeck2011} or variational uniform MPSs (VUMPS)~\cite{zauner-stauber2018a,vanderstraeten2019a}.
Note that we focus only on single-site unit cells in the main text: everything discussed here can be extended to multi-site unit cells, but at the cost of more cumbersome notation, which can be found in Appendix~\ref{app:multi-site-unit-cell}.
As is common in the tensor network literature, we will often use tensor network notation~\cite{bridgeman2017}, for example, we can write the iMPS expressed in Eq.~\eqref{eq:mps} as
\begin{equation}
    \ket{\Psi[A]} =
    \begin{tikzpicture}[tensornetwork]
        \node[atensor, label=\(A\)] (A1) at (1, 0) {};
        \node[atensor, label=\(A\)] (A2) at (2, 0) {};
        \node[atensor, label=\(A\)] (A3) at (3, 0) {};
        \node[atensor, label=\(A\)] (A4) at (4, 0) {};
        \node[atensor, label=\(A\)] (A5) at (5, 0) {};
        \draw (A1.south) -- +(0, -0.5);
        \draw (A2.south) -- +(0, -0.5);
        \draw (A3.south) -- +(0, -0.5);
        \draw (A4.south) -- +(0, -0.5);
        \draw (A5.south) -- +(0, -0.5);
        \draw (A1) -- (A2) -- (A3) -- (A4) -- (A5);
        \draw (A1.west) -- +(-0.5, 0) node[left] {\(\ldots\)};
        \draw (A5.east) -- +(0.5, 0) node[right] {\(\ldots\)};
    \end{tikzpicture}
    .
\end{equation}
We will usually omit the label of the physical index~\(s\) when writing three-index tensors like \(A^s\) unless it is needed for clarity, and we will sometimes omit tensor labels when they can be inferred from previous diagrams.

Given an iMPS approximation to the ground state, we can construct an excitation ansatz (EA) wave function to represent low-lying excited states, with the basic form~\cite{oestlund1995,haegeman2012}
\begin{subequations}\label{eq:ea}
\begin{align}
    &\phantom{=} \ket{\Phi_k[B_k]} \nonumber \\
    &= \sum_n \erm^{\irm kn} \cdots A^{s_{n-2}} A^{s_{n-1}} B_k^{s_n} \tilde{A}^{s_{n+1}} \tilde{A}^{s_{n+2}} \cdots \\
    &= \sum_n \erm^{\irm kn}
    \begin{tikzpicture}[tensornetwork]
        \node[atensor, label=\(A\)] (A1) at (1, 0) {};
        \node[atensor, label=\(A\)] (A2) at (2, 0) {};
        \node[atensor, fill=lightgray, label=\(B_k\)] (A3) at (3, 0) {};
        \node[atensor, label=\(\tilde{A}\)] (A4) at (4, 0) {};
        \node[atensor, label=\(\tilde{A}\)] (A5) at (5, 0) {};
        \draw (A1.south) -- +(0, -0.5);
        \draw (A2.south) -- +(0, -0.5);
        \draw (A3.south) -- +(0, -0.5) node[below] {\footnotesize \(n\)};
        \draw (A4.south) -- +(0, -0.5);
        \draw (A5.south) -- +(0, -0.5);
        \draw (A1) -- (A2) -- (A3) -- (A4) -- (A5);
        \draw (A1.west) -- +(-0.5, 0) node[left] {\(\ldots\)};
        \draw (A5.east) -- +(0.5, 0) node[right] {\(\ldots\)};
    \end{tikzpicture}
    ,
\end{align}
\end{subequations}
where \(\tilde{A}\) can either represent the same ground state as \(A\), or, in a symmetry-broken or topologically ordered phase, they can represent two different ground states.
In the former case, this state represents a plane wave of a local perturbation (described by the \(B_k\)~tensor) on top of the ground state with momentum~\(k\); in the latter case, this represents a domain-wall-type excitation between the two ground states.
(In the former case, we can also see this state as a \emph{tangent vector} to the manifold of uniform iMPSs~\(\ket{\Psi[A]}\)~\cite{haegeman2013a,vanderstraeten2019a}.)
To find the lowest-lying excited states for some momentum~\(k\), we can optimize the excitation energy of this state with respect to variational parameters in the tensor~\(B_k\), keeping \(A\) and \(\tilde{A}\) fixed.%
\footnote{It is important to note that the excitation ansatz is not a variational ansatz in the same sense as an (infinite) MPS, where the approximate ground-state energy is guaranteed to be greater than the exact energy.
Indeed, for the excitation ansatz, we can obtain an excitation energy \emph{lower} than the exact excitation energy.
This is owing to the fact that we are only using an approximation to the ground state in the first place: if the ground state were exact, then this would be a true variational ansatz and the numerical excitation energies will always be greater than the exact values.}
We write the tensor~\(B_k\) with a dependence on the momentum~\(k\) to emphasize that the form of this tensor will have to change with \(k\) in general.
In our tensor network diagrams, we use shaded shapes for these variational tensors to differentiate them from the background tensors, which we keep unshaded.

An equivalent way of writing an EA wave function~\eqref{eq:ea} is as an iMPS~\(\ket{\Psi[\mathcal{A}_k]}\) with a block-upper-triangular \(A\)~tensor
\begin{equation}\label{eq:ea-triangular}
    \mathcal{A}_k =
    \begin{pmatrix}
        \erm^{\irm k} A & B_k \\
        0 & \tilde{A} \\
    \end{pmatrix}
    .
\end{equation}
However, on a technical level, it is worth emphasizing that this state is noninjective, and has a norm that diverges linearly with system size~\(L\), and so it \emph{cannot} be treated in the same way as an injective iMPS\@.
Indeed, this noninjectivity is necessary to describe states with an intensive energy shift above the extensive ground state energy (otherwise, it would be impossible to separate from the choice of boundary: see Appendix~\ref{app:boundary}).
This form of an EA wave function is closely related to the matrix product operator (MPO) that creates a particle with momentum~\(k\)
\begin{equation}\label{eq:sma}
    \mathcal{W}_k[\hat{a}^\dagger] =
    \begin{pmatrix}
        \erm^{\irm k} \hat{I} & \hat{a}^\dagger \\
        0 & \hat{I} \\
    \end{pmatrix}
    ,
\end{equation}
where \(\hat{a}^\dagger\) is some generic creation operator.
This operator is the basis of the single-mode approximation (SMA)~\cite{bijl1941,feynman1954,feynman1956,girvin1986,arovas1988} for the dispersion relation of a single-particle excitation.
The single-particle EA wave function can thus be seen as a version of the SMA where the perturbation is instead made by directly varying a tensor in the iMPS, and this perturbation can be optimized independently for each \(k\).

This block-upper-triangular form can also be used to describe extensions of the excitation ansatz.
For instance, in order to increase the accuracy of an EA wave function, we can perturb a \emph{window} of \(N\) tensors~\cite{vanderstraeten2020} instead of using a single-tensor perturbation~\(B_k\): for example, using \(N = 3\),
\begin{equation}\label{eq:ea-multi-site-window}
    \ket{\Phi_k[\mathbf{B}]}
    = \sum_n \erm^{\irm kn}
    \begin{tikzpicture}[tensornetwork]
        \node[atensor, label=\(A\)] (A1) at (1, 0) {};
        \node[atensor, fill=lightgray, label=\(B_1\)] (A2) at (2, 0) {};
        \node[atensor, fill=lightgray, label=\(B_2\)] (A3) at (3, 0) {};
        \node[atensor, fill=lightgray, label=\(B_3\)] (A4) at (4, 0) {};
        \node[atensor, label=\(\tilde{A}\)] (A5) at (5, 0) {};
        \draw (A1.south) -- +(0, -0.5);
        \draw (A2.south) -- +(0, -0.5) node[below] {\footnotesize \(n\)};
        \draw (A3.south) -- +(0, -0.5);
        \draw (A4.south) -- +(0, -0.5);
        \draw (A5.south) -- +(0, -0.5);
        \draw (A1) -- (A2) -- (A3) -- (A4) -- (A5);
        \draw (A1.west) -- +(-0.5, 0) node[left] {\(\ldots\)};
        \draw (A5.east) -- +(0.5, 0) node[right] {\(\ldots\)};
    \end{tikzpicture}
    ,
\end{equation}
where we write \(\mathbf{B} = (B_1, B_2, B_3)\) to express the free parameters of the window with a single symbol (and we omit the explicit dependence of these tensors on momentum~\(k\) in place of the positional index).
In block-upper-triangular form, this state can be expressed as
\begin{equation}\label{eq:ea-multi-site-window-triangular}
    \mathcal{A}_k =
    \begin{pmatrix}
        \erm^{\irm k} A & B_1 \\
        & 0 & B_2 \\
        & & 0 & B_3 \\
        & & & \tilde{A} \\
    \end{pmatrix}
    .
\end{equation}
While this form still essentially describes a single-particle state, we can also describe multi-particle states using this block-triangular structure: given two single-particle EA wave functions \(\ket{\Phi_{k_1}[B_{k_1}]}\) and \(\ket{\Phi_{k_2}[C_{k_2}]}\), we can construct a two-particle ansatz of the form
\begin{align}
    &\phantom{=} \ket{\Upsilon_{k_1,k_2}[q_1,q_2,W]} \nonumber\\
    &= q_1 \sum_{m>n} \erm^{\irm (k_1n+k_2m)}
    \begin{tikzpicture}[tensornetwork, scale=0.75]
        \renewcommand{\tensorsize}{7.5pt}
        \node[atensor] (A1) at (1, 0) {};
        \node[atensor, fill=lightgray, label={\footnotesize\(B_{k_1}\)}] (A2) at (2, 0) {};
        \node[atensor] (A3) at (3, 0) {};
        \node[notensor] (A4) at (4, 0) {\footnotesize\(\ldots\)};
        \node[atensor] (A5) at (5, 0) {};
        \node[atensor, fill=lightgray, label={\footnotesize\(C_{k_2}\)}] (A6) at (6, 0) {};
        \node[atensor] (A7) at (7, 0) {};
        \draw (A1.south) -- +(0, -0.5);
        \draw (A2.south) -- +(0, -0.5) node[below] {\scriptsize\(n\)};
        \draw (A3.south) -- +(0, -0.5);
        \draw (A5.south) -- +(0, -0.5);
        \draw (A6.south) -- +(0, -0.5) node[below] {\scriptsize\(m\)};
        \draw (A7.south) -- +(0, -0.5);
        \draw (A1) -- (A2) -- (A3) -- (A4) -- (A5) -- (A6) -- (A7);
        \draw (A1.west) -- +(-0.5, 0) node[left] {\footnotesize\(\ldots\)};
        \draw (A7.east) -- +(0.5, 0) node[right] {\footnotesize\(\ldots\)};
    \end{tikzpicture}
    \nonumber\\
    &+ q_2 \sum_{m<n} \erm^{\irm (k_1n+k_2m)}
    \begin{tikzpicture}[tensornetwork, scale=0.75]
        \renewcommand{\tensorsize}{7.5pt}
        \node[atensor] (A1) at (1, 0) {};
        \node[atensor, fill=lightgray, label={\footnotesize\(C_{k_2}\)}] (A2) at (2, 0) {};
        \node[atensor] (A3) at (3, 0) {};
        \node[notensor] (A4) at (4, 0) {\footnotesize\(\ldots\)};
        \node[atensor] (A5) at (5, 0) {};
        \node[atensor, fill=lightgray, label={\footnotesize\(B_{k_1}\)}] (A6) at (6, 0) {};
        \node[atensor] (A7) at (7, 0) {};
        \draw (A1.south) -- +(0, -0.5);
        \draw (A2.south) -- +(0, -0.5) node[below] {\scriptsize\(m\)};
        \draw (A3.south) -- +(0, -0.5);
        \draw (A5.south) -- +(0, -0.5);
        \draw (A6.south) -- +(0, -0.5) node[below] {\scriptsize\(n\)};
        \draw (A7.south) -- +(0, -0.5);
        \draw (A1) -- (A2) -- (A3) -- (A4) -- (A5) -- (A6) -- (A7);
        \draw (A1.west) -- +(-0.5, 0) node[left] {\footnotesize\(\ldots\)};
        \draw (A7.east) -- +(0.5, 0) node[right] {\footnotesize\(\ldots\)};
    \end{tikzpicture}
    \nonumber\\
    &+ \sum_{n} \erm^{\irm (k_1+k_2)n}
    \begin{tikzpicture}[tensornetwork, scale=0.75]
        \renewcommand{\tensorsize}{7.5pt}
        \node[atensor] (A1) at (1, 0) {};
        \node[atensor] (A2) at (2, 0) {};
        \node[atensor] (A3) at (3, 0) {};
        \node[atensor, fill=lightgray, label={\footnotesize\(W\)}] (A4) at (4, 0) {};
        \node[atensor] (A5) at (5, 0) {};
        \node[atensor] (A6) at (6, 0) {};
        \node[atensor] (A7) at (7, 0) {};
        \draw (A1.south) -- +(0, -0.5);
        \draw (A2.south) -- +(0, -0.5);
        \draw (A3.south) -- +(0, -0.5);
        \draw (A4.south) -- +(0, -0.5) node[below] {\scriptsize\(n\)};
        \draw (A5.south) -- +(0, -0.5);
        \draw (A6.south) -- +(0, -0.5);
        \draw (A7.south) -- +(0, -0.5);
        \draw (A1) -- (A2) -- (A3) -- (A4) -- (A5) -- (A6) -- (A7);
        \draw (A1.west) -- +(-0.5, 0) node[left] {\footnotesize\(\ldots\)};
        \draw (A7.east) -- +(0.5, 0) node[right] {\footnotesize\(\ldots\)};
    \end{tikzpicture}
    ,
\end{align}
with the compact block-triangular representation%
\footnote{This is analogous to taking the product of two particle creation operators with momenta \(k_1\) and \(k_2\)~\eqref{eq:sma}, which is given by the matrix direct product of the two MPOs
\begin{equation*}
    \mathcal{W}_{k_1}[\hat{b}^\dagger] \otimes \mathcal{W}_{k_2}[\hat{c}^\dagger] =
    \begin{pmatrix}
        \erm^{\irm (k_1+k_2)} \hat{I} & \erm^{\irm k_2} \hat{b}^\dagger & \erm^{\irm k_1} \hat{c}^\dagger & \hat{b}^\dagger \hat{c}^\dagger \\
        & \erm^{\irm k_2} \hat{I} & & \hat{c}^\dagger \\
        & & \erm^{\irm k_1} \hat{I} & \hat{b}^\dagger \\
        & & & \hat{I} \\
    \end{pmatrix}
    .
\end{equation*}
}
\begin{equation}\label{eq:ea-2p-triangular}
    \mathcal{A}_{k_1,k_2} =
    \begin{pmatrix}
        \erm^{\irm (k_1+k_2)} A & q_1 B_{k_1} & q_2 C_{k_2} & W \\
        & \erm^{\irm k_2} A & & C_{k_2} \\
        & & \erm^{\irm k_1} A & B_{k_1} \\
        & & & A \\
    \end{pmatrix}
    .
\end{equation}
This is essentially a simplified version of the two-particle ansatz presented in Refs~\cite{vanderstraeten2014,vanderstraeten2015a}.
Here, the variational parameters to be optimized are the scalars \(q_1\) and \(q_2\), which describe the relative phase and amplitude for the components where \(B\) is on the left and \(C\) is on the right and vice versa, respectively, and the tensor~\(W\), which describes the ‘correction’ required to the wave function as the particles approach each other and start to deform, as the solutions \(B\) and \(C\) derived for isolated excitations will no longer be valid.
States describing three or more particles could also be constructed analogously.
Quite importantly, this similarity in the block-upper-triangular structure of EA wave functions to that of MPOs allows us to extend methods for efficiently applying MPOs onto regular iMPSs~\cite{michel2010} to the case of EA wave functions, which we discuss in Sec.~\ref{sec:mpo}.

\subsection{Gauge fixing}\label{sec:gauge-fixing}
When working with MPSs in practical applications, it is important to note that there is a form of gauge redundancy, in that there are a class of transformations acting on the tensors making up an MPS that leaves the overall state unchanged.
For instance, an iMPS~\(\ket{\Psi[A]}\) is unchanged under transformations of the form \(A^s \mapsto M A^s M^{-1}\), for some invertible matrix~\(M\).
The correct fixing of these gauge degrees of freedom is vital to ensure the stability and efficiency of MPS-based algorithms.
For an iMPS, we can partially fix the gauge degrees of freedom by enforcing either the \emph{left-} or the \emph{right-orthogonality} constraints:
\begin{subequations}\label{eq:orthogonality-conditions}
\begin{align}
    \sum_s A^{s\dagger} A^s &= I \quad\text{(left-orthogonal)}, \\
    \sum_s A^s A^{s\dagger} &= I \quad\text{(right-orthogonal)},
\end{align}
\end{subequations}
or, in diagrammatic notation (writing left- and right-orthogonal tensors with a semicircular shape),
\begin{subequations}
\begin{align}
    \begin{tikzpicture}[tensornetwork]
        \node[ltensor] (A1) at (0, 1) {};
        \node[ltensor] (A1conj) at (0, -1) {};
        \draw (A1.south) -- (A1conj.north);
        \draw[rounded corners] (A1.west) -- +(-0.5, 0) -- +(-0.5, -2) -- (A1conj.west);
        \draw (A1.east) -- +(0.5, 0);
        \draw (A1conj.east) -- +(0.5, 0);
    \end{tikzpicture}
    &=
    \begin{tikzpicture}[tensornetwork]
        \draw[rounded corners] (0, 1) -- +(-0.5, 0) -- +(-0.5, -2) -- +(0, -2);
    \end{tikzpicture}
    \quad\text{(left-orthogonal)}, \\
    \begin{tikzpicture}[tensornetwork]
        \node[rtensor] (A1) at (0, 1) {};
        \node[rtensor] (A1conj) at (0, -1) {};
        \draw (A1.south) -- (A1conj.north);
        \draw[rounded corners] (A1.east) -- +(0.5, 0) -- +(0.5, -2) -- (A1conj.east);
        \draw (A1.west) -- +(-0.5, 0);
        \draw (A1conj.west) -- +(-0.5, 0);
    \end{tikzpicture}
    &=
    \begin{tikzpicture}[tensornetwork]
        \draw[rounded corners] (0, 1) -- +(0.5, 0) -- +(0.5, -2) -- +(0, -2);
    \end{tikzpicture}
    \quad\text{(right-orthogonal)}.
\end{align}
\end{subequations}
There is also a \emph{mixed-orthogonal} form where we write all of the sites to the left of a certain tensor or bond in left-orthogonal form~\(A_L\), and all of the sites to the right in right-orthogonal form~\(A_R\), that is,
\begin{subequations}\label{eq:mixed-orthogonal}
\begin{align}
    \ket{\Psi} &=
    \begin{tikzpicture}[tensornetwork]
        \node[ltensor, label=\(A_L\)] (A1) at (1, 0) {};
        \node[ltensor, label=\(A_L\)] (A2) at (2, 0) {};
        \node[atensor, label=\(A_C\)] (A3) at (3, 0) {};
        \node[rtensor, label=\(A_R\)] (A4) at (4, 0) {};
        \node[rtensor, label=\(A_R\)] (A5) at (5, 0) {};
        \draw (A1.south) -- +(0, -0.5);
        \draw (A2.south) -- +(0, -0.5);
        \draw (A3.south) -- +(0, -0.5);
        \draw (A4.south) -- +(0, -0.5);
        \draw (A5.south) -- +(0, -0.5);
        \draw (A1) -- (A2) -- (A3) -- (A4) -- (A5);
        \draw (A1.west) -- +(-0.5, 0) node[left] {\(\ldots\)};
        \draw (A5.east) -- +(0.5, 0) node[right] {\(\ldots\)};
    \end{tikzpicture}
    \\
    &=
    \begin{tikzpicture}[tensornetwork]
        \node[ltensor, label=\(A_L\)] (A1) at (1, 0) {};
        \node[ltensor, label=\(A_L\)] (A2) at (2, 0) {};
        \node[ctensor, label=\(C\)] (A3) at (3, 0) {};
        \node[rtensor, label=\(A_R\)] (A4) at (4, 0) {};
        \node[rtensor, label=\(A_R\)] (A5) at (5, 0) {};
        \draw (A1.south) -- +(0, -0.5);
        \draw (A2.south) -- +(0, -0.5);
        \draw (A4.south) -- +(0, -0.5);
        \draw (A5.south) -- +(0, -0.5);
        \draw (A1) -- (A2) -- (A3) -- (A4) -- (A5);
        \draw (A1.west) -- +(-0.5, 0) node[left] {\(\ldots\)};
        \draw (A5.east) -- +(0.5, 0) node[right] {\(\ldots\)};
    \end{tikzpicture}
    .
\end{align}
\end{subequations}
we then call this central tensor~\(A_C\) or bond matrix~\(C\) the \emph{orthogonality centre} (the semicircular parts of the left- and right-orthogonal tensors point towards this centre).
Analogous to this mixed-orthogonal form, for a (single-particle) EA wave function~\eqref{eq:ea}, we use the left-orthogonal form for the left boundary tensor~\(A\), and the right-orthogonal form for the right boundary tensor~\(\tilde{A}\)~\cite{zauner-stauber2018b}.
(These orthogonal forms are sometimes called \emph{canonical forms} interchangeably in the MPS literature, but we instead define a \emph{canonical} form as a type of orthogonal form with additional constraints, which we explain further in Appendix~\ref{app:canonicality}: in our terminology, the individual components of an EA wave function are generally \emph{not} in a canonical form.)

As well as the above \emph{multiplicative} gauge freedom of the \(A\)~tensor in an iMPS, there is also an \emph{additive} gauge freedom in the \(B\)~tensor of the EA wave function~\eqref{eq:ea}, in that the wave function is unchanged under the transformation \(B_k^s \mapsto B_k^s + A^s M - \erm^{-\irm k} M \tilde{A}^s\) for some arbitrary matrix \(M\).
To fix this gauge degree of freedom, we can enforce the either the \emph{left-} or \emph{right-gauge-fixing conditions}~\cite{haegeman2013a}%
\footnote{For \(k = 0\) and when \(A\) and \(\tilde{A}\) represent the same wave function, these conditions also ensure that the state is orthogonal to the background wave function~\(\ket{\Psi[A]}\).}
\begin{equation}\label{eq:gauge-fixing}
    \begin{tikzpicture}[tensornetwork]
        \node[atensor, fill=lightgray, label=\(B\)] (A1) at (0, 1) {};
        \node[ltensor, label=below:\(A\)] (A1conj) at (0, -1) {};
        \draw (A1.south) -- (A1conj.north);
        \draw[rounded corners] (A1.west) -- +(-0.5, 0) -- +(-0.5, -2) -- (A1conj.west);
        \draw (A1.east) -- +(0.5, 0);
        \draw (A1conj.east) -- +(0.5, 0);
    \end{tikzpicture}
    = 0 \quad\text{or}\quad
    \begin{tikzpicture}[tensornetwork]
        \node[atensor, fill=lightgray, label=\(B\)] (A1) at (0, 1) {};
        \node[rtensor, label=below:\(\tilde{A}\)] (A1conj) at (0, -1) {};
        \draw (A1.south) -- (A1conj.north);
        \draw[rounded corners] (A1.east) -- +(0.5, 0) -- +(0.5, -2) -- (A1conj.east);
        \draw (A1.west) -- +(-0.5, 0);
        \draw (A1conj.west) -- +(-0.5, 0);
    \end{tikzpicture}
    = 0,
\end{equation}
respectively.
In order to explicitly enforce these conditions, we define the left and right null spaces \(N_L^s\) and \(N_R^s\) of \(A^s\) and \(\tilde{A}^s\), respectively.
To derive \(N_L^s\), we reshape \(A^s\) (which is in left-orthogonal form) into a \(dD\times D\) isometric matrix \(A\) by combining the physical index \(s\) with the left virtual index, then \(N_L\) is its orthogonal complement, that is the \(dD\times (d-1)D\) isometric matrix such that the matrix formed by combining the columns of \(A\) and \(N_L\), namely \(\begin{pmatrix} A & N_L \\ \end{pmatrix}\), is a \(dD\times dD\) unitary matrix (\(N_R\) is defined similarly, with left and right switched where appropriate).
To automatically satisfy the left- or right-gauge-fixing condition, we write
\begin{equation}\label{eq:gauge-fixing-null-space}
    B^s =
    \begin{tikzpicture}[tensornetwork]
        \node[ltensor, striped, label=\(N_L\)] (A1) at (1, 0) {};
        \node[atensor, fill=lightgray, label=\(X\)] (A2) at (2, 0) {};
        \draw (A1.south) -- +(0, -0.5);
        \draw (A1.west) -- +(-0.5, 0);
        \draw (A2.east) -- +(0.5, 0);
        \draw (A1) -- (A2);
    \end{tikzpicture}
    \quad\text{or}\quad
    B^s =
    \begin{tikzpicture}[tensornetwork]
        \node[atensor, fill=lightgray, label=\(Y\)] (A1) at (1, 0) {};
        \node[rtensor, striped, label=\(N_R\)] (A2) at (2, 0) {};
        \draw (A2.south) -- +(0, -0.5);
        \draw (A1.west) -- +(-0.5, 0);
        \draw (A2.east) -- +(0.5, 0);
        \draw (A1) -- (A2);
    \end{tikzpicture}
    ,
\end{equation}
respectively, for some \((d-1)D\times D\) matrix \(X\) or \(D\times (d-1)D\) matrix \(Y\) (note that we use a striped pattern to distinguish the fixed null-space tensors from the free \(X\) and \(Y\) matrices).
In what follows, without any loss of generality, we will use the left-gauge-fixing condition.

When we explicitly write the EA window in this form, keeping \(N_L\) fixed and only changing the entries of the \(X\) matrix, the left-gauge-fixing condition will always be satisfied, and so we generally do not need to worry about having to re-enforce this condition after doing some operation to the state.
In some other situations, we may need to transform an arbitrary EA wave function into one of these gauge conditions (e.g.\ we may want to transform from the left to the right gauge): we have outlined the procedure for doing this in Appendix~\ref{app:ea-gauge-transform}.

To see the utility of these gauge-fixing conditions, we calculate the inner product of two EA wave functions in the left gauge (with the same background wave functions) \(\braket{\Phi_k[\tilde{B}]|\Phi_k[B]}\), with \(B^s = N_L^s X\) and \(\tilde{B}^s = N_L^s \tilde{X}\).
This is a double sum over all possible positions of the \(B\)~tensors in the bra and ket wave functions, but because of the left-gauge-fixing condition \eqref{eq:gauge-fixing}, any summand where the bra and ket \(B\)~tensors are on different sites will be zero.
And so because of the left- and right-orthogonality of the boundaries, this inner product reduces to
\begin{equation}\label{eq:ea-inner-prod}
    \braket{\Phi_k[\tilde{B}]|\Phi_k[B]} = \sum_n
    \begin{tikzpicture}[tensornetwork]
        \node[ltensor, striped, label=\(N_L\)] (A1) at (1, 1) {};
        \node[atensor, fill=lightgray, label=\(X\)] (A2) at (2, 1) {};
        \node[ltensor, striped, label=below:\(N_L\)] (A1conj) at (1, -1) {};
        \node[atensor, fill=lightgray, label=below:\(\tilde{X}\)] (A2conj) at (2, -1) {};
        \draw (A1) -- (A2);
        \draw (A1conj) -- (A2conj);
        \draw (A1) -- (A1conj);
        \draw[rounded corners] (A1.west) -- +(-0.5, 0) -- +(-0.5, -2) -- (A1conj.west);
        \draw[rounded corners] (A2.east) -- +(0.5, 0) -- +(0.5, -2) -- (A2conj.east);
    \end{tikzpicture}
    = \sum_n
    \begin{tikzpicture}[tensornetwork]
        \node[atensor, fill=lightgray, label=\(X\)] (A2) at (2, 1) {};
        \node[atensor, fill=lightgray, label=below:\(\tilde{X}\)] (A2conj) at (2, -1) {};
        \draw[rounded corners] (A2.west) -- +(-0.5, 0) -- +(-0.5, -2) -- (A2conj.west);
        \draw[rounded corners] (A2.east) -- +(0.5, 0) -- +(0.5, -2) -- (A2conj.east);
    \end{tikzpicture}
\end{equation}
(using the fact that \(N_L\) is an isometry).
That is to say, the inner product \(\braket{\Phi_k[\tilde{B}]|\Phi_k[B]}\) is proportional to the Frobenius inner product of \(X\) and \(\tilde{X}\), and we call a state ‘normalized’ when \(X\) has a Frobenius norm of one (although the squared norm of the state itself will diverge linearly with system size~\(L\)).
This is quite important, as it makes the effective eigenvalue problem for the lowest-energy states in terms of the variational parameters in \(X\) into an ordinary eigenvalue problem, which is typically easier to solve numerically than a generalized eigenvalue problem with an arbitrary norm matrix.

For EA wave functions with \(N\)-site windows~\eqref{eq:ea-multi-site-window}, to enforce the left-gauge-fixing condition we only need to ensure that the first site in the window is written in the form \(B_1^{s_1} = N_L^{s_1} X\).
The \(X\)~matrix can then be incorporated into the second tensor~\(B^{s_2}_2\), so we can write \(N_L^{s_1} B_2^{s_2} \cdots B_N^{s_N}\) (for the right-gauge-fixing condition, we would replace the last \(B\)~tensor with \(N_R\) instead)~\cite{haegeman2013a,vanderstraeten2020}.
As this \(N_L\)~tensor is fixed, for an \(N\)-site window the remaining \(N-1\) tensors contain all the degrees of freedom of the ansatz; and for a single-site window, the degrees of freedom are contained in a \((d-1)D \times D\) matrix~\(X\)~\eqref{eq:gauge-fixing-null-space}.
These remaining tensors are then written in mixed-orthogonal form~\eqref{eq:mixed-orthogonal} with the orthogonality centre somewhere within the window; for example, for \(N = 5\) with the orthogonality centre on site~\(4\), the window takes the form
\begin{equation}
    \begin{tikzpicture}[tensornetwork]
        \node[ltensor, striped, label=\(N_L\)] (A1) at (1, 0) {};
        \node[ltensor, fill=lightgray, label=\(B_2\)] (A2) at (2, 0) {};
        \node[ltensor, fill=lightgray, label=\(B_3\)] (A3) at (3, 0) {};
        \node[atensor, fill=lightgray, label=\(B_4\)] (A4) at (4, 0) {};
        \node[rtensor, fill=lightgray, label=\(B_5\)] (A5) at (5, 0) {};
        \draw (A1.south) -- +(0, -0.5);
        \draw (A2.south) -- +(0, -0.5);
        \draw (A3.south) -- +(0, -0.5);
        \draw (A4.south) -- +(0, -0.5);
        \draw (A5.south) -- +(0, -0.5);
        \draw (A1) -- (A2) -- (A3) -- (A4) -- (A5);
        \draw (A1.west) -- +(-0.5, 0);
        \draw (A5.east) -- +(0.5, 0);
    \end{tikzpicture}
    .
\end{equation}
Using this left-gauge-fixing condition, the only nonzero terms in the inner product of two EA wave functions are the ones where the null-space matrices of the bra and ket are on the same site, even if the windows have different lengths; for example, for a ket with a five-site window and a bra with a three-site window,
\begin{subequations}
\begin{align}
    \braket{\Phi_k[\tilde{B}]|\Phi_k[B]} &= \sum_n
    \begin{tikzpicture}[tensornetwork]
        \node[ltensor, striped] (A1) at (1, 1) {};
        \node[ltensor, fill=lightgray] (A2) at (2, 1) {};
        \node[ltensor, fill=lightgray] (A3) at (3, 1) {};
        \node[ltensor, fill=lightgray] (A4) at (4, 1) {};
        \node[atensor, fill=lightgray] (A5) at (5, 1) {};
        \node[ltensor, striped] (A1conj) at (1, -1) {};
        \node[ltensor, fill=lightgray] (A2conj) at (2, -1) {};
        \node[atensor, fill=lightgray] (A3conj) at (3, -1) {};
        \node[rtensor] (A4conj) at (4, -1) {};
        \node[rtensor] (A5conj) at (5, -1) {};
        \draw (A1) -- (A2) -- (A3) -- (A4) -- (A5);
        \draw (A1conj) -- (A2conj) -- (A3conj) -- (A4conj) -- (A5conj);
        \draw (A1) -- (A1conj);
        \draw (A2) -- (A2conj);
        \draw (A3) -- (A3conj);
        \draw (A4) -- (A4conj);
        \draw (A5) -- (A5conj);
        \draw[rounded corners] (A1.west) -- +(-0.5, 0) -- +(-0.5, -2) -- (A1conj.west);
        \draw[rounded corners] (A5.east) -- +(0.5, 0) -- +(0.5, -2) -- (A5conj.east);
    \end{tikzpicture}
    \\
    &= \sum_n
    \begin{tikzpicture}[tensornetwork]
        \node[ltensor, fill=lightgray] (A2) at (2, 1) {};
        \node[ltensor, fill=lightgray] (A3) at (3, 1) {};
        \node[ltensor, fill=lightgray] (A4) at (4, 1) {};
        \node[atensor, fill=lightgray] (A5) at (5, 1) {};
        \node[ltensor, fill=lightgray] (A2conj) at (2, -1) {};
        \node[atensor, fill=lightgray] (A3conj) at (3, -1) {};
        \node[rtensor] (A4conj) at (4, -1) {};
        \node[rtensor] (A5conj) at (5, -1) {};
        \draw (A2) -- (A3) -- (A4) -- (A5);
        \draw (A2conj) -- (A3conj) -- (A4conj) -- (A5conj);
        \draw (A2) -- (A2conj);
        \draw (A3) -- (A3conj);
        \draw (A4) -- (A4conj);
        \draw (A5) -- (A5conj);
        \draw[rounded corners] (A2.west) -- +(-0.5, 0) -- +(-0.5, -2) -- (A2conj.west);
        \draw[rounded corners] (A5.east) -- +(0.5, 0) -- +(0.5, -2) -- (A5conj.east);
    \end{tikzpicture}
    .
\end{align}
\end{subequations}

\section{MPO fixed-point equations}\label{sec:mpo}
A wave function by itself is not of much use to us if we cannot calculate expectation values of operators.
For ordinary MPSs, we can efficiently apply operators by writing them as \emph{matrix product operators} (MPOs), which can either have a finite support, or for an infinite system, we can have a translation-invariant MPO, in which case the expectation values can be calculated efficiently using a recursive algorithm by writing the MPO in a triangular form~\cite{michel2010}.
As we have shown in Sec.~\ref{sec:ea}, EA wave functions can also be expressed in a triangular form \eqref{eq:ea-triangular}, which allows us to extend this algorithm to calculate expectation values of EA wave functions.

Given an EA wave function in block-upper-triangular form~\(\mathcal{A}\)~\eqref{eq:ea-triangular} and an upper-triangular MPO~\(\mathcal{W}\), we define the generalized transfer operator%
\footnote{For a mixed expectation value between two different wave functions, we can replace the \(\mathcal{A}\) or \(\mathcal{A}^*\) tensor with that of a different EA wave function.}%
\begin{subequations}\label{eq:transfer-operator}
\begin{align}
    T^{\omega\alpha\beta}_{\omega'\alpha'\beta'} &= \sum_{ss'} (\mathcal{A}^{\beta s'}_{\beta'})^* \otimes \mathcal{W}^{\omega ss'}_{\omega'} \otimes \mathcal{A}^{\alpha s}_{\alpha'} \\
    &=
    \begin{tikzpicture}[tensornetwork]
        \node[atensor, label=above:\(\mathcal{A}\)]      (A) at (0, 1) {};
        \node[wtensor, label={right, yshift=5pt}:\(\mathcal{W}\)] (W) at (0, 0) {};
        \node[atensor, label=below:\(\mathcal{A}^*\)]    (Aconj) at (0, -1) {};
        \draw[thick] (A.west) -- +(-0.5, 0)     node[left, text height=0.5ex] {\(\scriptstyle\alpha'\)};
        \draw[thick] (A.east) -- +(0.5, 0)      node[right, text height=0.5ex] {\(\scriptstyle\alpha\)};
        \draw (W.west) -- +(-0.5, 0)            node[left, text height=0.5ex] {\(\scriptstyle\omega'\)};
        \draw (W.east) -- +(0.5, 0)             node[right, text height=0.5ex] {\(\scriptstyle\omega\)};
        \draw[thick] (Aconj.west) -- +(-0.5, 0) node[left, text height=0.5ex] {\(\scriptstyle\beta'\)};
        \draw[thick] (Aconj.east) -- +(0.5, 0)  node[right, text height=0.5ex] {\(\scriptstyle\beta\)};
        \draw (A) -- (W) -- (Aconj);
    \end{tikzpicture}
    .
\end{align}
\end{subequations}
Here, we use Greek indices to represent the virtual indices of the MPO and the block indices of the triangular form of the EA wave functions, where the upper (lower) indices correspond to bonds pointing right (left) in the tensor network diagram, so that, for example, \(\mathcal{A}^{1s}_0 = B^s\) in Eq.~\eqref{eq:ea-triangular}.
This generalized transfer operator~\(T^{\omega\alpha\beta}_{\omega'\alpha'\beta'}\) acts on the \(D^2\)-dimensional space given by the tensor product of the \(D\)-dimensional bra and ket MPS virtual spaces.
In order to find the expectation value for this operator, we will need to solve for the left or right fixed point of this transfer operator, which will satisfy the equations
\begin{subequations}
\begin{align}
    \pbra{E^{\omega\alpha\beta}(n+1)} &= \sum_{\omega'\alpha'\beta'} \pbra{E^{\omega'\alpha'\beta'}(n)} T^{\omega\alpha\beta}_{\omega'\alpha'\beta'} \\
    \begin{tikzpicture}[tensornetwork]
        \coordinate (A) at (0, 1) {};
        \coordinate (W) at (0, 0) {};
        \coordinate (Aconj) at (0, -1) {};
        \node[etensor, label={left, text height=0.5em}:\(\pbra{E(n+1)}\)] (E) at (-1, 0) {};
        \draw[thick] (E.east |- A) -- +(0.5, 0) node[right, text height=0.5ex] {\(\scriptstyle\alpha\)};
        \draw (E.east |- W) -- +(0.5, 0) node[right, text height=0.5ex] {\(\scriptstyle\omega\)};
        \draw[thick] (E.east |- Aconj) -- +(0.5, 0) node[right, text height=0.5ex] {\(\scriptstyle\beta\)};
    \end{tikzpicture}
    &=
    \begin{tikzpicture}[tensornetwork]
        \node[atensor] (A) at (0, 1) {};
        \node[wtensor] (W) at (0, 0) {};
        \node[atensor] (Aconj) at (0, -1) {};
        \node[etensor, label={left, text height=0.5em}:\(\pbra{E(n)}\)] (E) at (-1, 0) {};
        \draw[thick] (A) -- (E.east |- A);
        \draw[thick] (A.east) -- +(0.5, 0)      node[right, text height=0.5ex] {\(\scriptstyle\alpha\)};
        \draw (W) -- (E.east |- W);
        \draw (W.east) -- +(0.5, 0)             node[right, text height=0.5ex] {\(\scriptstyle\omega\)};
        \draw[thick] (Aconj) -- (E.east |- Aconj);
        \draw[thick] (Aconj.east) -- +(0.5, 0)  node[right, text height=0.5ex] {\(\scriptstyle\beta\)};
        \draw (A) -- (W) -- (Aconj);
    \end{tikzpicture}
\end{align}
\end{subequations}
and
\begin{subequations}
\begin{align}
    \pket{F_{\omega'\alpha'\beta'}(n-1)} &= \sum_{\omega\alpha\beta} T^{\omega\alpha\beta}_{\omega'\alpha'\beta'} \pket{F_{\omega\alpha\beta}(n)} \\
    \begin{tikzpicture}[tensornetwork]
        \coordinate (A) at (0, 1) {};
        \coordinate (W) at (0, 0) {};
        \coordinate (Aconj) at (0, -1) {};
        \node[etensor, label={right, text height=0.5em}:\(\pket{F(n-1)}\)] (F) at (1, 0) {};
        \draw[thick] (F.west |- A) -- +(-0.5, 0) node[left, text height=0.5ex] {\(\scriptstyle\alpha'\)};
        \draw (F.west |- W) -- +(-0.5, 0) node[left, text height=0.5ex] {\(\scriptstyle\omega'\)};
        \draw[thick] (F.west |- Aconj) -- +(-0.5, 0) node[left, text height=0.5ex] {\(\scriptstyle\beta'\)};
    \end{tikzpicture}
    &=
    \begin{tikzpicture}[tensornetwork]
        \node[atensor] (A) at (0, 1) {};
        \node[wtensor] (W) at (0, 0) {};
        \node[atensor] (Aconj) at (0, -1) {};
        \node[etensor, label={right, text height=0.5em}:\(\pket{F(n)}\)] (F) at (1, 0) {};
        \draw[thick] (A.west) -- +(-0.5, 0)     node[left, text height=0.5ex] {\(\scriptstyle\alpha'\)};
        \draw[thick] (A) -- (F.west |- A);
        \draw (W.west) -- +(-0.5, 0)            node[left, text height=0.5ex] {\(\scriptstyle\omega'\)};
        \draw (W) -- (F.west |- W);
        \draw[thick] (Aconj.west) -- +(-0.5, 0) node[left, text height=0.5ex] {\(\scriptstyle\beta'\)};
        \draw[thick] (Aconj) -- (F.west |- Aconj);
        \draw (A) -- (W) -- (Aconj);
    \end{tikzpicture}
    ,
\end{align}
\end{subequations}
respectively.%
\footnote{\(\pbra{E^{\omega\alpha\beta}}\) and \(\pket{F_{\omega\alpha\beta}}\) are (co)vectors in the \(D^2\)-dimensional dual virtual space, hence the similarity in our notation here to the bra–ket notation, although we will often interpret them as \(D\times D\) matrices mapping between the virtual spaces for the bra and ket wave functions.}
To obtain the expectation value, we only need to calculate one of \(\pbra{E}\) or \(\pket{F}\), but for calculating the action of the effective Hamiltonian for the EA algorithm, we need partial solutions for both \(\pbra{E}\) and \(\pket{F}\).
In the rest of this section, we will focus on the method for solving \(\pbra{E}\); solving \(\pket{F}\) follows a similar procedure, with the obvious modifications.

Since we write the MPO and the \(\mathcal{A}\)~tensor in block-upper-triangular form, the transfer operator is only nonzero when \(\omega' \leq \omega\), \(\alpha' \leq \alpha\), and \(\beta' \leq \beta\), so we can write
\begin{subequations}
\begin{align}
    \pbra{E^{\omega\alpha\beta}(n+1)} &= \sum_{\omega'\leq\omega} \sum_{\alpha'\leq\alpha} \sum_{\beta'\leq\beta} \pbra{E^{\omega'\alpha'\beta'}(n)} T^{\omega'\alpha'\beta'}_{\omega\alpha\beta} \\
    &= \pbra{E^{\omega\alpha\beta}(n)} T^{\omega\alpha\beta}_{\omega\alpha\beta} + \pbra{C^{\omega\alpha\beta}(n)}, \label{eq:diag}
\end{align}
\end{subequations}
where the off-diagonal terms are collected into
\begin{equation}\label{eq:off-diag}
    \pbra{C^{\omega\alpha\beta}(n)} = \underbrace{\sum_{\omega'\leq\omega} \sum_{\alpha'\leq\alpha} \sum_{\beta'\leq\beta}}_{\substack{\text{except for}\\(\omega',\alpha',\beta') = (\omega,\alpha,\beta)}} \pbra{E^{\omega'\alpha'\beta'}(n)} T^{\omega\alpha\beta}_{\omega'\alpha'\beta'}.
\end{equation}
Following these equations, we can solve for \(\pbra{E^{\omega\alpha\beta}(n)}\) recursively, starting with \(\pbra{E^{111}(n)}\)\footnote{Using 1 for the initial index of \(\omega\), \(\alpha\), and \(\beta\).}, which we set to be the left eigenvector of the transfer operator~\(T^{111}_{111}\) (this eigenvector will be the \(D\times D\) identity matrix if the left boundary is in left-orthogonal form).%
\footnote{This initial choice is helpful, but not required for the algorithm; see Appendix~\ref{app:boundary}.}

This \(\pbra{E^{\omega\alpha\beta}(n)}\)~matrix will be a polynomial in \(n\), whose maximum degree~\(P\) is bounded by the number of diagonal transfer operator terms~\(T^{\omega\alpha\beta}_{\omega\alpha\beta}\) whose principal eigenvalue is on the unit circle.
Furthermore, if there is a principal eigenvalue on the unit circle with a value \(\erm^{\irm q}\neq1\), then we will need to write \(E(n)\) as a Fourier series including a mode at this \(q\).
Thus, the general form of \(\pbra{E^{\omega\alpha\beta}(n)}\) is
\begin{equation}
    \pbra{E^{\omega\alpha\beta}(n)} = \sum_q \erm^{\irm qn} \sum_{p=0}^P n^p \pbra{E^{\omega\alpha\beta pq}},
\end{equation}
summing over all principal eigenvalues \(\erm^{\irm q}\) on the unit circle.
In terms of these polynomial and Fourier components, we can directly calculate the off-diagonal term~\eqref{eq:off-diag} as
\begin{equation}\label{eq:off-diag-pq}
    \pbra{C^{\omega\alpha\beta pq}} = \underbrace{\sum_{\omega'\leq\omega} \sum_{\alpha'\leq\alpha} \sum_{\beta'\leq\beta}}_{\substack{\text{except for}\\(\omega',\alpha',\beta') = (\omega,\alpha,\beta)}} \pbra{E^{\omega'\alpha'\beta'pq}} T^{\omega\alpha\beta}_{\omega'\alpha'\beta'}.
\end{equation}
The way that we handle the diagonal term in the fixed-point equation~\eqref{eq:diag} will depend on the nature of the largest eigenvalue of the transfer operator~\(T^{\omega\alpha\beta}_{\omega\alpha\beta}\) (since we are focusing on a fixed value of the indices \(\omega\), \(\alpha\), and \(\beta\), we omit them from the rest of this section for brevity).

The first and simplest case is when \(T = 0\), and so we simply have \(\pbra{E(n+1)}\) = \(\pbra{C(n)}\), hence, in terms of the polynomial and Fourier components \(p\) and \(q\), we have
\begin{equation}\label{eq:case-1}
    \pbra{E^{pq}} = \erm^{-\irm q} \pbra{C^{pq}} - \sum_{r=p+1}^P \binom{r}{p} \pbra{E^{rq}},
\end{equation}
which can be solved for each \(q\) independently, starting with \(p = P\) and iterating downwards to \(p = 0\).

The second case is when \(T\) is nonzero and has a spectral radius less than one, in which case we have \(\pbra{E(n+1)} = \pbra{E(n)} T + \pbra{C(n)}\), and in the \(pq\) components we obtain
\begin{equation}\label{eq:case-2}
    \pbra{E^{pq}} (I - \erm^{-\irm q} T) = \erm^{-\irm q} \pbra{C^{pq}} - \sum_{r=p+1}^P \binom{r}{p} \pbra{E^{rq}},
\end{equation}
which is again solved for each \(q\) by starting with \(p = P\) and iterating to \(p = 0\), using a linear solver such as the generalized minimal residual method (GMRES)~\cite{saad1986} to solve for \(\pbra{E^{pq}}\) (in Appendix~\ref{app:gmres}, we discuss some advanced considerations for optimizing the linear solver).

The third case is when \(T\) has a spectral radius of one, with a principal eigenvalue of \(\erm^{iQ}\) (and we assume that all other eigenvalues lie strictly inside the unit circle\footnote{The procedure for the case where there are multiple eigenvalues on the unit circle is similar to the one presented here, but we need to orthogonalize against all of the corresponding eigenvectors and keep track of multiple parallel parts.}).
We cannot simply use the linear equations from the second case~\eqref{eq:case-2} here, since the operator acting on the left-hand side would be singular.
In order to avoid this singularity, we start by finding the left and right eigenvectors \(\pbra{L}\) and \(\pket{R}\) corresponding to this transfer matrix eigenvalue, and decompose \(\pbra{C}\) into components parallel and perpendicular to these eigenvectors,%
\footnote{As \(\pbra{C^{pq}}\) is a covector and \(\pket{R}\) is a vector, their inner product \(\pbraket{C^{pq}|R}\) is a scalar; alternatively, interpreting \(\pbra{C^{pq}}\) and \(\pket{R}\) as \(D\times D\) matrices, \(\pbraket{C^{pq}|R}\) denotes their Frobenius inner product.}
\begin{subequations}
\begin{gather}\label{eq:case-3-decomposition}
    c_\parallel^{pq} = \pbraket{C^{pq}|R},
    \qquad
    \pbra{C_\parallel^{pq}} = c_\parallel^{pq} \pbra{L}, \\
    \pbra{C_\perp^{pq}} = \pbra{C^{pq}} - \pbra{C_\parallel^{pq}}.
\end{gather}
\end{subequations}
For the perpendicular components, we can solve the linear equations as in the second case above
\begin{equation}\label{eq:case-3-perp}
    \pbra{E_\perp^{pq}} (I - \erm^{-\irm q} T) = \erm^{-\irm q} \pbra{C_\perp^{pq}} - \sum_{r=p+1}^P \binom{r}{p} \pbra{E_\perp^{rq}},
\end{equation}
since the parts parallel to the eigenvector have been removed, making this linear problem nonsingular.
For the parallel components with \(q' \neq Q\), we have
\begin{equation}\label{eq:case-3-parallel-1}
    e_\parallel^{pq} = \frac{c_{\parallel}^{pq}}{\erm^{\irm q} - \erm^{\irm Q}},
\end{equation}
while for the component with \(q = Q\), the order of the polynomial is increased from \(P\) to \(P+1\), and we have
\begin{multline}\label{eq:case-3-parallel-2}
    e_\parallel^{(p+1)Q} = \frac{1}{p+1} \left[ \erm^{-\irm Q} c_\parallel^{pQ} - \sum_{r=p+2}^{P+1} \binom{r}{p} e^{rQ}_\parallel \right] \\
    - \sum_{q\neq Q} \frac{c_{\parallel}^{(p+1)q}}{\erm^{\irm q} - \erm^{\irm Q}},
\end{multline}
and we set the zero-degree component to be%
\footnote{Note that there is some freedom in the choice of this term, which we discuss in Appendix~\ref{app:boundary}.}
\begin{equation}\label{eq:case-3-parallel-3}
    e^{0Q}_\parallel = - \sum_{q\neq Q} \frac{c_{\parallel}^{0q}}{\erm^{\irm q} - \erm^{\irm Q}}.
\end{equation}
We can then combine the parallel and perpendicular results to get \(\pbra{E^{pq}} = e_\parallel^{pq}\pbra{L} + \pbra{E_\perp^{pq}}\).

\begin{algorithm}[t]
    \caption{Solving the MPO fixed-point equations for an excitation ansatz wave function.}
    \label{alg:fixed-point}
    \SetKwInOut{Input}{Input}\SetKwInOut{Output}{Output}
    \Input{EA wave function described by \(\mathcal{A}^{\alpha'\alpha s}\)~\eqref{eq:ea-triangular}, \\ MPO described by \(\mathcal{W}^{\omega'\omega ss'}\).}
    \Output{Left fixed-point polynomial \(\pbra{E^{\omega\alpha\beta pq}}\)~\eqref{eq:diag}.}
    \BlankLine
    Initialize \(\pbra{E^{11101}}\) to be left eigenvector \(\pbra{L}\) of \(T^{111}_{111}\). \\
    \For{\(\alpha, \beta = 1, \ldots, \dim\mathcal{A}\), and \(\omega = 1, \ldots, \dim \mathcal{W}\)}{
        Calculate \(\pbra{C^{\omega\alpha\beta pq}}\) \eqref{eq:off-diag-pq} \(\forall p, q\). \\
        \uIf{\(T^{\omega\alpha\beta}_{\omega\alpha\beta} = 0\)}{
            Evaluate \(\pbra{E^{\omega\alpha\beta pq}}\) \eqref{eq:case-1} \(\forall p, q\) (\(p\) descending). \\
        }
        \uElseIf{\(T\)’s spectral radius \(< 1\)}{
            Evaluate \(\pbra{E^{\omega\alpha\beta pq}}\) \eqref{eq:case-2} \(\forall p, q\) (\(p\) descending). \\
        }
        \uElseIf{\(T\)’s principal eigenvalue \(\erm^{\irm Q}\) on unit circle}{
            Calculate its eigenvectors \(\pbra{L}\) and \(\pket{R}\). \\
            Decompose \(\pbra{C^{\omega\alpha\beta pq}}\) into \(\pbra{C_\parallel^{\omega\alpha\beta pq}}\), \(\pbra{C_\perp^{\omega\alpha\beta pq}}\) \eqref{eq:case-3-decomposition} \(\forall p, q\). \\
            Evaluate \(\pbra{E_\perp^{\omega\alpha\beta pq}}\) \eqref{eq:case-3-perp} \(\forall p, q\) (\(p\) descending). \\
            Evaluate \(e_\parallel^{\omega\alpha\beta pq}\) \eqref{eq:case-3-parallel-1} \(\forall p, q\) (\(q \neq Q\)). \\
            Evaluate \(e_\parallel^{\omega\alpha\beta pQ}\) \eqref{eq:case-3-parallel-2}, \eqref{eq:case-3-parallel-3} \(\forall p\) (descending). \\
            Evaluate \(\pbra{E^{\omega\alpha\beta pq}} = e_\parallel^{\omega\alpha\beta pq} \pbra{L} + \pbra{E_\perp^{\omega\alpha\beta pq}}\) \(\forall p, q\).
        }
    }
\end{algorithm}

This method for calculating \(\pbra{E}\) is summarized in Algorithm~\ref{alg:fixed-point}.
We note that this algorithm reduces to the one presented in Ref.~\cite{michel2010} if we use a regular iMPS \(\mathcal{A} = A\) instead of an EA wave function.
It is also worth emphasizing that this algorithm is generic, and works for \emph{any} generalization of an EA wave function that can be written in a block-upper-triangular form, such as wave functions with multi-site windows~\eqref{eq:ea-multi-site-window-triangular}, or multi-particle wave functions with multiple windows~\eqref{eq:ea-2p-triangular}.

\subsection{Moments and cumulants}
Once we have calculated the final components of the fixed-point polynomial~\(\pbra{E^{\omega\alpha\beta}(n)}\) for \(\omega = \dim \mathcal{W}\) and \(\alpha, \beta = \dim\mathcal{A}\), then the part parallel to the left eigenvector~\(e_\parallel^{\omega\alpha\beta}(n)\) will give the expectation value of the operator represented by the MPO~\(\mathcal{W}\).%
\footnote{Note that if we are only interested in this expectation value, then we do not need to solve the equations for the parallel components \eqref{eq:case-3-perp} for this final element, which would otherwise be wasted computational effort.}
If the operator is a linearly extensive Hamiltonian \(\hat{H}\), then this (unnormalized) expectation value will have the form (for system size \(L\))
\begin{equation}
    e_\parallel^{\omega\alpha\beta}(L) = E L^2 + \Delta L,
\end{equation}
where \(E\) is the ground state energy density and \(\Delta\) is the energy offset owing to the excitation.
If the operator is the identity, then this will give the (squared) norm of the state, which will be \(L\) for a properly normalized state (by Eq.~\eqref{eq:ea-inner-prod}).
And so, the properly normalized expectation value of \(\hat{H}\) will be
\begin{equation}\label{eq:h-exp}
    \braket{\hat{H}} \equiv \frac{\braket{\Phi|\hat{H}|\Phi}}{\braket{\Phi|\Phi}} = EL + \Delta.
\end{equation}
The normalized expectation value of \(\hat{H}^2\) will give us the variance density of the ground state~\(\sigma_E^2\) and the variance offset owing to the excitation~\(\sigma_\Delta^2\),
\begin{subequations}
\begin{align}
    \braket{\hat{H}^2} &= \sigma_E^2 L + \sigma_\Delta^2 + \braket{\hat{H}}{}^2 \\
    &= E^2 L^2 + (2E\Delta + \sigma_E^2) L + (\Delta^2 + \sigma_\Delta^2).
\end{align}
\end{subequations}

In general, the \(n\)th moment~\(\mu_n\) (given by the normalized expectation value of \(\hat{H}^n\)) is the \(n\)th complete Bell polynomial of the first \(n\) cumulants \(\kappa_1, \ldots, \kappa_n\), which are linear polynomials in the system size~\(L\),
\begin{equation}
    \kappa_n \equiv \kappa^\text{bg}_n L + \kappa^\text{ex}_n.
\end{equation}
For instance, \(\kappa^\text{bg}_1\) is the background energy density~\(E\), \(\kappa^\text{ex}_1\) is the excitation energy~\(\Delta\), and \(\kappa^\text{bg}_2\) and \(\kappa^\text{ex}_2\) are the linear~\(\sigma^2_E\) and constant~\(\sigma^2_\Delta\) contributions to the variance, respectively.
Higher-order cumulants, such the variance, can be used to assess the error of the state, or (for the case of the ground state in particular) determine the position of critical points with much greater precision than the order parameter alone~\cite{west2015,pillay2019}.

Here, the linearly extensive terms~\(\kappa^\text{bg}_n L\) are the same as those for the background wave function itself (hence the superscript label)~\cite{west2015,pillay2019}, as the EA window cannot cause a change in the extensive properties of the state.
Hence, the background cumulants~\(\kappa^\text{bg}_n\) can be obtained just by considering the background by itself as an iMPS, in which case, the \(n\)th cumulant is simply the linear coefficient of the \(n\)th moment of the background wave function~\(\mu^\text{bg}_n\).
To extract the excitation cumulants~\(\kappa^\text{ex}_n\), it is sufficient to consider the constant component of the \(n\)th moment~\(\mu_n\),%
\footnote{In general, the higher-degree terms will combinations of the background and excitation cumulants, but since we already know the background cumulants, we will not gain anything more by looking at these higher-degree terms that we cannot get from the constant terms.}
which will be the complete Bell polynomial of the first \(n\) excitation cumulants \(\kappa^\text{ex}_1, \ldots, \kappa^\text{ex}_n\) (and hence we call this constant component the \emph{excitation moment}~\(\mu^\text{ex}_n\)).
We can obtain the \(n\)th excitation cumulant from the first \(n-1\) moments and cumulants and the \(n\)th moment by the recursion relation
\begin{equation}
    \kappa^\text{ex}_n = \mu^\text{ex}_n - \sum_{m=1}^{n-1} \binom{n-1}{m-1} \kappa^\text{ex}_m \mu^\text{ex}_{n-m}.
\end{equation}
Alternatively, since the \(n\)th moment~\(\mu_n\) is a degree-\(n\) polynomial whose degree-\(m\) term only depends on the background cumulants and the first \(n-m\) excitation cumulants, we can obtain all of the excitation cumulants recursively from a single moment \(\mu_n\) by starting from the highest-degree term and iterating downwards.%
\footnote{Something to note when using topologically nontrivial excitations, where the left and right backgrounds are different states, is that the two backgrounds will generally have slight differences in energy, variance and other quantities, which we would not expect in the exact states, but appear because we are using MPS approximations with a finite bond dimension.
This will result in slight differences in these quantities when calculating moments using either the \(\pbra{E}\) or \(\pket{F}\) matrices.
In principle, as we converge the background wave functions, these differences should shrink as well.
It is reasonable to use the average of the quantities obtained using the \(\pbra{E}\) and \(\pket{F}\) matrices to get a better estimate for finite bond dimension.}

\subsection{The excitation ansatz algorithm}\label{sec:ea-algorithm}
In addition to obtaining expectation values, we can also use the fixed points \(\pbra{E}\) and \(\pket{F}\) to express the effective Hamiltonian~\(\mathcal{H}_\text{eff}^k\) in the excitation ansatz algorithm~\cite{haegeman2012} in a very compact form.

The EA algorithm entails minimizing the excitation energy \(\braket{\Phi^k[B]|\hat{H}|\Phi^k[B]}\) for a normalized EA wave function~\(\ket{\Phi^k[B]}\) with momentum~\(k\).
By using the left-gauge-fixing condition and writing \(B^s = N_L^s X\), this generalized eigenvalue problem becomes an ordinary eigenvalue problem in terms of the variational parameters in \(X\) (as the inner product \(\braket{\Phi^k[\tilde{B}]|\Phi^k[B]}\) becomes proportional to the Frobenius inner product of \(X\) and \(\tilde{X}\)~\eqref{eq:ea-inner-prod}).
To solve this ordinary eigenvalue problem numerically, we need a way to calculate the effective action of the Hamiltonian~\(\hat{H}\) on \(B\), parameterized by \(X\), namely
\begin{subequations}
\begin{align}
    \mathcal{H}_\text{eff}^k B
    &= \frac{\partial}{\partial\tilde{B}} \lim_{L\rightarrow\infty} \frac{1}{L} \braket{\Phi^k[\tilde{B}]|\hat{H}|\Phi^k[B]} \\
    &=
    \begin{tikzpicture}[tensornetwork]
        \node[etensor, label={left, text height=0.5em}:\(\pbra{E^{\omega\alpha 1}}\)] (E) at (-1, 0) {};
        \node[atensor, label=above:\(\mathcal{A}\)] (A) at (0, 1) {};
        \node[wtensor, label={right, yshift=5pt, xshift=-2pt}:\(\mathcal{W}\)] (W) at (0, 0) {};
        \node[notensor] (Z) at (0, -1) {};
        \node[etensor, label={right, text height=0.5em}:\(\pket{F_{\omega'\alpha'2}}\)] (F) at (1, 0) {};
        \draw (E) -- (W) -- (F);
        \draw (A) -- (W) -- (Z);
        \draw[thick] (A) -- (E.east |- A.west);
        \draw[thick] (A) -- (F.west |- A.east);
        \draw (Z) -- (E.east |- Z.west);
        \draw (Z) -- (F.west |- Z.east);
    \end{tikzpicture}
    . \label{eq:heff}
\end{align}
\end{subequations}
We then find the lowest-energy eigenvector (i.e.\ \(B\)~tensor) using a numerical eigensolver, such as the Arnoldi algorithm (or we can look at the lowest \(n\) eigenvectors if we are interested in more than one state at that momentum).
Here, we write \(\pbra{E^{\omega\alpha\beta}}\) and \(\pket{F_{\omega'\alpha'\beta'}}\) to mean the constant zero-momentum components of the full polynomial expressions (which need to be corrected to remove any spurious constant shifts from the eigenvalues, as discussed in Appendix~\ref{app:heff-boundary}).

We will generally want to solve for the excitations over some range of momentum~\(k\), discretized in steps of \(\Delta k\).
We could either solve for each desired value of \(k\) in parallel, or we can do the calculations serially, using the solution at \(k\) as the initial guess for \(k + \Delta k\), which can significantly speed up the eigensolver, as the optimal \(B\)~tensor for momentum \(k\) should already be close to the optimum at \(k+\Delta k\) if the step~\(\Delta k\) is sufficiently small.

\subsection{EA DMRG}\label{sec:ea-dmrg}
To find the optimum EA wave function with a multi-site window, the original proposed method is to coarse-grain the entire window into a single tensor, which is then optimized as in the case of a single-site window~\cite{haegeman2013a}.
However, for local dimension~\(d\) and window size~\(N\), the computational cost of this approach scales exponentially \(O(d^N)\).
As proposed in Ref.~\cite{vanderstraeten2020}, this can be improved by decomposing this single window tensor into an MPS, and updating each tensor sequentially by sweeping from one end of the window to the other in a fashion similar to finite DMRG.
In this case, the computational cost per sweep scales linearly with the window size.
To update the \(n\)th tensor in the window~\(B_n\), we find the lowest eigenvector of the effective Hamiltonian (which has a similar form to the single-site effective Hamiltonian)
\begin{equation}\label{eq:ea-dmrg-heff}
    \mathcal{H}_\text{eff}^{n,k} B_n
    =
    \begin{tikzpicture}[tensornetwork]
        \node[etensor, label={left, text height=0.5em}:\(\pbra{E^{\omega\alpha n}}\)] (E) at (-1, 0) {};
        \node[atensor] (A) at (0, 1) {};
        \node[wtensor] (W) at (0, 0) {};
        \node[notensor] (Z) at (0, -1) {};
        \node[etensor, label={right, text height=0.5em}:\(\pket{F_{\omega'\alpha'(n+1)}}\)] (F) at (1, 0) {};
        \draw (E) -- (W) -- (F);
        \draw (A) -- (W) -- (Z);
        \draw[thick] (A) -- (E.east |- A.west);
        \draw[thick] (A) -- (F.west |- A.east);
        \draw (Z) -- (E.east |- Z.west);
        \draw (Z) -- (F.west |- Z.east);
    \end{tikzpicture}
    .
\end{equation}
We can perform this numerical optimization using the Arnoldi algorithm, as in the single-site EA algorithm.
In most other respects, this algorithm is identical to finite DMRG~\cite{schollwoeck2011}.
Since we enforce the left-gauge-fixing condition by replacing the first tensor in the window with the null-space tensor~\(N_L\), we only need to update sites \(2\) to \(N\) in the window.
With this gauge-fixing condition, we can find an initial wave function with window size~\(N\) by using a solution for a smaller window size (or a single-site window from the original EA algorithm) and incorporate sites into the right-hand side of the window from the right background wave function.
We give a summary of performing a single sweep in Algorithm~\ref{alg:ea-dmrg}.
We perform enough sweeps until we are satisfied with the convergence of the wave function, or we can increase the size of the window and continue sweeping if we want to increase the accuracy further.

\begin{algorithm}[t]
    \caption{A single EA DMRG sweep.}
    \label{alg:ea-dmrg}
    \SetKwInOut{Input}{Input}
    \Input{EA wave function with momentum \(k\) and an \(N\)-site window obeying the left-gauge-fixing condition \((N_L, B_2, B_3, \ldots, B_N)\).}
    \BlankLine
    Start with the window in mixed-orthogonal form~\eqref{eq:mixed-orthogonal} with the orthogonality centre on site \(N\). \\
    \For{\(n = N,N-1,\ldots,3\)}{
        Optimize \(B_n\) using \(\mathcal{H}_\text{eff}^{n,k}\)~\eqref{eq:ea-dmrg-heff}. \\
        Calculate SVD \(B_n^s = U D (V^\dagger)^s\). \\
        Move the orthogonality centre one site left: \(B_n^s \leftarrow (V^\dagger)^s\), \(B_{n-1}^s \leftarrow B_{n-1}^s U D\).
    }
    Optimize \(B_2\) using \(\mathcal{H}_\text{eff}^{2,k}\)~\eqref{eq:ea-dmrg-heff}. \\
    \For{\(n = 3,4,\ldots,N\)}{
        Calculate SVD \(B_{n-1}^s = U^s D V^\dagger\). \\
        Move the orthogonality centre one site right: \(B_{n-1}^s \leftarrow U^s\), \(B_n^s \leftarrow D V^\dagger B_n^s\). \\
        Optimize \(B_n\) using \(\mathcal{H}_\text{eff}^{n,k}\)~\eqref{eq:ea-dmrg-heff}.
    }
\end{algorithm}

\subsection{Variance minimization}\label{sec:variance-minimization}
Instead of minimizing the energy itself as discussed in the previous sections, it is also possible to find a state that minimizes some other quantity.
This is useful, for instance, when optimizing a single-particle excitation branch contained inside of a multi-particle continuum.
In some cases, we can distinguish this single-particle state from the multi-particle states by some conserved quantum number in the MPS or by some symmetry, which can be fixed by adding an energy penalty to the Hamiltonian proportional to the symmetry operator.
But generally, minimizing the energy will result in attempting to find the lowest-energy multi-particle states rather than the desired single-particle state.
However, the single-particle EA wave function is less apt in describing multi-particle states, and will typically have a much higher excitation variance than when the state being described has a single-particle character.
We can use this property to our advantage by minimizing the excitation variance itself rather than the energy to obtain these single-particle excitations, as proposed in Ref.~\cite{zauner-stauber2018b}.

In practical terms, we can just use the same algorithms for the excitation ansatz (Sec.~\ref{sec:ea-algorithm}) and EA DMRG (Sec.~\ref{sec:ea-dmrg}), but instead of minimizing the expectation value of the Hamiltonian, we instead minimize the variance
\begin{equation}
    (\hat{H} - (EL + \Delta))^2 = (\hat{H} - EL)^2 - 2 \Delta (\hat{H} - EL) + \Delta^2.
\end{equation}
Here, \(E\) is the background energy density, which is already known, but we need a value for the excitation energy~\(\Delta\).
In order to perform this optimization, we use some initial guess for \(\Delta\), which we refine as we converge the wave function.
If we are scanning over \(k\)~space, then we can use the wave function at the previous \(k\)~step to get an estimate for the energy for the current \(k\)~step, or we could try to scan over a range of values of \(\Delta\) to attempt to find some possible excitation with a single-particle character.
In order to converge the wave function, we first optimize the variance using this initial value of \(\Delta\), then once we have found the optimum for this \(\Delta\), we update \(\Delta\) to be the excitation energy of this optimum state, and we restart the optimization with this new \(\Delta\).
In this way, we should be able to continue to approach the desired minimal-variance state (provided it exists) to some reasonable degree of convergence, which we can quantify by the value of the energy variance.
Nevertheless, variance minimization will generally be much more tricky than energy minimization, as the optimization landscape is more complicated, and we may often get stuck in false minima or move away from the desired state.
So in practice, some manual adjustments and fine-tuning may be needed to obtain the best results.

This approach is similar in spirit to methods such as DMRG-X~\cite{khemani2016} and DMRG-S~\cite{zhang2023} for finding excited states by optimizing a quantity other than the energy.

\subsection{Momentum derivatives}\label{sec:momentum-derivatives}
Another interesting application of the block-triangular form of an EA wave function is in calculating derivatives of expectation values with respect to momentum (which can be used, for instance, to get the group velocity of the excitations).
For an EA wave function~\(\ket{\Phi_k}\), we have
\begin{equation}
    \frac{\drm}{\drm k} \braket{\Phi_k|\hat{O}|\Phi_k}
    = 2\operatorname{Re}\Braket{\Phi_k|\hat{O}|\frac{\drm\Phi_k}{\drm k}},
\end{equation}
where we evaluate
\begin{subequations}
\begin{align}
    \Ket{\frac{\drm\Phi_k}{\drm k}} &= \sum_n \cdots \mathcal{A}^{s_{n-1}} \frac{\drm \mathcal{A}^{s_n}}{\drm k} \mathcal{A}^{s_{n+1}} \cdots \\
    &= \cdots \mathcal{B}^{s_{-1}} \mathcal{B}^{s_{0}} \mathcal{B}^{s_{1}} \cdots,
\end{align}
\end{subequations}
defining the \(\mathcal{B}\)~tensor as
\begin{equation}
    \mathcal{B} =
    \begin{pmatrix}
        \mathcal{A} & \frac{\drm\mathcal{A}}{\drm k} \\
        & \mathcal{A} \\
    \end{pmatrix}
    .
\end{equation}
For a single-site window~\eqref{eq:ea-triangular}, we have
\begin{equation}
    \mathcal{B} =
    \begin{pmatrix}
        \erm^{\irm k} A & B & \irm \erm^{\irm k} A & \frac{\drm B}{\drm k} \\
        & \tilde{A} \\
        && \erm^{\irm k} A & B \\
        &&& \tilde{A} \\
    \end{pmatrix}
    ,
\end{equation}
and by removing redundant rows and columns, we can simplify this to obtain
\begin{equation}\label{eq:derivative-mps}
    \mathcal{B} =
    \begin{pmatrix}
        \erm^{\irm k} A & \irm \erm^{\irm k} A & \frac{\drm B}{\drm k} \\
        & \erm^{\irm k} A & B \\
        && \tilde{A} \\
    \end{pmatrix}
    .
\end{equation}
We note that this is similar to the form of a two-particle EA wave function~\eqref{eq:ea-multi-site-window-triangular}.

In general, the \(n\)th derivative can be calculated by
\begin{equation}
    \frac{\drm^n}{\drm k^n} \braket{\Phi_k|\hat{O}|\Phi_k}
    = \sum_{m=0}^n \binom{n}{m} \Braket{\frac{\drm^{n-m}\Phi_k}{\drm k^{n-m}}|\hat{O}|\frac{\drm^m\Phi_k}{\drm k^m}},
\end{equation}
with \(\ket{\frac{\drm^n\Phi_k}{\drm k^n}} = \ket{\Psi[\mathcal{B}_n]}\), where%
\footnote{Compare the first \(n+1\) columns with with the MPO expression for the \(n\)th power of an operator \(\hat{O} = \sum_i \hat{O}_i\) in Ref.~\cite{lin2017}, Eq.~(32).}
\begin{align}\label{eq:derivative-mps-general}
    &\mathcal{B}_n = \\
    &\footnotesize
    \begin{pmatrix}
        \binom{n}{n} \erm^{\irm k} A & \cdots & \binom{n}{2} \irm^{n-2} \erm^{\irm k} A & \binom{n}{1} \irm^{n-1} \erm^{\irm k} A & \binom{n}{0} \irm^n \erm^{\irm k} A & \frac{\drm^n B}{\drm k^n} \\
        & \ddots &&&& \vdots \\
        && \binom{2}{2} \erm^{\irm k} A & \binom{2}{1} \irm \erm^{\irm k} A & \binom{2}{0} \irm^2 \erm^{\irm k} A & \frac{\drm^2 B}{\drm k^2} \\
        &&& \erm^{\irm k} A & \irm \erm^{\irm k} A & \frac{\drm B}{\drm k} \\
        &&&& \erm^{\irm k} A & B \\
        &&&&& \tilde{A} \\
    \end{pmatrix}
    \normalsize
    .
    \nonumber
\end{align}
Calculating the momentum-derivatives of EA wave functions with a multi-site window or multiple windows can be done using a similar construction.

In practice, the difficulty in constructing these momentum-derivative wave functions comes from the derivatives of the EA window~\(B\) with respect to \(k\), which appear in the rightmost column.
Since we typically obtain these \(B_k\)~tensors for a discrete grid of values in \(k\)~space with step size~\(\Delta k\), a naive approach would be to calculate the derivatives using a finite-difference method: \(\drm B_k/\drm k = (B_{k+\Delta k}-B_{k-\Delta k})/2\Delta k\), \(\drm^2 B_k/\drm k^2 = (B_{k+\Delta k}-2B_{k}+B_{k-\Delta k})/ \Delta k^2\), etc.
This approach is complicated by the freedom in the phase of the \(B\)~tensor, \(B_k \rightarrow \erm^{\irm\theta} B_k\), which leaves the overall wave function unchanged.
We can try to fix this phase by finding the factor~\(\erm^{\irm \theta}\) that minimizes the Frobenius norm of the difference \(B_k-\erm^{\irm \theta} B_{k+\Delta k}\), and we can thus use the above expressions to obtain the derivatives of \(B\).
However, this can require quite a small step~\(\Delta k\) in regions where \(B\) is changing quickly; in fact, neglecting \(\drm B/\drm k\) when calculating the first derivative can sometimes give more accurate results than this approach in such regions (see Fig.~\ref{fig:heisenberg-derivatives} below and the surrounding discussion).
In order to more reliably obtain derivatives of the \(B\)~tensor, instead of using a discrete grid, we could use try to construct a continuous representation~\(B(k)\) by means of a Fourier series, for instance: this would give us easy access to the derivatives of \(B\) as well as possibly being more computationally efficient overall (supposing we can accurately represent \(B(k)\) as a Fourier series of perhaps less than ten terms, as opposed to hundreds or thousands of points in momentum space).

\section{Results and Discussion}\label{sec:results}
In order to better understand how the excitation ansatz works in practice, we look at the calculation of the low-lying excitations in two well-studied models: the spin-1 Heisenberg chain and the one-dimensional Hubbard model.
We focus on the accuracy of the wave function as we change the bond dimension and EA window size, and we illustrate how the new techniques discussed in the previous section can help to better analyse these excitations.

\subsection{Spin-1 Heisenberg chain}\label{sec:heisenberg}
First, we look at the spin-1 Heisenberg chain with antiferromagnetic interactions, following the papers originally introducing the excitation ansatz~\cite{oestlund1995,haegeman2012}, with the Hamiltonian
\begin{equation}\label{eq:heisenberg}
    \hat{H} = J \sum_i \hat{\mathbf{S}}_i \cdot \hat{\mathbf{S}}_{i+1},
\end{equation}
where \(\hat{\mathbf{S}}_i\) is the spin-1 operator acting at site~\(i\), and we set the interaction energy \(J > 0\) to be unity.
The ground state of this model is in a gapped phase with symmetry-protected topological order, where the low-lying collective excitations are spin-1 magnons~\cite{haldane1983a,haldane1983b}.

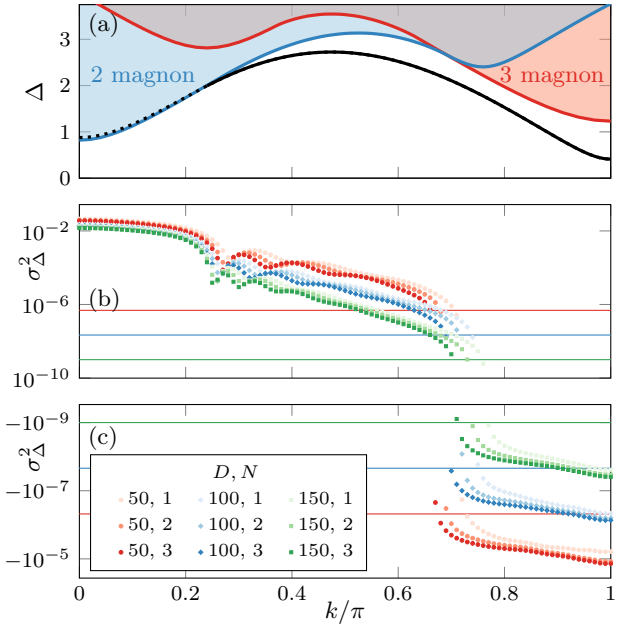
\begin{figure}[t]
    \input{figures/heisenberg}
    \caption{(a)~A plot of the low-lying excitation spectrum of the spin-1 Heisenberg model~\eqref{eq:heisenberg}.
    The black curve shows the single-magnon energies obtained using the excitation ansatz (EA) algorithm with bond dimension \(D = 150\) and window size \(N = 1\); the blue and red regions show the two- and three-magnon continua, respectively, derived using the single-magnon energies.
    At \(k \approx 0.24\pi\), the single-magnon band enters the two-magnon continuum; the dotted black curve shows the lowest energies obtained in this region with the excitation ansatz.
    (b),~(c)~The excitation energy variance \(\sigma_\Delta^2\) (positive and negative) for the optimal EA wave functions calculated for a range of different \(D\) and \(N\) (the horizontal lines show the respective variances per site of the background wave functions).}
    \label{fig:heisenberg}
\end{figure}

In Fig.~\ref{fig:heisenberg}(a), we plot the lowest-lying excitation energies obtained using the excitation ansatz for \(k \in [0, \pi]\), using a step size~\(\Delta k = 0.01\pi\).
For \(k \gtrsim 0.24\pi\), this corresponds to a single-magnon excitation, while for \(k \lesssim 0.24\pi\), the single-magnon branch enters into the two-magnon continuum, and is unstable against decay into two-magnon states in this region: this can be seen in the broadening of the spectral function as it enters the continuum~\cite{white2004,white2008}.
In Figs~\ref{fig:heisenberg}(b) and \ref{fig:heisenberg}(c) we plot the excitation variances for these states as we tune the bond
dimension~\(D\) and window size~\(N\).%
\footnote{We note that the bond dimension~\(D\) here is that of the background wave function MPS, not the ‘full’ bond dimension of the block-triangular MPS~\eqref{eq:ea-triangular}, \eqref{eq:ea-multi-site-window-triangular}, as the latter would grow with window size~\(N\).
Indeed, since we never explicitly contract over the full block-triangular tensor, the background bond dimension~\(D\) is a better indication of the performance scaling.}
It is important to note here that the excitation variance~\(\sigma_\Delta^2\) is the constant offset in the full expression for the variance \(\sigma_E^2 L + \sigma_\Delta^2\) for system size \(L\), where \(\sigma_E^2\) is the variance density of the background wave function, and as a result, this excitation variance can turn out to be negative.%
\footnote{There is no special physical significance when the sign of the excitation variance changes from positive to negative.
Indeed, as long as the MPS representation of the ground state is approximate, we can always in principle obtain a negative variance by using a sufficiently large window, as discussed below.}
As we have plotted the variance on a logarithmic scale, we separate it into two panels (b) and (c) showing positive and negative values, respectively.
When \(k \lesssim 0.24\pi\), the excitation variance is relatively high, and does not decrease very significantly as \(D\) increases, while for \(k \gtrsim 0.24\pi\), as we increase \(D\) the excitation variance jumps down at a similar magnitude to the decrease in the background variance density.
This is a result of how the single-particle excitation ansatz is adept at describing the wave function of a single-magnon state, but struggles to properly describe a two-magnon state (where a multi-particle generalization of the excitation ansatz is needed, as in Eq.~\eqref{eq:ea-2p-triangular} and Ref.~\cite{vanderstraeten2015a}).%
\footnote{Although we can get the energies of the two- or three-magnon continua (shown in Fig.~\ref{fig:heisenberg}) by summing the single-magnon energies, if we want other properties, such as spectral weights or \(S\)~matrix elements, then we need to solve for the multi-particle wave functions themselves.}

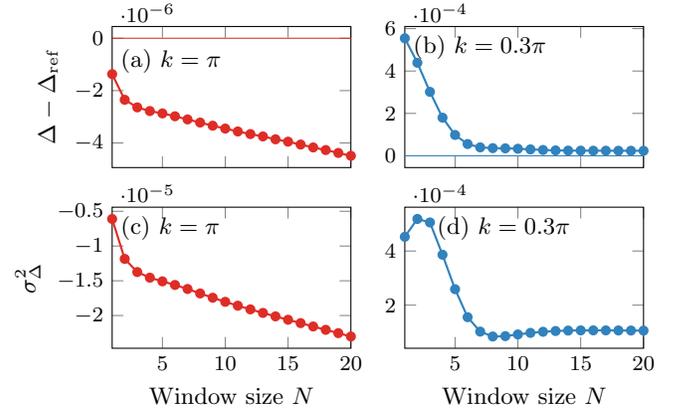
\begin{figure}[t]
    \input{figures/heisenberg-window-size}
    \caption{A more detailed plot of how the excitation energy~\(\Delta\) (a),~(b) and variance~\(\sigma_\Delta^2\) (c),~(d) of an EA wave function change as we increase its window size~\(N\).
    We consider the lowest-energy excitations for bond dimension \(D = 50\) at \(k = \pi\) (a,~c), where the energy gap is at its minimum, and \(k = 0.3\pi\) (b),~(d), where the single-magnon band is close to entering the two-magnon continuum.
    For these two momenta, we use the reference energies \(\Delta_\text{ref} = 0.410479248463\) from Ref.~\cite{haegeman2012}, and \(\Delta_\text{ref} = 2.31283\) from the \(D = 150\) calculation, respectively (which we show as horizontal lines in (a),~(b)).
    We note that the vertical axis scales in panels (b) and (d) differ from those in (a) and (c).}
    \label{fig:heisenberg-window-size}
\end{figure}

Focusing now on the single-magnon states, as we increase \(k\) from \(k \approx 0.24 \pi\), the excitation variance at a fixed bond dimension and window size steadily decreases, which is because of the magnon at \(k \sim 0.24\pi\) being spatially ‘broader’ than the magnon at \(k \sim \pi\), so a larger window is required to describe it with the same accuracy~\cite{haegeman2013a,vanderstraeten2020}.
This is shown in more detail in Fig.~\ref{fig:heisenberg-window-size}, where we fix the bond dimension and tune the window size for \(k = \pi\) and \(0.3\pi\).
For \(k = \pi\) (Figs~\ref{fig:heisenberg-window-size}(a) and \ref{fig:heisenberg-window-size}(c)), both the excitation energy and variance settle into a linearly decreasing dependence on the window size, with the excitation energy drifting away from the actual value.
At first glance, it seems counter-intuitive that increasing the size of the window could \emph{decrease} the accuracy of the results, but what is actually happening is that after increasing beyond a certain window size, we have optimized the excitation as much as we can so that the added sites start optimizing the background instead.
As more sites are added into the window, the ‘corrected’ portion of the background increases linearly, hence the linear decrease in the energy and variance.
As the values presented here as the excitation energy and variance are only the constant offsets from the values of the background wave function, we inherently cannot distinguish between the ‘actual’ excitation energy and variance and this correction to the background values.
For instances such as this, it is much more beneficial to increase the bond dimension of the background itself to get a better excitation wave function.

On the other hand, for \(k = 0.3\pi\) (Figs~\ref{fig:heisenberg-window-size}(b) and \ref{fig:heisenberg-window-size}(d)), the excitation energy and variance initially decrease much more dramatically with window size, such that the spurious linear decay that dominates at larger \(N\) is not visible on the same axis scale.
As shown in Ref.~\cite{haegeman2013b}, for isolated excitation branches, the error of the excitation itself decreases exponentially with window size, at a rate dependent on the gap below and above to other states at the same momentum.
In this case—as we have a variational approximation to the ground state—this exponential decrease in the error is valid until the point where the error in the background wave function becomes dominant.

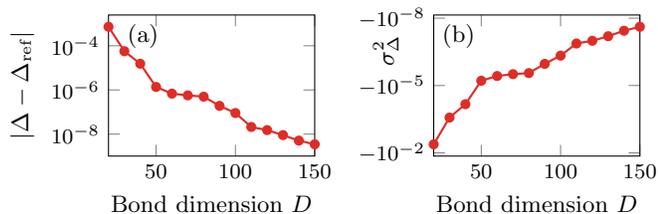
\begin{figure}[t]
    \input{figures/heisenberg-bond-dimension}
    \caption{A more detailed plot of how the excitation energy~\(\Delta\) (a) and variance~\(\sigma_\Delta^2\) (b) of an EA wave function change as we increase the MPS bond dimension~\(D\) of the background wave function.
    Calculated at \(k = \pi\) (using the same reference energy as Fig.~\ref{fig:heisenberg-window-size}) using a single-site window.}
    \label{fig:heisenberg-bond-dimension}
\end{figure}

In Fig.~\ref{fig:heisenberg-bond-dimension}, we now fix the window size to a single site, and look at how the excitation energy error and variance at \(k = \pi\) depend on the background bond dimension.
The error and magnitude of the variance both decrease steadily at similar rates as the bond dimension is increased.
It is important to note that as we have a negative excitation variance, its value is actually \emph{increasing} (i.e.\ approaching zero from below) as the wave function becomes more accurate.
The full expression for the variance, containing the contributions from both the background and the excitation \(\sigma^2_E L + \sigma^2_\Delta\), will be decreasing overall due to the decreasing linear contribution from the background: as we converge the wave function, we expect both parts of this to go to zero.
For single-magnon wave functions at other momenta, we see similar behaviour in the excitation energy, though there may be some zero crossings in the variance, which can be seen in Fig.~\ref{fig:heisenberg}.

\begin{figure}[t]
    \input{figures/heisenberg-derivatives}
    \caption{(a),~(b)~The first two derivatives (with respect to momentum) of the single-magnon excitation energy in the spin-1 Heisenberg model~\eqref{eq:heisenberg}, calculated using the momentum-derivative MPS~\eqref{eq:derivative-mps-general} for step sizes \(\Delta k = 0.01\pi\) and \(\Delta k = 0.005\pi\), and a ‘simplified’ version neglecting the terms depending on derivatives of \(B\).
    We use a bond dimension \(D = 150\), and benchmark against the finite-difference method applied to the original energy data from the excitation ansatz.
    The insets give a closer view of the region around \(k \sim 0.24\pi\) with the most variability.
    (c)~The Frobenius norm of the first two derivatives of the \(B\)~tensor calculated with the step size \(\Delta k = 0.01\pi\).}
    \label{fig:heisenberg-derivatives}
\end{figure}
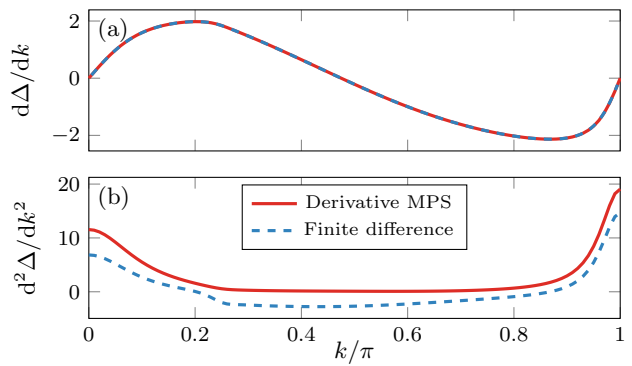

In Figs~\ref{fig:heisenberg-derivatives}(a), and \ref{fig:heisenberg-derivatives}(b), we show the first two derivatives of the lowest excitation energy calculated using the momentum-derivative MPS~\eqref{eq:derivative-mps-general} discussed in Sec.~\ref{sec:momentum-derivatives}, calculating the derivatives of the \(B\)~tensor using the finite-difference method discussed there.
We compare against the finite-difference method applied to the dispersion relation from Fig.~\ref{fig:heisenberg}(a), and observe generally excellent agreement.
However, around \(k \sim 0.24\pi\) where the single-magnon mode enters the two-magnon continuum, the derivative MPS results diverge from the benchmark results, but this is improved as we decrease the step size from \(\Delta k = 0.01\pi\) to \(0.005\pi\).
In Fig.~\ref{fig:heisenberg-derivatives}(c), we show the Frobenius norms of the first two derivatives of \(B\), which spike at around \(k \sim 0.24\pi\): this is to be expected, as the excitations are changing from single-magnon to two-magnon states.

We also compare against a ‘simplified’ momentum wave function where we neglect the terms dependent on derivatives of the \(B\)~tensor.
The results that we obtain for the first derivative (Fig.~\ref{fig:heisenberg-derivatives}(a)) are visually identical to what we obtain using the finite-difference method on the excitation energy data.
In fact, this agrees even better than when we calculate \(\drm B/\drm k\) using the finite-difference method, indicating that the error in using a poor approximation to \(\drm B/\drm k\) here is greater than the error of ignoring this term entirely!
However, for the second derivative (Fig.~\ref{fig:heisenberg-derivatives}(b)), this simplified approach does not match the expected results at all.%
\footnote{Indeed, it is expected that the higher derivatives of \(B\) must become significant at some order, since otherwise we would have a fixed \(B\)~tensor independent of momentum (as in the basic version of the single-mode approximation~\eqref{eq:sma}), whereas we would expect in general that the EA window will deform as we change momentum.
It is also interesting to note that the second derivative is always positive in the ‘simplified’ approach neglecting the derivatives of \(B\): this is since this \(B\)~tensor is optimal at the momentum at which it was optimized, so that if we tried to use this \(B\)~tensor for the full \(k\)~space, the energy will be convex around the momentum at which it was found, hence the positive second derivative.}
As discussed at the end of Sec.~\ref{sec:momentum-derivatives}, a potentially more robust way to obtain these derivatives may be if we used a continuous parametrization to represent \(B(k)\), such as a Fourier series.

\subsection{Hubbard model}
Next, we look at the one-dimensional Hubbard model, describing itinerant spin-1/2 fermions on a chain with repulsive on-site interactions, described by the Hamiltonian
\begin{multline}\label{eq:hubbard}
    \hat{H} = -t \sum_{\sigma,i} \left( \hat{c}_{\sigma,i} \hat{c}_{\sigma,i+1}^\dagger + \text{H.c.} \right) \\
    + U \sum_i \left( \hat{n}_{\uparrow,i} - \tfrac{1}{2} \right) \left( \hat{n}_{\downarrow,i} - \tfrac{1}{2} \right),
\end{multline}
where \(\hat{c}_{\sigma,i}^\dagger\), \(\hat{c}_{\sigma,i}\), and \(\hat{n}_{\sigma,i}\) are the fermionic creation, annihilation, and number operators, respectively, for spin \(\sigma \in \{\uparrow, \downarrow\}\) at site~\(i\).
We set the hopping energy~\(t\) to be unity, use the interaction strength \(U = 5\), and work at half filling with an equal number of up and down spins.
This model is Bethe ansatz integrable~\cite{lieb1968,essler2005}, which we can use to obtain exact excitation energies to compare with our numerical results.%
\footnote{We stress that even though this model is integrable, it is still highly nontrivial to treat with MPS numerics, and the methodology used here can also be used after adding a term that breaks integrability, or for a similar model that is nonintegrable.}
In this model, the elementary excitations are spinons and chargons (or holons), with only spin and charge degrees of freedom, respectively, as a result of spin–charge separation~\cite{tomonaga1950,luttinger1963,haldane1981}.

The nature of these excitations was previously explored using the MPS excitation ansatz in Ref.~\cite{zauner-stauber2018b}.
Writing the MPS in terms of conserved spin projection and particle number quantum numbers~\cite{mcculloch2007}, we need a two-site unit cell to represent the ground state at half filling.
If we were to directly construct an EA wave function on top of this MPS approximation to the ground state~\(\ket{\Psi}\), we would only be able to represent excitations that are integer combinations of the underlying fermions, and not individual spinons or chargons.
In order to represent these fractionalized excitations, we need to make use of a technical property of this MPS approximation: for this model, the finite-bond-dimension ground state~\(\ket{\Psi}\) with a two-site unit cell artificially breaks translation symmetry, in that the fidelity per site of this state after applying a single-site translation~\(\hat{T}\)—namely, \(\braket{\Psi|\hat{T}|\Psi}{}^{1/L}\) (for system size~\(L\))—will be slightly less than one (as we improve the accuracy of the state by using a higher bond dimension, this fidelity density will get closer to one).
The spinon and chargon excitations are then topologically \emph{nontrivial} excitations between the ground state approximation~\(\ket{\Psi}\) and its translated counterpart~\(\hat{T}\ket{\Psi}\), which allows us to specify excitations with fractional quantum numbers (see Ref.~\cite{zauner-stauber2018b}, or Appendix~\ref{app:fractionalized-qns} for an alternative formulation).

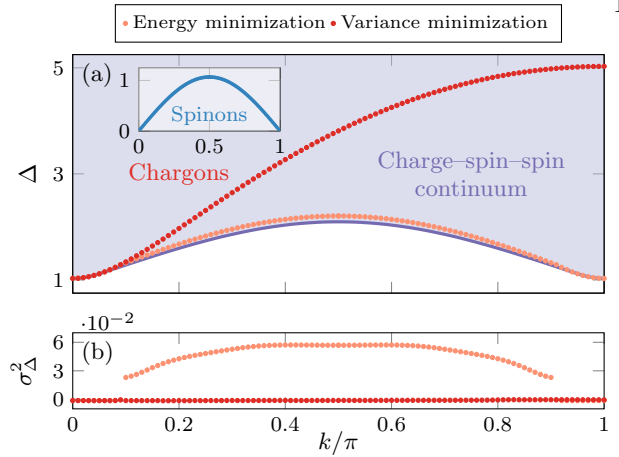
\begin{figure}[t]
    \input{figures/hubbard}
    \caption{(a)~The chargon excitation spectrum in the Hubbard model~\eqref{eq:hubbard} at unpolarized half filling with \(U = 5t\), obtained using the excitation ansatz on the zero-spin-projection sector with an MPS bond dimension \(D = 50\).
    We show the excitation energies obtained by minimizing the energy or by minimizing the energy variance (as described in the main text).
    The inset plot shows the spinon excitation energies calculated using energy minimization.
    The chargons enter the charge–spin–spin continuum at \(k \approx 0.1\pi\); after this point, energy minimization will target the bottom of this continuum, so we use variance minimization instead to obtain the desired chargon energies.
    Both the numerical spinon and chargon energies are visually identical the exact Bethe ansatz energies (which we do not show to avoid cluttering the figure).
    (b)~The corresponding excitation energy variances~\(\sigma_\Delta^2\) for the chargon EA wave functions obtained using energy and variance minimization.}
    \label{fig:hubbard}
\end{figure}

In Fig.~\ref{fig:hubbard}(a) we plot the excitations energies found using the MPS excitation ansatz: by setting the particle number and spin projection quantum numbers for the EA window, we can control whether this excitation is a spinon or a chargon (as described in Appendix~\ref{app:fractionalized-qns}).
The spinon excitations (shown in the inset) are gapless with an approximately linear dispersion at \(k = 0\) and \(\pi\), and the chargon excitations (in the main panel) are gapped, with an approximately quadratic dispersion around the minimum.
Because we are only specifying the quantum number of the spin projection, we cannot distinguish between a isolated charge excitation and a composite charge–spin–spin excitation with zero spin projection.
Around \(k = 0\), the lowest-energy charge excitations are isolated chargons, but after about \(k = 0.1\pi\), the isolated spinon lies within the charge–spin–spin continuum.
As a result, if we attempt to find the lowest-energy excitations beyond this value using the excitation ansatz, we will be targeting these three-particle states; and since the single-particle EA wave function is not good at representing multi-particle states, it is not surprising that the excitation energy variance jumps significantly at this point, as we can see in Fig.~\ref{fig:hubbard}(b).%
\footnote{Although the variance \emph{does} jump down again towards \(k = \pi\), but this because we are using an MPS representation of the ground state with a two-site unit cell, which restricts us to reduced Brillouin zone of \([0,\pi)\), so the chargon dispersion relation ‘wraps around’ from \(\pi\) to 0, and reaches its minimum again at around \(\pi\), rather \(2\pi\) as would be the case in the full Brillouin zone.
We also note that another symptom of the poor accuracy of the minimum energy wave functions within this region is that there is a visible gap between their energies and the actual lower bound of the charge–spin–spin continuum in Fig.~\ref{fig:hubbard}(a), whereas the wave functions representing chargon excitations lie directly on top of the Bethe ansatz results.}
Thus, the isolated charge excitations with energies within this continuum are inaccessible using energy minimization alone, as noted in Ref.~\cite{zauner-stauber2018b}.

To circumvent this issue, instead of finding the state with the lowest energy, we can find the state with the lowest energy \emph{variance}, as suggested in Ref.~\cite{zauner-stauber2018b} (we describe the procedure to do so in Sec.~\ref{sec:variance-minimization}).
We sweep over \(k\), starting at \(k = 0.1\pi\) and increasing to \(\pi\), so that when we optimize the state at momentum~\(k\), we use the optimized state from the previous step \(k-\Delta k\) as an initial guess.
As we are using a relatively small step in momentum (\(\Delta k = 0.01\pi\)), the change in the wave function from one step to the next is fairly minimal, so we will have a good initial condition for the minimum-variance state.
The energies of these minimum-variance states are shown in Fig.~\ref{fig:hubbard}(a), and are visually identical with the exact Bethe ansatz results using a modest MPS bond dimension of 50.
The variances of these states (shown in Fig.~\ref{fig:hubbard}(b)) do not significantly differ from the states in the region \(k \leq 0.09\pi\) that we obtained with energy minimization: this allows us to verify the accuracy of these wave functions without needing to compare with the exact energies.

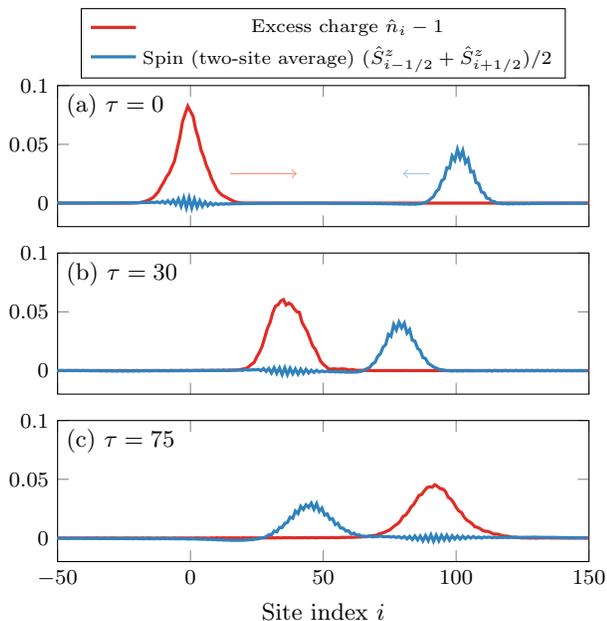
\begin{figure}[t]
    \input{figures/hubbard-wave-packets}
    \caption{The time evolution of a charge and a spin wave packet in the Hubbard model (initially on the left and right, respectively), constructed from the EA wave functions obtained previously (Fig.~\ref{fig:hubbard}) using the method presented in Ref.~\cite{van-damme2021a}.
    We show snapshots at three points in time \(\tau\), including the initial state, the state just before collision, and the state after collision.
    We plot the excess charge (on top of an even distribution of one charge per site) and the two-site average of the \(z\) projection of the spin (in order to remove the even–odd fluctuation; see Fig.~\ref{fig:hubbard-wave-packets-domain-wall} below).
    (An animation of the time evolution is available within the Supplemental Material~\cite{animation}.)}
    \label{fig:hubbard-wave-packets}
\end{figure}

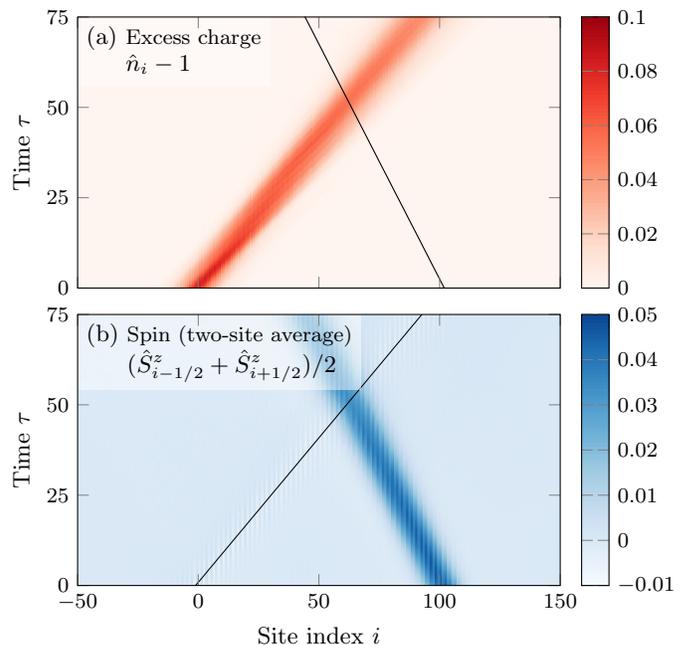
\begin{figure}[t]
    \input{figures/hubbard-wave-packets-full}
    \caption{Heat-map plots showing the full time evolution of the charge (a) and spin (b) wave packets from Fig.~\ref{fig:hubbard-wave-packets}.
    The superimposed guidelines show the approximate location of the other wave packet.}
    \label{fig:hubbard-wave-packets-full}
\end{figure}

To further demonstrate the usefulness of having access to not only the chargon energies for the full momentum range, but also their wave functions, we use these—as well as the spinon wave functions—to construct two real-space spinon and chargon wave packets with coherent momenta, following the protocol described in Ref.~\cite{van-damme2021a}.
We construct an initial state containing these two wave packets placed opposite each other, whose momenta are chosen such that they will propagate towards each other over time.
We then calculate the time evolution of this initial state using the time-dependent variational principle (TDVP) algorithm~\cite{haegeman2016}, using single-site updates with environment expansion~\cite{mcculloch2024,mcculloch2024b}.
We use a wave function with infinite boundary conditions (IBC)~\cite{phien2012,phien2013,milsted2013,zauner2015}, using a finite window of sites that we grow with time, connected to a time-invariant background on either side.
Figure~\ref{fig:hubbard-wave-packets} shows three snapshots from this time evolution, and Fig.~\ref{fig:hubbard-wave-packets-full} shows the full evolution in a heat-map plot.
When the two wave packets pass through each other, they do not scatter off each other, as a consequence of spin–charge separation in the Hubbard model (there is some slight broadening of the wave packets over time, but this would happen even if each wave packet evolved on its own without colliding with the other).
Although it may seem trivial to see two separate particles passing through each other without any scattering, when we consider that these are emergent fractionalized excitations on top of a strongly coupled background, it is impressive that we can obtain the expected physical behaviour from an approximate numerical simulation.

\begin{figure}[t]
    \input{figures/hubbard-wave-packets-domain-wall}
    \caption{The initial state of the time evolution in Fig.~\ref{fig:hubbard-wave-packets}, plotting the spin for each individual site rather than a two-site average.
    This shows the numerical translation-symmetry breaking in the even–odd spin fluctuation.}
    \label{fig:hubbard-wave-packets-domain-wall}
\end{figure}
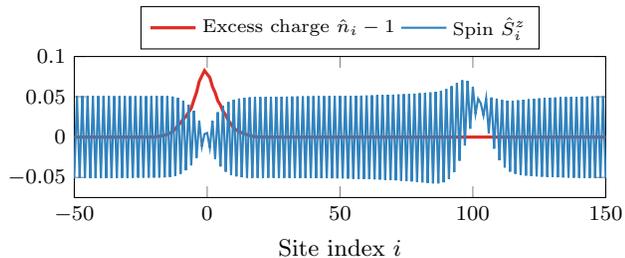

In Figs~\ref{fig:hubbard-wave-packets} and \ref{fig:hubbard-wave-packets-full} we plotted the two-site average of the spin: this averaging is because the MPS approximation to the ground state will artificially break translation symmetry, as touched upon earlier, causing there to be an imbalance in the spin between odd and even sites.
In Fig.~\ref{fig:hubbard-wave-packets-domain-wall}, we plot the initial state showing the spin of each site rather than the two-site average.
Although this fluctuation in the spin is just a numerical artefact, there is some physical significance to it, since the spinon and chargon wave packets are domain walls in these odd–even fluctuations (as a result of their being fractionalized excitations): this can be seen in Fig.~\ref{fig:hubbard-wave-packets-domain-wall} under close inspection.

\section{Conclusions}\label{sec:conclusion}
The MPS excitation ansatz has already proved to be a very useful numerical tool for studying low-lying excitations in quantum many-body systems.
We have rounded out the existing literature on this topic by presenting a more general method for calculating expectation values for arbitrary observables.
This is based on the newly-presented interpretation of the excitation ansatz as an MPS with a block-triangular structure, which can concisely describe extensions to the original formulation, such as increasing the spatial support of the excitations, or describing multi-particle states.
Our technique encapsulates established methods for optimizing the excitation ansatz, and allows us to formulate new ones, such as variance minimization.

We have shown the versatility of this method by calculating the energy variance of the elementary magnon excitations in the spin-1 Heisenberg model, analysing the effect of tuning the bond dimension and EA window size, as well as directly calculating the derivative of the state with respect to momentum.
We have further demonstrated how we can find stable excitations within a continuum in the Hubbard model by optimizing the energy variance of the excited states, and we showcased how the excitation ansatz and dynamical time-evolution methods complement each other by performing a time-evolution simulation of the scattering of spin and charge wave packets~\cite{van-damme2021a}.

There are many further directions to take the study presented here.
One interesting prospect is to look at the multi-particle excitation ansatz~\cite{vanderstraeten2014,vanderstraeten2015a}.
Our methods to calculate observables already work with multi-particle wave functions, however, we need a method to optimize the wave function itself, similar to that in Ref.~\cite{vanderstraeten2015a}.
This is necessary, for instance, for calculating the \(S\)~matrix and spectral function weights.

Another possible extension would be representing excited states over the whole momentum range by using a series expansion rather than having separate wave functions over a grid in momentum space.
This would have a reduced computational cost (as we could represent the full momentum range with perhaps tens of terms in the series expansion instead of hundreds of different momentum points), which would be beneficial for constructing multi-particle states from these wave functions, as also allow us to easily calculate the full expression for momentum derivatives~\eqref{eq:derivative-mps-general}.

While it is fairly straightforward for us to find multiple energy eigenstates for a single-site window EA wave function, it is more nontrivial to simultaneously optimize multiple eigenstates if we increase the spatial support of this window.
To solve this issue, we could look at a sort of ‘multi-target EA wave function’, in the spirit of Ref.~\cite{li2024}, where we simultaneously represent and optimize multiple excited states in a single wave function.

It would also be interesting to see whether this technique can be extended to generalized forms of the excitation ansatz in other tensor networks, such as PEPS\@.
This would be helpful in studying excitations in systems with a higher spatial dimensionality, where MPS methods suffer from the requirement of an exponentially growing bond dimension in the system size along the additional dimensions.

Recently, a similar idea has been investigated for systematically calculating expectation values for EA wave functions by means of generating functions~\cite{liao2019,tu2021}.
For instance, a basic EA wave function~\eqref{eq:ea} can be generated from the MPS with the (site-dependent) \(A\) tensor \(A_n(\lambda) = A + \lambda \erm^{\irm kn} B\).
We take the derivative of this MPS with respect to \(\lambda\) and take the limit \(\lambda \rightarrow 0\) to obtain the EA wave function: this can be done efficiently with automatic differentiation.
While this is fairly straightforward for a finite-size system, for an infinite system, we need a method for evaluating the fixed-point equation (such as the one presented in this work).
An appealing feature of this approach is its natural generalization to arbitrary tensor networks, allowing efficient calculation of expectation values for the iPEPS excitation ansatz~\cite{ponsioen2022,ponsioen2023,tu2024}.

\begin{acknowledgments}
We acknowledge helpful discussions with Maarten Van Damme and Valentin Zauner-Stauber.
We thank the anonymous referee for providing inspiration for the method of phase fixing for momentum-derivative states discussed in Sec.~\ref{sec:momentum-derivatives}.
I.P.M. acknowledges funding from the National Science and Technology Council of Taiwan (NSTC) Grants No.~112-2811-M-007-044 and No.~113-2112-M-007-MY2.
The calculations in this work were performed using the Matrix Product Toolkit~\cite{mptoolkit}.
\end{acknowledgments}

\appendix

\section{Generalization to multi-site unit cells}\label{app:multi-site-unit-cell}
In the main text, for the sake of clarity and brevity, we restricted ourselves to iMPSs with only a single-site unit cell.
Everything discussed there can also be applied to states with multi-site unit cells, with some necessary modifications, which we outline here.

For a iMPS with a single-site unit cell, we explicitly enforce translation invariance by repeating the same \(A\) tensor at each site; for an iMPS with an \(m\)-site unit cell, we instead repeat a sequence of \(m\) tensors \(\{A_1, A_2, \ldots, A_m\}\), that is,
\begin{equation}
    \ket{\Psi[\mathbf{A}]} =
    \begin{tikzpicture}[tensornetwork]
        \node[atensor, label=\(A_m\)] (A1) at (1, 0) {};
        \node[atensor, label=\(A_1\)] (A2) at (2, 0) {};
        \node[atensor, label=\(A_2\)] (A3) at (3, 0) {};
        \node[notensor] (A4) at (4, 0) {\(\ldots\)};
        \node[atensor, label=\(A_m\)] (A5) at (5, 0) {};
        \node[atensor, label=\(A_1\)] (A6) at (6, 0) {};
        \draw (A1.south) -- +(0, -0.5);
        \draw (A2.south) -- +(0, -0.5);
        \draw (A3.south) -- +(0, -0.5);
        \draw (A5.south) -- +(0, -0.5);
        \draw (A6.south) -- +(0, -0.5);
        \draw (A1) -- (A2) -- (A3) -- (A4) -- (A5) -- (A6);
        \draw (A1.west) -- +(-0.5, 0) node[left] {\(\ldots\)};
        \draw (A6.east) -- +(0.5, 0) node[right] {\(\ldots\)};
    \end{tikzpicture}
    .
\end{equation}
We need to use this type of state if we have an inhomogeneous lattice, where, for instance, interaction strengths over even and odd bonds could be different, or if we are trying to represent a state on a higher-dimensional lattice, such as an infinite ladder or cylinder.
Even if we have a homogeneous one-dimensional chain, we will still need a multi-site unit cell if we want to describe a state that spontaneously breaks translation symmetry, such as an antiferromagnetic state.

To form an EA wave function on top of a background state with a \(m\)-site unit cell, instead of having a single-window tensor~\(B\), we have \(m\) separate windows \(B_1, B_2, \ldots, B_m\) for each site in the unit cell, that is,%
\footnote{Naively, we could try to write a single \(m\)-site window covering the whole unit cell, but this causes issues when trying to apply the gauge-fixing conditions, as explained in the following subsection (Appendix~\ref{app:gauge-fixing}).}
\begin{align}\label{eq:ea-multi-site-uc}
    \ket{\Phi_k[\mathbf{B}]} = &\sum_{n\,\text{s.t.}\,m\mid n} \erm^{\irm kn} \nonumber\\
    \Big[ &
    \begin{tikzpicture}[tensornetwork, scale=0.75]
        \renewcommand{\tensorsize}{7.5pt}
        \node[atensor, label={\footnotesize\(A_m\)}] (A1) at (1, 0) {};
        \node[atensor, fill=lightgray, label={\footnotesize\(B_1\)}] (A2) at (2, 0) {};
        \node[atensor, label={\footnotesize\(\tilde{A}_2\)}] (A3) at (3, 0) {};
        \node[notensor] (A4) at (4, 0) {\(\ldots\)};
        \node[atensor, label={\footnotesize\(\tilde{A}_m\)}] (A5) at (5, 0) {};
        \node[atensor, label={\footnotesize\(\tilde{A}_1\)}] (A6) at (6, 0) {};
        \draw (A1.south) -- +(0, -0.5);
        \draw (A2.south) -- +(0, -0.5);
        \draw (A3.south) -- +(0, -0.5);
        \draw (A5.south) -- +(0, -0.5);
        \draw (A6.south) -- +(0, -0.5);
        \draw (A1) -- (A2) -- (A3) -- (A4) -- (A5) -- (A6);
        \draw (A1.west) -- +(-0.5, 0) node[left] {\(\ldots\)};
        \draw (A6.east) -- +(0.5, 0) node[right] {\(\ldots\)};
    \end{tikzpicture}
    \nonumber\\
    + &
    \begin{tikzpicture}[tensornetwork, scale=0.75]
        \renewcommand{\tensorsize}{7.5pt}
        \node[atensor, label={\footnotesize\(A_m\)}] (A1) at (1, 0) {};
        \node[atensor, label={\footnotesize\(A_1\)}] (A2) at (2, 0) {};
        \node[atensor, fill=lightgray, label={\footnotesize\(B_2\)}] (A3) at (3, 0) {};
        \node[notensor] (A4) at (4, 0) {\(\ldots\)};
        \node[atensor, label={\footnotesize\(\tilde{A}_m\)}] (A5) at (5, 0) {};
        \node[atensor, label={\footnotesize\(\tilde{A}_1\)}] (A6) at (6, 0) {};
        \draw (A1.south) -- +(0, -0.5);
        \draw (A2.south) -- +(0, -0.5);
        \draw (A3.south) -- +(0, -0.5);
        \draw (A5.south) -- +(0, -0.5);
        \draw (A6.south) -- +(0, -0.5);
        \draw (A1) -- (A2) -- (A3) -- (A4) -- (A5) -- (A6);
        \draw (A1.west) -- +(-0.5, 0) node[left] {\(\ldots\)};
        \draw (A6.east) -- +(0.5, 0) node[right] {\(\ldots\)};
    \end{tikzpicture}
    \nonumber\\
    &\vdots \nonumber\\
    + &
    \begin{tikzpicture}[tensornetwork, scale=0.75]
        \renewcommand{\tensorsize}{7.5pt}
        \node[atensor, label={\footnotesize\(A_m\)}] (A1) at (1, 0) {};
        \node[atensor, label={\footnotesize\(A_1\)}] (A2) at (2, 0) {};
        \node[notensor] (A3) at (3, 0) {\(\ldots\)};
        \node[atensor, label={\footnotesize\(A_{m-1}\)}] (A4) at (4, 0) {};
        \node[atensor, fill=lightgray, label={\footnotesize\(B_m\)}] (A5) at (5, 0) {};
        \node[atensor, label={\footnotesize\(\tilde{A}_1\)}] (A6) at (6, 0) {};
        \draw (A1.south) -- +(0, -0.5);
        \draw (A2.south) -- +(0, -0.5);
        \draw (A4.south) -- +(0, -0.5);
        \draw (A5.south) -- +(0, -0.5);
        \draw (A6.south) -- +(0, -0.5);
        \draw (A1) -- (A2) -- (A3) -- (A4) -- (A5) -- (A6);
        \draw (A1.west) -- +(-0.5, 0) node[left] {\(\ldots\)};
        \draw (A6.east) -- +(0.5, 0) node[right] {\(\ldots\)};
    \end{tikzpicture}
    \Big].
\end{align}
We can write this in a block-upper-triangular form with an \(m\)-site unit cell as
\begin{subequations}
\begin{align}
    \ket{\Phi_k[\mathbf{B}]} &= \cdots \mathcal{A}_m^{s_0} \mathcal{A}_1^{s_1} \mathcal{A}_2^{s_2} \cdots \mathcal{A}_m^{s_m} \mathcal{A}_1^{s_{m+1}} \cdots, \\
    \mathcal{A}_1 &=
    \begin{pmatrix}
        \erm^{\irm km} A_1 & B_1 \\
        0 & \tilde{A}_1 \\
    \end{pmatrix}
    , \\
    \mathcal{A}_i &=
    \begin{pmatrix}
        A_i & B_i \\
        0 & \tilde{A}_i \\
    \end{pmatrix}
    , \qquad i > 1.
\end{align}
\end{subequations}
It is important to note that the phase factor~\(\erm^{\irm km}\) only occurs once per unit cell here: even if we wrote a factor~\(\erm^{\irm k}\) in every \(\mathcal{A}\)~tensor, this is equivalent to the above form if we shift the phase of the window tensors~\(B_i\) for \(i>1\).
In effect, by increasing the size of the unit cell, we decrease the range of the first Brillouin zone to \([0, 2\pi/m)\).
As a result of this, for instance, if \(m = 2\), then \(k = 0\) and \(k = \pi\) would be indistinguishable, as \(\erm^{\irm km} = 1\) in both cases.%
\footnote{If we had a wave function with a single-site unit cell~\(A\) that we artificially extended to have a two-site unit cell with \(A_1 = A_2 = A\), then we would have a \(k = 0\) excitation if \(B_1 = B_2\), and a \(k = \pi\) excitation if \(B_1 = -B_2\).
In such a case as this, we can recover the full range of the first Brillouin zone by directly inspecting the behaviour of the \(B\)~tensors under single-site translations.}
Though it should be noted that if we are using an inhomogeneous lattice, such as a cylinder, with a ‘physical’ unit cell of \(M > 1\) sites, then we should write this phase factor as \(\erm^{\irm km/M}\) instead, so that if the MPS unit cell size~\(m\) is the same as the physical unit cell size~\(M\), we recover the full range of the first Brillouin zone \([0, 2\pi)\) as expected.

Generalizations of the basic form of the EA wave function, such as multi-site windows and multi-particle excitations, can be handled in a similar way.
For example, for multi-site windows, we would have \(m\) separate windows of length~\(N\), each of which starts on a different site in the unit cell (in principle, each window could have a different length as well, but it is conceptually simpler to keep them all the same length).

\subsection{Gauge fixing}\label{app:gauge-fixing}
For an \(m\)-site unit cell, we can apply the same left- or right-gauge-fixing conditions~\eqref{eq:gauge-fixing} to each window individually, that is, \(B_i = N_{L,i} X_i\) for each \(i\), where \(N_{L,i}\) is the left null space of \(A_i\) (using the left gauge).
The inner product~\eqref{eq:ea-inner-prod} is then the sum of the Frobenius product of each window,
\begin{equation}
    \braket{\Phi_k[\tilde{\mathbf{B}}]|\Phi_k[\mathbf{B}]} = \sum_{n\,\text{s.t.}\,m\mid n} \sum_{i=1}^m
    \begin{tikzpicture}[tensornetwork]
        \node[atensor, fill=lightgray, label=\(X_i\)] (A2) at (2, 1) {};
        \node[atensor, fill=lightgray, label=below:\(\tilde{X}_i\)] (A2conj) at (2, -1) {};
        \draw[rounded corners] (A2.west) -- +(-0.5, 0) -- +(-0.5, -2) -- (A2conj.west);
        \draw[rounded corners] (A2.east) -- +(0.5, 0) -- +(0.5, -2) -- (A2conj.east);
    \end{tikzpicture}
    .
\end{equation}
In this case, we consider a state to be normalized if the sum of the squares of the Frobenius norms of each \(X_i\) in the unit cell is one.
Similarly, for multi-site windows, using the left-gauge-fixing condition, the inner product is the sum of the separate inner products of each window.

An important consequence of applying gauge-fixing conditions to a wave function with a multi-site unit cell is that each of the windows starting on different sites of the unit cell will be orthogonal to each other.
This is why we have one window for each site in the unit cell as in Eq.~\eqref{eq:ea-multi-site-uc}, rather than naively writing an single \(m\)-site window covering the whole unit cell: if we used the gauge-fixing conditions in the latter case, we would be discarding components where the window starts on a different site in the unit cell.
It is also worth noting that each window need not have the same norm: in general, some may have a larger norm than others, and some may even have a zero norm.

\subsection{MPO fixed-point equations}
For an \(m\)-site unit cell, the first point equations for \(\pbra{E}\) and \(\pket{F}\) are the same as for a single-site unit cell (described in Sec.~\ref{sec:mpo} in the main text), but we need to use the full unit-cell transfer operator, which is the product of transfer operators for each individual site in the unit cell,
\begin{multline}
    (T_\text{full})^{\omega_m\alpha_m\beta_m}_{\omega_0\alpha_0\beta_0} \\
    = (T_1)^{\omega_1\alpha_1\beta_1}_{\omega_0\alpha_0\beta_0} (T_2)^{\omega_2\alpha_2\beta_2}_{\omega_1\alpha_1\beta_1} \cdots (T_m)^{\omega_m\alpha_m\beta_m}_{\omega_{m-1}\alpha_{m-1}\beta_{m-1}},
\end{multline}
with each \(T_i\) being defined analogously to the transfer operator for a single-site unit cell~\eqref{eq:transfer-operator} using \(\mathcal{A}_i\).
With this ‘full’ transfer operator, \(\pbra{E^{\omega\alpha\beta}}\) and \(\pket{F_{\omega\alpha\beta}}\) will be \(D\times D\) matrices acting on the bra and ket unit cell boundaries, and the expectation values the we obtain at the end will be per unit cell, rather than per site.
We can also obtain the ‘internal’ \(\pbra{E}_i\) and \(\pket{F}_i\) tensors for some position~\(i\) within the unit cell by
\begin{subequations}
\begin{align}
    \pbra{E}_i &= \pbra{E} T_1 T_2 \cdots T_i, \\
    \pket{F}_i &= T_i T_{i+1} \cdots T_m \pket{F},
\end{align}
\end{subequations}
where we implicitly sum over the \(\omega\alpha\beta\) indices each time we multiply an internal transfer operator: for example, \(\pbra{E^{\omega_1\alpha_1\beta_1}}_1 = \pbra{E^{\omega_0\alpha_0\beta_0}} (T_1)^{\omega_1\alpha_1\beta_1}_{\omega_0\alpha_0\beta_0}\).
We also define \(\pbra{E}_0 = \pbra{E}\) and \(\pket{F}_{m+1} = \pket{F}\) for the following subsection.

\subsection{The excitation ansatz algorithm}
When we construct the effective Hamiltonian for an \(m\)-site unit cell, we have \(m\) effective Hamiltonian matrices for each window tensor~\(B_i\), of the form
\begin{equation}
    (\mathcal{H}_\text{eff}^k)_i B_i
    =
    \begin{tikzpicture}[tensornetwork]
        \node[etensor, label={left, text height=0.5em}:\(\pbra{E^{\omega\alpha 1}}_{i-1}\)] (E) at (-1, 0) {};
        \node[atensor, label=above:\(\mathcal{A}_i\)] (A) at (0, 1) {};
        \node[wtensor, label={right, yshift=5pt, xshift=-2pt}:\(\mathcal{W}\)] (W) at (0, 0) {};
        \node[notensor] (Z) at (0, -1) {};
        \node[etensor, label={right, text height=0.5em}:\(\pket{F_{\omega'\alpha'2}}_{i+1}\)] (F) at (1, 0) {};
        \draw (E) -- (W) -- (F);
        \draw (A) -- (W) -- (Z);
        \draw[thick] (A) -- (E.east |- A.west);
        \draw[thick] (A) -- (F.west |- A.east);
        \draw (Z) -- (E.east |- Z.west);
        \draw (Z) -- (F.west |- Z.east);
    \end{tikzpicture}
    ,
\end{equation}
which we need to solve simultaneously.
In other words, we want to find the lowest-energy eigenvectors of the form \(B_1 \otimes B_2 \otimes \cdots \otimes B_m\) of the direct product of the effective Hamiltonians \((\mathcal{H}_\text{eff}^k)_1 \otimes (\mathcal{H}_\text{eff}^k)_2 \otimes \cdots \otimes (\mathcal{H}_\text{eff}^k)_m\).

For multi-site windows, the EA DMRG algorithm (Sec.~\ref{sec:ea-dmrg} in the main text) functions similarly in that we sweep over all of the windows simultaneously.
We start at the rightmost site in each window, and sweep from the end of each window to the other, simultaneously updating the \(n\)th tensor in each window at each step.

\section{Canonical forms for MPSs}\label{app:canonicality}
Although it is common in the MPS literature to see the terms \emph{orthogonal} and \emph{canonical} used interchangeably, here we define these two terms distinctly in order to distinguish two different concepts.
By an \emph{orthogonal} form, we mean that the MPS is written in terms of left- or right-orthogonal tensors satisfying the conditions described in Eq.~\eqref{eq:orthogonality-conditions} in the main text—we may have all of the tensors in left-orthogonal form or all of them in right-orthogonal form, or we can have a mixed-orthogonal form~\eqref{eq:mixed-orthogonal} where all tensors to the left (right) of some central site or bond are in left-(right-)orthogonal form.

Imposing these orthogonality constraints fixes the gauge degrees of freedom of the MPS, but only partially, since \(A^s\) can still undergo unitary gauge transformations \(A^s \mapsto U A^s U^\dagger\) or be multiplied by a phase factor \(A^s \mapsto \erm^{\irm\phi} A^s\) while still obeying these constraints.
We can further fix these degrees of freedom by bringing the MPSs into a \emph{canonical} form.
For a translation-invariant state in left-orthogonal form, the principal right eigenvector of the transfer operator will be the reduced density matrix~\(\rho\),
\begin{equation}
    \begin{tikzpicture}[tensornetwork]
        \node[ltensor] (A1) at (0, 1) {};
        \node[ltensor] (A1conj) at (0, -1) {};
        \node[ctensor, label={right, text depth=}:\(\rho\)] (rho) at (0.75, 0) {};
        \draw (A1.south) -- (A1conj.north);
        \draw[rounded corners] (A1.east) -- (A1.east -| rho.north) -- (rho) -- (rho.south |- A1conj.east) -- (A1conj.east);
        \draw (A1.west) -- +(-0.5, 0);
        \draw (A1conj.west) -- +(-0.5, 0);
    \end{tikzpicture}
    =
    \begin{tikzpicture}[tensornetwork]
        \coordinate (A1) at (0, 1) {};
        \coordinate (A1conj) at (0, -1) {};
        \node[ctensor, label={right, text depth=}:\(\rho\)] (rho) at (0.75, 0) {};
        \draw[rounded corners] (rho.north) -- (rho.north |- A1.east) -- +(-0.5, 0);
        \draw[rounded corners] (rho.south) -- (rho.south |- A1conj.east) -- +(-0.5, 0);
    \end{tikzpicture}
    .
\end{equation}
If the MPS is injective, then the transfer operator is a completely positive map, and so by the Perron–Frobenius theorem, \(\rho\) will be positive-definite and Hermitian.
We can then unitarily diagonalize \(\rho\) as \(U \Lambda^2 U^\dagger\).
To bring the state into its canonical form, we conjugate the \(A_L\)~tensor by \(U\) (\(\tilde{A}_L^s = U^\dagger A_L^sU\)) to make its principal right eigenvector the positive diagonal matrix \(\Lambda^2\).
We then say that the state is in \emph{left-canonical} form when we have the \(A_L\)~tensor in this form with a positive diagonal \(\Lambda\)~matrix.%
\footnote{Strictly speaking, we do not need to incorporate \(U\) into \(A_L\): we could conjugate \(\Lambda\) by \(U\) instead (\(\tilde{\Lambda} = U \Lambda U^\dagger\)) and define canonicality as having a positive-definite Hermitian \(\Lambda\)~matrix rather than a positive diagonal \(\Lambda\) matrix.}
We can similarly define a right-canonical form, which is related to the left-canonical form by \(A_L \Lambda = \Lambda A_R\): the mixed-canonical form will simply be the mixed-orthogonal form~\eqref{eq:mixed-orthogonal} using \(C = \Lambda\).

The key distinction we make between a canonical form and the more general orthogonal form has to do with knowing the \(\Lambda\)~matrix.
For a translation-invariant state, each bond is associated with the same \(\Lambda\)~matrix, and so the only change we need to make to bring a state into canonical form is conjugating the \(A\)~tensor by a unitary matrix.
However, for a non-translation-invariant state, we have in general a different \(\Lambda\)~matrix associated with each bond.
For a finite state, we can canonicalize the state by sweeping from one end to the other and perform a singular value decomposition (SVD) at each site, which is usually already done in MPS algorithms such as DMRG and TDVP\@.
But for an \emph{infinite} state breaking translation invariance, this is not so straightforward.
As an illustrative example, given a translation-invariant state in mixed-canonical form, we apply an operator acting only on the centre site, breaking translation symmetry,
\begin{equation}\label{eq:o-psi}
    \hat{O} \ket{\Psi} =
    \begin{tikzpicture}[tensornetwork]
        \node[ltensor, label=\(A_L\)] (A1) at (1, 0) {};
        \node[ltensor, label=\(A_L\)] (A2) at (2, 0) {};
        \node[atensor, label=\(\tilde{A}_C^0\), fill=lightgray] (A3) at (3, 0) {};a
        \node[rtensor, label=\(A_R\)] (A4) at (4, 0) {};
        \node[rtensor, label=\(A_R\)] (A5) at (5, 0) {};
        \draw (A1.south) -- +(0, -0.5);
        \draw (A2.south) -- +(0, -0.5);
        \draw (A3.south) -- +(0, -0.5);
        \draw (A4.south) -- +(0, -0.5);
        \draw (A5.south) -- +(0, -0.5);
        \draw (A1) -- (A2) -- (A3) -- (A4) -- (A5);
        \draw (A1.west) -- +(-0.5, 0) node[left] {\(\ldots\)};
        \draw (A5.east) -- +(0.5, 0) node[right] {\(\ldots\)};
    \end{tikzpicture}
    .
\end{equation}
We call this sort of state an \emph{infinite boundary condition} (IBC) wave function~\cite{phien2012,phien2013,milsted2013,zauner2015}: there is a single-site window (\(\tilde{A}_C^0\)) connected to two semi-infinite boundaries.
(Note that each component of an EA wave function~\eqref{eq:ea} with the window on site \(n\) considered by itself is essentially an IBC wave function.)
While it is still reasonable to say the state is in a mixed-orthogonal form, we do not say that it is mixed \emph{canonical}, since the \(\Lambda\)~matrix corresponding to the original translation invariant state will not be valid, except far from the window.

If we wanted to canonicalize this state, we can consider moving the orthogonality centre one site to the right: we take the SVD \(\tilde{A}_C^0= UDV^\dagger\), and left-multiply \(DV^\dagger\) onto the following site~\(A_R\) to become \(\tilde{A}_C^1\), and the original centre site becomes \(\tilde{A}_L^0 = U\), so that
\begin{equation}
    \hat{O} \ket{\Psi} =
    \begin{tikzpicture}[tensornetwork]
        \node[ltensor, label=\(A_L\)] (A1) at (1, 0) {};
        \node[ltensor, label=\(A_L\)] (A2) at (2, 0) {};
        \node[ltensor, label=\(\tilde{A}_L^0\), fill=lightgray] (A3) at (3, 0) {};a
        \node[atensor, label=\(\tilde{A}_C^1\), fill=lightgray] (A4) at (4, 0) {};
        \node[rtensor, label=\(A_R\)] (A5) at (5, 0) {};
        \draw (A1.south) -- +(0, -0.5);
        \draw (A2.south) -- +(0, -0.5);
        \draw (A3.south) -- +(0, -0.5);
        \draw (A4.south) -- +(0, -0.5);
        \draw (A5.south) -- +(0, -0.5);
        \draw (A1) -- (A2) -- (A3) -- (A4) -- (A5);
        \draw (A1.west) -- +(-0.5, 0) node[left] {\(\ldots\)};
        \draw (A5.east) -- +(0.5, 0) node[right] {\(\ldots\)};
    \end{tikzpicture}
    .
\end{equation}
In general, this \(\tilde{A}_L^0\)~tensor will \emph{not} be the same as the \(A_L\)~tensor in the original state, and \(\tilde{A}_C^1\) will be different from both the \(A_C\) in the original state and the \(\tilde{A}_C^0\) from the previous step.
As we continue to move the orthogonality centre further to the right in this fashion, we expect \(\tilde{A}_L^n\) and \(\tilde{A}_C^n\) to approach their forms for the original translation-invariant state at some distance determined by the correlation length.
We can also move the orthogonality centre to the left, and repeat the same process in that direction.
In the end, this procedure results in a state of the form
\begin{equation}
    \hat{O} \ket{\Psi} =
    \begin{tikzpicture}[tensornetwork]
        \node[ltensor, label=\(\tilde{A}_L^{-2}\), fill=lightgray] (A1) at (1, 0) {};
        \node[ltensor, label=\(\tilde{A}_L^{-1}\), fill=lightgray] (A2) at (2, 0) {};
        \node[atensor, label=\(\tilde{A}_C^0\), fill=lightgray] (A3) at (3, 0) {};a
        \node[rtensor, label=\(\tilde{A}_R^1\), fill=lightgray] (A4) at (4, 0) {};
        \node[rtensor, label=\(\tilde{A}_R^2\), fill=lightgray] (A5) at (5, 0) {};
        \draw (A1.south) -- +(0, -0.5);
        \draw (A2.south) -- +(0, -0.5);
        \draw (A3.south) -- +(0, -0.5);
        \draw (A4.south) -- +(0, -0.5);
        \draw (A5.south) -- +(0, -0.5);
        \draw (A1) -- (A2) -- (A3) -- (A4) -- (A5);
        \draw (A1.west) -- +(-0.5, 0) node[left] {\(\ldots\)};
        \draw (A5.east) -- +(0.5, 0) node[right] {\(\ldots\)};
    \end{tikzpicture}
    ,
\end{equation}
where the window extends to the left and right for as many sites as it takes for \(\tilde{A}_L^{-n}\) and \(\tilde{A}_R^n\) to be sufficiently close to the translation invariant background.

This final canonicalized window will need to be much larger than the single-site window in Eq.~\eqref{eq:o-psi}.
This is owing to the way in which an MPS automatically encodes entanglement: modifying a single tensor in an MPS create a local disturbance in the wave function around this site, the size of which depends on the correlation length of the state.
As such, it is usually more efficient to work with a smaller (sometimes single-site) window with IBC and EA wave functions, since this small window can encode information about the wave function well beyond its boundaries.
The trade-off with using a small window is that the state will no longer be in a canonical form, so that we need to be careful about how we handle the boundaries.
In time evolution, for example, we can end up with spurious effects at the window boundaries if we are not careful.

\subsection{Splicing two IBC wave functions}
For another concrete example where this distinction between orthogonality and canonicality is relevant, we consider the situation where we have two IBC wave functions describing two different wave packets, and we splice them together into a single wave function to study their time evolution, as in Fig.~\ref{fig:hubbard-wave-packets} in the main text and in Ref.~\cite{van-damme2021a}.
We start with a left and a right IBC wave function with one orthogonality centre each, and we want to create a single IBC wave function containing both wave packets.
This new wave function will need to have single orthogonality centre, which we will somehow need to form from the initial two.

A naive approach would be to move the orthogonality centre of the left window to the bond at the right-hand site of the window by left-orthogonalizing the window, obtaining the matrix~\(\Lambda_L\), and vice versa for the right window getting \(\Lambda_R\).
The problem here is that in general \(\Lambda_L\) and \(\Lambda_R\) will be different from each other, and both different from the \(\Lambda\)~matrix of the boundary wave function in the thermodynamic limit.
So if we were to form the combined IBC window by using the boundary’s \(\Lambda\)~matrix as the orthogonality centre, with the left-orthogonal tensors of the left window to its left, and the right-orthogonal tensors of the right window to its right, the wave function around the centre of the window may end up distorted.
For a more sophisticated guess, we could use \(\Lambda_L \Lambda^{-1} \Lambda_R\), which is exact in the case that the left and right windows can be expressed as disjoint actions on the physical degrees of freedom.

If we wanted to splice these windows safely, we can incorporate a number of tensors from the central boundary into the left window, moving the \(\Lambda_L\)~matrix past these tensors until it is sufficiently close to the boundary \(\Lambda\)~matrix to some degree of tolerance, and the same with the right window.
We can then form the single window using the orthogonality centre \(\Lambda\) (which \(\Lambda_L \Lambda^{-1} \Lambda_R\) will also be approximately equal to anyway): this new window should have significantly reduced distortion.

\section{Transforming an EA wave function into a particular gauge}\label{app:ea-gauge-transform}
In Sec.~\ref{sec:gauge-fixing} in the main text, we discussed how the window tensor in an EA wave function has redundant gauge degrees of freedom, which can be partially fixed by enforcing either the left- or right-gauge-fixing conditions.
These can be explicitly enforced by writing the \(B\)~tensor in the form \eqref{eq:gauge-fixing-null-space}
\begin{equation}
    B^s =
    \begin{tikzpicture}[tensornetwork]
        \node[ltensor, striped, label=\(N_L\)] (A1) at (1, 0) {};
        \node[atensor, fill=lightgray, label=\(X\)] (A2) at (2, 0) {};
        \draw (A1.south) -- +(0, -0.5);
        \draw (A1.west) -- +(-0.5, 0);
        \draw (A2.east) -- +(0.5, 0);
        \draw (A1) -- (A2);
    \end{tikzpicture}
    \quad\text{or}\quad
    B^s =
    \begin{tikzpicture}[tensornetwork]
        \node[atensor, fill=lightgray, label=\(Y\)] (A1) at (1, 0) {};
        \node[rtensor, striped, label=\(N_R\)] (A2) at (2, 0) {};
        \draw (A2.south) -- +(0, -0.5);
        \draw (A1.west) -- +(-0.5, 0);
        \draw (A2.east) -- +(0.5, 0);
        \draw (A1) -- (A2);
    \end{tikzpicture}
    ,
\end{equation}
respectively.
We will usually write the EA window directly in this form, and only modify the degrees of freedom in the \(X\) or \(Y\)~matrix, so that the wave function will always automatically obey this gauge-fixing condition.
However, there may be situations where we have some arbitrary EA wave function that we want to make obey one of these gauge-fixing conditions, or we could want to change a wave function from the left gauge to the right gauge.
If we have some EA wave function with momentum~\(k\) and window tensor~\(B\), then we can transform it into the right gauge with window \(B' = Y N_R\) by enforcing that the inner product of the original wave function with the wave function in the right gauge is equal to one, that is
\begin{equation}
    \begin{tikzpicture}[baseline=-0.25em]
        \node[atensor, fill=lightgray, label=above:\(B\)] (B) at (0, 0.5) {};
        \node[atensor, fill=lightgray, label=below:\(Y\)] (Y) at (-0.5, -0.5) {};
        \node[rtensor, striped, label=below:\(N_R\)] (NR) at (0, -0.5) {};
        \draw (Y) -- (NR);
        \draw (B) -- (NR);
        \draw[rounded corners=5pt] (Y.west) -- +(-0.25, 0) -- +(-0.25, 1) -- (B.west);
        \draw[rounded corners=5pt] (B.east) -- +(0.25, 0) -- +(0.25, -1) -- (NR.east);
    \end{tikzpicture}
    + \erm^{\irm k}
    \begin{tikzpicture}[baseline=-0.25em]
        \node[ltensor, label=above:\(A\)] (A) at (0, 0.5) {};
        \node[atensor, fill=lightgray, label=below:\(Y\)] (Y) at (-0.5, -0.5) {};
        \node[rtensor, striped, label=below:\(N_R\)] (NR) at (0, -0.5) {};
        \draw (Y) -- (NR);
        \draw (A) -- (NR);
        \draw[rounded corners=5pt] (Y.west) -- +(-0.25, 0) -- +(-0.25, 1) -- (A.west);
        \draw (A.east) -- +(0.25, 0);
        \draw (NR.east) -- (NR.east -| A.east) -- +(0.25, 0);
    \end{tikzpicture}
    \left[I - \erm^{\irm k}
    \begin{tikzpicture}[baseline=-0.25em]
        \node[ltensor, label=above:\(A\)] (A) at (0, 0.5) {};
        \node[rtensor, label=below:\(\tilde{A}\)] (C) at (0, -0.5) {};
        \draw (A) -- (C);
        \draw (A.west) -- +(-0.25, 0);
        \draw (C.west) -- +(-0.25, 0);
        \draw (A.east) -- +(0.25, 0);
        \draw (C.east) -- +(0.25, 0);
    \end{tikzpicture}
    \right]^{-1}
    \begin{tikzpicture}[baseline=-0.25em]
        \node[atensor, fill=lightgray, label=above:\(B\)] (B) at (0, 0.5) {};
        \node[rtensor, label=below:\(\tilde{A}\)] (C) at (0, -0.5) {};
        \draw (B) -- (C);
        \draw (B.west) -- +(-0.25, 0);
        \draw (C.west) -- +(-0.25, 0);
        \draw[rounded corners=5pt] (B.east) -- +(0.25, 0) -- +(0.25, -1) -- (C.east);
    \end{tikzpicture}
    = 1
\end{equation}
(noting that any term with \(B\) to the left of \(Y N_R\) is zero owing to the right-gauge-fixing condition).
By solving for \(Y\), we obtain
\begin{equation}
    Y =
    \begin{tikzpicture}[baseline=-0.25em]
        \node[atensor, fill=lightgray, label=above:\(B\)] (B) at (0, 0.5) {};
        \node[notensor] (Y) at (-0.5, -0.5) {};
        \node[rtensor, striped, label=below:\(N_R\)] (NR) at (0, -0.5) {};
        \draw (Y) -- (NR);
        \draw (B) -- (NR);
        \draw[rounded corners=5pt] (Y.west) -- +(-0.25, 0) -- +(-0.25, 1) -- (B.west);
        \draw[rounded corners=5pt] (B.east) -- +(0.25, 0) -- +(0.25, -1) -- (NR.east);
    \end{tikzpicture}
    + \erm^{\irm k}
    \begin{tikzpicture}[baseline=-0.25em]
        \node[ltensor, label=above:\(A\)] (A) at (0, 0.5) {};
        \node[notensor] (Y) at (-0.5, -0.5) {};
        \node[rtensor, striped, label=below:\(N_R\)] (NR) at (0, -0.5) {};
        \draw (Y) -- (NR);
        \draw (A) -- (NR);
        \draw[rounded corners=5pt] (Y.west) -- +(-0.25, 0) -- +(-0.25, 1) -- (A.west);
        \draw (A.east) -- +(0.25, 0);
        \draw (NR.east) -- (NR.east -| A.east) -- +(0.25, 0);
    \end{tikzpicture}
    \left[I - \erm^{\irm k}
    \begin{tikzpicture}[baseline=-0.25em]
        \node[ltensor, label=above:\(A\)] (A) at (0, 0.5) {};
        \node[rtensor, label=below:\(\tilde{A}\)] (C) at (0, -0.5) {};
        \draw (A) -- (C);
        \draw (A.west) -- +(-0.25, 0);
        \draw (C.west) -- +(-0.25, 0);
        \draw (A.east) -- +(0.25, 0);
        \draw (C.east) -- +(0.25, 0);
    \end{tikzpicture}
    \right]^{-1}
    \begin{tikzpicture}[baseline=-0.25em]
        \node[atensor, fill=lightgray, label=above:\(B\)] (B) at (0, 0.5) {};
        \node[rtensor, label=below:\(\tilde{A}\)] (C) at (0, -0.5) {};
        \draw (B) -- (C);
        \draw (B.west) -- +(-0.25, 0);
        \draw (C.west) -- +(-0.25, 0);
        \draw[rounded corners=5pt] (B.east) -- +(0.25, 0) -- +(0.25, -1) -- (C.east);
    \end{tikzpicture}
    .
\end{equation}
Here, the expression \((I-T)^{-1}\) (where \(T\) is the mixed transfer matrix) is the closed form of the geometric series \(I + T + T^2 + \ldots\).
We do not need to calculate this inverse explicitly, but only its action on the terms on its right, which can be done efficiently using a linear solver such as GMRES~\cite{saad1986}.
Note that if there is any component in the original EA wave function that cannot be written in the right gauge (i.e.\ any part proportional to the background state), it will be projected out in this procedure.

On a slightly tangential—but still important—note, although an EA wave function can be equivalently written in the left- or right-gauge-fixing conditions, the individual components of the wave function where the window is on site~\(n\) will \emph{not} be equivalent for both gauge-fixing conditions.
With the left-gauge-fixing condition, the area of this particular component that is affected by the window is biased to the left of the window at site~\(n\) (similarly, the right gauge mainly affects the area to the right of the window).
For this reason, Ref.~\cite{van-damme2021a} introduced a ‘central gauge’ that aims to make the make the part of the wave function component affected by the window centred around the window: this makes it so that the derived real-space wave packets discussed in that work are more symmetric.

\section{Fractionalized excitations and quantum numbers}\label{app:fractionalized-qns}
In Ref.~\cite{zauner-stauber2018b}, the authors showed that it is possible to represent excited states with fractionalized quantum numbers by using a topologically nontrivial MPS excitation ansatz—that is, one with different left and right background states.
This method requires us to write the MPS tensors in terms of conserved quantum numbers (QNs)~\cite{mcculloch2007}.
If we want to write a translation-invariant iMPS with conserved QNs, we use a repeating unit cell, which is identical up to some possible shift in the QNs per unit cell.
For instance, for the Hubbard model~\eqref{eq:hubbard} at unpolarized half filling, if we conserve QNs for the \(\mathrm{U}(1)\) particle number symmetry and \(\mathrm{U}(1)\) spin-projection symmetry, we need to use a two-site unit cell with a particle QN shift of \(2\) and spin-projection QN shift of \(0\) per unit cell (we denote this combined quantum number as an ordered tuple \((2, 0)\)).
Importantly, we \emph{cannot} use a single-site unit cell here if we enforce these symmetries, since we would need a QN shift of \((1, 0)\) per single-site unit cell, but the local basis states for each site \(\ket{0}\), \(\ket{\uparrow}\), \(\ket{\downarrow}\), and \(\ket{\uparrow\downarrow}\) have QNs \((0, 0)\), \((1, 1/2)\), \((1, -1/2)\), and \((2, 0)\) respectively, which cannot make up the required \((1, 0)\) per unit cell.

If we have a state with a nontrivial QN shift as the above example, the MPS tensors will be translation invariant up to shifting the QN labels by this value per unit cell.
In Ref.~\cite{zauner-stauber2018b}, an alternative formulation is presented where the MPS tensors are made fully translation invariant, but the QN labels for the local indices are shifted instead.
These two formulations are equivalent, but would need to be implemented in different ways.

For an EA wave function, if we want to specify the QN of the excitation, we need to add a fourth ‘dummy’ index to the EA window tensor (this would be the \(X\)~matrix when applying the gauge-fixing condition \eqref{eq:gauge-fixing-null-space}; for a multi-site window, this would typically be the tensor being used as the orthogonality centre): the QN of this index then describes the QN of the excitation (compare Ref.~\cite{vanderstraeten2020}).
If we have only Abelian QNs (or if the excitation QN is non-Abelian but has a quantum dimension of one), then we can merge this dummy index with the left virtual index, effectively shifting the QNs of the left boundary by the excitation QN\@.
(For non-Abelian QNs, we need to keep this dummy index explicitly: we leave the details of how to do this to another study.)

The excitation QN can only take certain values depending on the QNs of the virtual bases of the left and right boundary wave functions: for a topologically trivial excitation, this will generally only correspond to ‘integer’ excitations.
However, we can also represent \emph{fractionalized} excitations if we form a topologically nontrivial excitation where we rotate the unit cell of the right boundary by \(n\) sites (such that the \(m\)th (single-site) window occurs between site \(m-1\) of the left boundary and site \(n+m+1\) of the right boundary, modulo the unit cell size).
Intuitively, we can understand this as effectively shifting the excitation QN by the background’s QN shift per unit cell ‘divided’ by \(n\).
Considering this unpolarized half-filing state in the Hubbard model with a two-site unit cell, if we rotate the right boundary by a single site, we effectively shift the QN of the excitation by \((1, 0)\), allowing us to represent chargon and spinon excitations with QNs \((1, 0)\) and \((0, 1/2)\), by using window QNs of \((0, 0)\) and \((-1, 1/2)\), respectively.

\section{Note on boundary terms}\label{app:boundary}
When we are working with infinite states, in order to calculate the expectation value of an MPO on a section of the lattice of length~\(n\), we need to enforce a condition for terms that cross the boundary~\cite{michel2010}.
For example, for a Hamiltonian with nearest-neighbour and on-site terms, the on-site term will occur \(n\) times in this section, and the nearest-neighbour term will appear \(n-1\) times if we ignore the contributions crossing the boundary, or \(n\) or \(n+1\) times if we include this term for one or both boundaries, respectively.
These conditions are somewhat arbitrary, and have no effect on the final asymptotic results, but are nevertheless vital when considering the expectation value over a small region.
We implicitly fix these conditions when we choose the initial element of the \(E\)~matrix \(\pbra{E^{111}(n)}\) to be a constant equal to the left eigenvector of the transfer matrix.
Furthermore, whenever the transfer matrix has an eigenvalue~\(\erm^{\irm Q}\) on the unit circle, there is freedom in the choice of the parallel zero-degree component with momentum~\(Q\), which we fix by Eq.~\eqref{eq:case-3-parallel-3} in order to remove spurious subleading boundary contributions~\cite{michel2010}.

\subsection{Fixing the boundary terms for the effective Hamiltonian}\label{app:heff-boundary}
When constructing the effective Hamiltonian acting on the window tensor of an EA wave function \eqref{eq:heff}, we need to choose these boundary terms appropriately such that the eigenvalues of this effective Hamiltonian correspond to the constant shift~\(\Delta\) in the expectation value owing to the excitation~\eqref{eq:h-exp}.
To do this, we can correct the boundary contributions to the final MPO element for either the left-boundary term of the \(E\)~matrix \(\pbra{E^{w11}(n)}\) or the right-boundary term of the \(F\)~matrix \(\pket{F_{1NN}(n)}\), where \(w\) is the MPO bond dimension, and \(N\) is the number of blocks in the block-triangular \(\mathcal{A}\) tensor.
We choose to correct \(\pbra{E^{w11}(n)}\).
These boundary contributions can be corrected by subtracting two contributions from the parallel part of the constant zero-momentum component~\(e^{w1101}_\parallel\): the energy per unit cell, given by the linear component~\(e^{w1111}_\parallel\), and the ‘bond energy’, which is the energy contribution for all terms that cross the unit cell boundary.
The simplest way to calculate this bond energy is to calculate the \(F\)~matrix for the left boundary \(\pket{\tilde{F}_{\omega}(n)}\) (that is, the \(F\)~matrix for left boundary wave function considered by itself as an iMPS, \emph{not} \(\pket{F_{\omega11}(n)}\)).
The bond energy is then given by the inner product of the constant zero-momentum components \(\sum_\omega \pbraket{E^{\omega 1101}|\tilde{F}_{\omega01}}\) (provided that we fix the boundary term for the zero-degree components as in Eq.~\eqref{eq:case-3-parallel-3}).

\section{Floating-point precision of linear solvers}\label{app:gmres}

The key component of the MPO fixed point is solving the set of linear equations of Eq.~\eqref{eq:case-2}, which can be cast in the form
\begin{equation}\label{eq:linear}
   (I - T)(x) = C ,
\end{equation}
solving for the matrix~$x$, given the (generalized) transfer operator~$T$ and right-hand-side matrix~$C$. This is equivalent to the limit of the geometric series
\begin{equation}\label{eq:sequence}
   x = C + T(C) + T(T(C)) + T(T(T(C))) + \cdots .
\end{equation}
For a given floating-point precision with unit round-off~$u$, it is possible, in principle, to evaluate $x$ to essentially full precision using Eq.~\eqref{eq:sequence} iteratively, via $x_{n+1} = T(x_n) + C$ with $x_0 = C$, however the convergence will be rather slow. The convergence rate is given by $\lambda_1$, being the largest magnitude eigenvalue of $T$ smaller than 1, with the number of iterations required to converge to precision $\epsilon$ being $N \simeq \ln \epsilon / \ln \lambda_1 = - \xi \ln \epsilon$, where $\xi = -1/\ln \lambda_1$ is the physical correlation length per wave function unit cell. It is much more efficient to instead compute the solution of Eq.~\eqref{eq:linear}
directly, using a linear solver. The restarted GMRES algorithm~\cite{saad1986} is ideal for this task, since it has the best convergence, and the memory requirements of storing a Krylov sequence of matrices is not excessive. However, it is not possible to obtain the solution to full floating-point precision using GMRES alone. Given a residual norm of $\epsilon_r$, the numerical approximation~$\hat{x}$ differs from the exact solution~$x$ by $\frac{\norm{x - \hat{x}}}{\norm{x}} \lesssim \kappa_\infty\epsilon_r$, where $\kappa_\infty$ is the matrix condition number in the infinity norm~\cite{carson2020}. Hence, the accuracy of GMRES for floating-point unit round-off~$u$ is bounded by $k_\infty u$.

\begin{algorithm}[t]
    \caption{Linear solver with iterative refinement.}
    \label{alg:gmres}
    \SetKwInOut{Input}{Input}\SetKwInOut{Output}{Output}
    \Input{Generalized transfer operator~$T$, fixed right-hand-side matrix~$C$, and desired precision~$\epsilon_r$}
    \Output{Solution~$x$ of $(I-T)(x) = C$}
    \BlankLine
    Initialize $n=0$, $x_0 = 0$, $r_0 = C$. \\
    \While{$\norm{r_n} > \epsilon_r$}
    {
      Solve for $d_n$ using GMRES $(I-T)(d_n) = r_n$. \\
      Compute $x_{n+1} = x_n + d_n$. \\
      Compute $r_{n+1} = C - (I-T)(x_{n+1})$. \\
      Update $n \leftarrow n+1$.
   }
\end{algorithm}

This accuracy bound can be improved using iterative refinement. The basic idea is that starting with an initial solution~$x_1$, the linear solver is used a second time, using the residual vector itself as the right-hand side. This gives a correction term that can be added to $x_1$ to obtain a refined solution~$x_2$. This procedure is described in Algorithm~\ref{alg:gmres}.
Even for fixed precision~$u$, iterative refinement improves the forward error bound to $\kappa u$, where $\kappa$ is the condition number of $(I-T)$ in the 2-norm. Since the largest eigenvalue of $(I-T)$ will be $\simeq 1$, and the smallest eigenvalue is $\simeq 1 - |\lambda_1|$, this gives $\kappa \simeq 1 - |\lambda_1| \simeq -1 / \ln \lambda_1 = \xi$. This error bound of $\xi u$ is a general improvement, since it is always the case~\cite{higham2002} that $\kappa < \kappa_\infty$, especially when $(I-T)$ has badly scaled rows. In practice, we find this iterative refinement helpful.

It is possible to improve this error bound further using mixed precision arithmetic. The key step is calculating the residual vector $r_n = C - (I-T)(x_n)$, where cancellation can occur that destroys the accuracy. Using extra precision in calculating $r_n$ ensures that there are enough correct digits to give a correction~$d_n$ that improves the solution~$x_{n+1}$. Typically this is done using precision~$u^2$ (so if the overall calculation uses double precision then the residual is calculated using quadruple precision). This is relatively cheap as this only needs to be done once per iterative refinement loop. Remarkably, it is possible to go further and use \emph{three} precisions~\cite{carson2020,amestoy2024}, where the bulk of the computations of the contraction $(I-T)(x)$ can be done in single-precision,\footnote{In fact it could even be done in half precision, however the convergence is only guaranteed if $\kappa_\infty$ is smaller than $1/u$, which might not be satisfied for such a small precision.} which makes this scheme particularly attractive for modern hardware such as GPUs where there are significant speedups for single-precision arithmetic.

\bibliography{ref}
\end{document}

%% file: figures/heisenberg.tex
\begin{tikzpicture}
    \begin{axis}[width=\linewidth, height=0.45\linewidth, yshift=0.90\linewidth-70,
                 xmin=0, xmax=1, ymin=0, ymax=3.75,
                 xtick=\empty, ytick=\empty]
    \end{axis}
    \begin{axis}[width=\linewidth, height=0.45\linewidth, yshift=0.90\linewidth-70,
                 xmin=0, xmax=1, ymin=0, ymax=3.75,
                 xticklabels={},
                 ylabel=\(\Delta\),
                 cycle list/Reds-3, cycle list/Blues-3, cycle list/Greens-3]
        \addplot+[name path=3A, very thick, no markers, index of colormap=2 of Reds-3] table[y expr=\thisrowno{4}] {data/heisenberg/continua.dat};
        \addplot+[name path=3B, very thick, no markers, index of colormap=2 of Reds-3] table[y expr=\thisrowno{3}] {data/heisenberg/continua.dat};
        \addplot+[index of colormap=1 of Reds-3, opacity=0.5] fill between[of=3A and 3B];
        \addplot+[name path=2A, very thick, no markers, index of colormap=2 of Blues-3] table[y expr=\thisrowno{2}] {data/heisenberg/continua.dat};
        \addplot+[name path=2B, very thick, no markers, index of colormap=2 of Blues-3] table[y expr=\thisrowno{1}] {data/heisenberg/continua.dat};
        \addplot+[index of colormap=1 of Blues-3, opacity=0.5] fill between[of=2A and 2B];
        \addplot+[very thick, dotted, no markers, black] table {data/heisenberg/data-m150-w1.dat};
        \addplot+[very thick, no markers, black] table[skip first n=24] {data/heisenberg/data-m150-w1.dat};
        \node[anchor=north west] at (rel axis cs: 0, 1) {(a)};
        \node[index of colormap=2 of Reds-3] at (0.89, 2.2) {3 magnon};
        \node[index of colormap=2 of Blues-3] at (0.12, 2.2) {2 magnon};
    \end{axis}
    \begin{semilogyaxis}[width=\linewidth, height=0.45\linewidth, yshift=0.45\linewidth-35,
                 xmin=0, xmax=1,
                 xticklabels={},
                 ylabel=\(\sigma_\Delta^2\),
                 every axis y label/.style={at={(ticklabel* cs:0.65)}, anchor=south, xshift=-10pt, rotate=90},
                 cycle list/Reds-3, cycle list/Blues-3, cycle list/Greens-3]
        \addplot+[only marks, mark size=0.7pt, index of colormap=0 of Reds-3] table[y expr=\thisrowno{2}] {data/heisenberg/data-m50-w1.dat};
        \addplot+[only marks, mark size=0.7pt, index of colormap=1 of Reds-3] table[y expr=\thisrowno{2}] {data/heisenberg/data-m50-w2.dat};
        \addplot+[only marks, mark size=0.7pt, index of colormap=2 of Reds-3] table[y expr=\thisrowno{2}] {data/heisenberg/data-m50-w3.dat};
        \addplot+[only marks, mark=square*, mark options={rotate=45}, mark size=0.566pt, index of colormap=0 of Blues-3] table[y expr=\thisrowno{2}] {data/heisenberg/data-m100-w1.dat};
        \addplot+[only marks, mark=square*, mark options={rotate=45}, mark size=0.566pt, index of colormap=1 of Blues-3] table[y expr=\thisrowno{2}] {data/heisenberg/data-m100-w2.dat};
        \addplot+[only marks, mark=square*, mark options={rotate=45}, mark size=0.566pt, index of colormap=2 of Blues-3] table[y expr=\thisrowno{2}] {data/heisenberg/data-m100-w3.dat};
        \addplot+[only marks, mark=square*, mark size=0.566pt, index of colormap=0 of Greens-3] table[y expr=\thisrowno{2}] {data/heisenberg/data-m150-w1.dat};
        \addplot+[only marks, mark=square*, mark size=0.566pt, index of colormap=1 of Greens-3] table[y expr=\thisrowno{2}] {data/heisenberg/data-m150-w2.dat};
        \addplot+[only marks, mark=square*, mark size=0.566pt, index of colormap=2 of Greens-3] table[y expr=\thisrowno{2}] {data/heisenberg/data-m150-w3.dat};
        \addplot+[no markers, index of colormap=2 of Reds-3] {4.7962156490833e-07};
        \addplot+[no markers, index of colormap=2 of Blues-3] {2.1808573613669e-08};
        \addplot+[no markers, index of colormap=2 of Greens-3] {9.9515817808538e-10};
        \node[anchor=north west] at (rel axis cs: 0, 0.5665) {(b)};
    \end{semilogyaxis}
    \begin{semilogyaxis}[width=\linewidth, height=0.45\linewidth,
                 xmin=0, xmax=1, y dir=reverse,
                 yticklabels={\(-10^{-9}\), \(-10^{-7}\), \(-10^{-5}\)},
                 xlabel=\(k/\pi\),
                 ylabel=\(\sigma_\Delta^2\),
                 every axis y label/.style={at={(ticklabel* cs:0.7)}, anchor=south, xshift=-10pt, rotate=90},
                 every axis x label/.style={at={(ticklabel* cs:0.5)}, anchor=north, yshift=-5pt},
                 cycle list/Reds-3, cycle list/Blues-3, cycle list/Greens-3,
                 legend style={at={(0.02, 0.04)}, anchor=south west, legend columns=4, legend transposed=true, font=\scriptsize}]
        \addlegendimage{empty legend}
        \addplot+[only marks, mark size=0.7pt, index of colormap=0 of Reds-3] table[y expr=-\thisrowno{2}] {data/heisenberg/data-m50-w1.dat};
        \addplot+[only marks, mark size=0.7pt, index of colormap=1 of Reds-3] table[y expr=-\thisrowno{2}] {data/heisenberg/data-m50-w2.dat};
        \addplot+[only marks, mark size=0.7pt, index of colormap=2 of Reds-3] table[y expr=-\thisrowno{2}] {data/heisenberg/data-m50-w3.dat};
        \addlegendimage{empty legend}
        \addplot+[only marks, mark=square*, mark options={rotate=45}, mark size=0.566pt, index of colormap=0 of Blues-3] table[y expr=-\thisrowno{2}] {data/heisenberg/data-m100-w1.dat};
        \addplot+[only marks, mark=square*, mark options={rotate=45}, mark size=0.566pt, index of colormap=1 of Blues-3] table[y expr=-\thisrowno{2}] {data/heisenberg/data-m100-w2.dat};
        \addplot+[only marks, mark=square*, mark options={rotate=45}, mark size=0.566pt, index of colormap=2 of Blues-3] table[y expr=-\thisrowno{2}] {data/heisenberg/data-m100-w3.dat};
        \addlegendimage{empty legend}
        \addplot+[only marks, mark=square*, mark size=0.566pt, index of colormap=0 of Greens-3] table[y expr=-\thisrowno{2}] {data/heisenberg/data-m150-w1.dat};
        \addplot+[only marks, mark=square*, mark size=0.566pt, index of colormap=1 of Greens-3] table[y expr=-\thisrowno{2}] {data/heisenberg/data-m150-w2.dat};
        \addplot+[only marks, mark=square*, mark size=0.566pt, index of colormap=2 of Greens-3] table[y expr=-\thisrowno{2}] {data/heisenberg/data-m150-w3.dat};
        \addplot+[no markers, index of colormap=2 of Reds-3] {4.7962156490833e-07};
        \addplot+[no markers, index of colormap=2 of Blues-3] {2.1808573613669e-08};
        \addplot+[no markers, index of colormap=2 of Greens-3] {9.9515817808538e-10};
        \legend{~, {50, 1}, {50, 2}, {50, 3}, {\(D, N\)}, {100, 1}, {100, 2}, {100, 3}, ~, {150, 1}, {150, 2}, {150, 3}};
        \node[anchor=north west] at (rel axis cs: 0, 0.925) {(c)};
    \end{semilogyaxis}
\end{tikzpicture}

%% file: figures/heisenberg-window-size.tex
\begin{tikzpicture}
    \begin{axis}[width=0.55\linewidth, height=0.4\linewidth, yshift=0.8\linewidth-45,
                 xmin=1, xmax=20,
                 xticklabels={},
                 ylabel=\(\Delta-\Delta_\text{ref}\),
                 cycle list/Reds-3]
        \addplot+[thick, mark=*, mark size=1.5pt, index of colormap=2 of Reds-3] table[y expr=\thisrowno{1}-0.410479248463] {data/heisenberg/data-k1.00-m50.dat};
        \addplot+[no markers, index of colormap=2 of Reds-3, domain=1:20] {0};
        \node[anchor=north west] at (rel axis cs: 0, 0.9) {(a) \(k = \pi\)};
    \end{axis}
    \begin{axis}[width=0.55\linewidth, height=0.4\linewidth, yshift=0.8\linewidth-45, xshift=0.45\linewidth,
                 xmin=1, xmax=20,
                 xticklabels={},
                 cycle list/Blues-3]
        \addplot+[thick, mark=*, mark size=1.5pt, index of colormap=2 of Blues-3] table[y expr=\thisrowno{1}-2.31283] {data/heisenberg/data-k0.30-m50.dat};
        \addplot+[no markers, index of colormap=2 of Blues-3, domain=1:20] {0};
        \node[anchor=north west] at (rel axis cs: 0, 1) {(b) \(k = 0.3\pi\)};
    \end{axis}
    \begin{axis}[width=0.55\linewidth, height=0.4\linewidth, yshift=0.4\linewidth-15,
                 xmin=1, xmax=20,
                 xlabel=Window size \(N\),
                 ylabel=\(\sigma_\Delta^2\),
                 cycle list/Reds-3]
        \addplot+[thick, mark=*, mark size=1.5pt, index of colormap=2 of Reds-3] table[y expr=\thisrowno{2}] {data/heisenberg/data-k1.00-m50.dat};
        \node[anchor=north west] at (rel axis cs: 0, 1) {(c) \(k = \pi\)};
    \end{axis}
    \begin{axis}[width=0.55\linewidth, height=0.4\linewidth, yshift=0.4\linewidth-15, xshift=0.45\linewidth,
                 xmin=1, xmax=20,
                 xlabel=Window size \(N\),
                 cycle list/Blues-3]
        \addplot+[thick, mark=*, mark size=1.5pt, index of colormap=2 of Blues-3] table[y expr=\thisrowno{2}] {data/heisenberg/data-k0.30-m50.dat};
        \node[anchor=north west] at (rel axis cs: 0.1, 1) {(d) \(k = 0.3\pi\)};
    \end{axis}
\end{tikzpicture}

%% file: figures/heisenberg-bond-dimension.tex
\begin{tikzpicture}
    \begin{semilogyaxis}[width=0.5\linewidth, height=0.4\linewidth,
                 xmin=20, xmax=150,
                 xlabel=Bond dimension \(D\),
                 ylabel=\(|\Delta-\Delta_\text{ref}|\),
                 cycle list/Reds-3]
        \addplot+[thick, mark=*, mark size=1.5pt, index of colormap=2 of Reds-3] table[y expr=\thisrowno{2}] {data/heisenberg/data-k1.00-w1.dat};
        \node[anchor=north west] at (rel axis cs: 0.05, 1) {(a)};
    \end{semilogyaxis}
    \begin{semilogyaxis}[width=0.5\linewidth, height=0.4\linewidth, xshift=0.5\linewidth,
                 xmin=20, xmax=150, y dir=reverse,
                 yticklabels={\(-10^{-8}\), \(-10^{-5}\), \(-10^{-2}\)},
                 xlabel=Bond dimension \(D\),
                 every axis y label/.style={at={(ticklabel* cs:0.75)}, anchor=south, xshift=-10pt, rotate=90},
                 ylabel=\(\sigma_\Delta^2\),
                 cycle list/Reds-3]
        \addplot+[thick, mark=*, mark size=1.5pt, index of colormap=2 of Reds-3] table[y expr=-\thisrowno{3}] {data/heisenberg/data-k1.00-w1.dat};
        \node[anchor=north west] at (rel axis cs: 0, 1) {(b)};
    \end{semilogyaxis}
\end{tikzpicture}

%% file: figures/heisenberg-derivatives.tex
\begin{tikzpicture}
    \begin{axis}[width=\linewidth, height=0.40\linewidth, yshift=0.40\linewidth-35,
                 xmin=0, xmax=1,
                 xticklabels={},
                 ylabel=\(\drm\Delta/\drm k\),
                 cycle list/Reds-3, cycle list/Blues-3, cycle list/Greens-3,
                 legend style={at={(0.5, 1.05)}, anchor=south, legend columns=2, legend transposed=false, font=\scriptsize}]
        \addplot+[very thick, no markers, index of colormap=2 of Reds-3] table {data/heisenberg/derivative-dk0.01-m150.dat};
        \addplot+[very thick, no markers, index of colormap=2 of Blues-3] table {data/heisenberg/derivative-dk0.005-m150.dat};
        \addplot+[thick, no markers, index of colormap=2 of Greens-3] table {data/heisenberg/derivative-m150.dat};
        \addplot+[thick, no markers, dashed, black] table {data/heisenberg/derivative-fd-m150.dat};
        \legend{\(\Delta k = 0.01\pi\), \(\Delta k = 0.005\pi\), Simplified, Finite difference};
        \node[anchor=north west] at (rel axis cs: 0, 1) {(a)};
    \end{axis}
    \begin{axis}[width=0.35\linewidth, height=0.30\linewidth, yshift=0.43\linewidth-35, xshift=0.075\linewidth,
                 xmin=0.2, xmax=0.3, xtick={0.20, 0.25, 0.30},
                 every tick label/.append style={font=\tiny},
                 cycle list/Reds-3, cycle list/Blues-3, cycle list/Greens-3]
        \addplot+[thick, no markers, index of colormap=2 of Reds-3] table {data/heisenberg/derivative-dk0.01-m150.dat};
        \addplot+[thick, no markers, index of colormap=2 of Blues-3] table {data/heisenberg/derivative-dk0.005-m150.dat};
        \addplot+[thick, no markers, index of colormap=2 of Greens-3] table {data/heisenberg/derivative-m150.dat};
        \addplot+[thick, no markers, dashed, black] table {data/heisenberg/derivative-fd-m150.dat};
    \end{axis}
    \begin{axis}[width=\linewidth, height=0.40\linewidth,
                 xmin=0, xmax=1,
                 xticklabels={},
                 ylabel=\(\drm^2\Delta/\drm k^2\),
                 cycle list/Reds-3, cycle list/Blues-3, cycle list/Greens-3]
        \addplot+[very thick, no markers, index of colormap=2 of Reds-3] table {data/heisenberg/derivative2-dk0.01-m150.dat};
        \addplot+[very thick, no markers, index of colormap=2 of Blues-3] table {data/heisenberg/derivative2-dk0.005-m150.dat};
        \addplot+[thick, no markers, index of colormap=2 of Greens-3] table {data/heisenberg/derivative2-m150.dat};
        \addplot+[thick, no markers, dashed, black] table {data/heisenberg/derivative2-fd-m150.dat};
        \node[anchor=north west] at (rel axis cs: 0, 1) {(b)};
    \end{axis}
    \begin{axis}[width=0.35\linewidth, height=0.30\linewidth, yshift=0.09\linewidth, xshift=0.2\linewidth,
                 xmin=0.2, xmax=0.3, ymax=0.2, xtick={0.20, 0.25, 0.30},
                 every tick label/.append style={font=\tiny},
                 cycle list/Reds-3, cycle list/Blues-3, cycle list/Greens-3]
        \addplot+[thick, no markers, index of colormap=2 of Reds-3] table {data/heisenberg/derivative2-dk0.01-m150.dat};
        \addplot+[thick, no markers, index of colormap=2 of Blues-3] table {data/heisenberg/derivative2-dk0.005-m150.dat};
        \addplot+[thick, no markers, index of colormap=2 of Greens-3] table {data/heisenberg/derivative2-m150.dat};
        \addplot+[thick, no markers, dashed, black] table {data/heisenberg/derivative2-fd-m150.dat};
    \end{axis}
    \begin{axis}[width=\linewidth, height=0.40\linewidth, yshift=-0.40\linewidth+35,
                 xmin=0, xmax=1, ymin=0,
                 xlabel=\(k/\pi\),
                 ylabel=Norm,
                 every axis x label/.style={at={(ticklabel* cs:0.5)}, anchor=north, yshift=-5pt},
                 cycle list/Purples-3, cycle list/Oranges-3,
                 legend style={at={(0.98, 0.95)}, anchor=north east, legend columns=2, legend transposed=true, font=\scriptsize}]
        \addplot+[very thick, no markers, index of colormap=2 of Purples-3] table[y expr=\thisrowno{1}] {data/heisenberg/derivativeB-dk0.01-m150.dat};
        \addplot+[very thick, no markers, index of colormap=2 of Oranges-3] table[y expr=\thisrowno{2}] {data/heisenberg/derivativeB-dk0.01-m150.dat};
        \legend{\(\|\drm B/\drm k\|\), \(\|\drm^2B/\drm k^2\|\)};
        \node[anchor=north west] at (rel axis cs: 0, 1) {(c)};
    \end{axis}
\end{tikzpicture}

%% file: figures/hubbard.tex
\begin{tikzpicture}
    \begin{axis}[width=\linewidth, height=0.55\linewidth, yshift=0.30\linewidth-30,
                 xmin=0, xmax=1, ymin=0.75, ymax=5.25,
                 xtick=\empty, ytick=\empty]
    \end{axis}
    \node at (0, 0.615\linewidth) {};
    \begin{axis}[width=\linewidth, height=0.55\linewidth, yshift=0.30\linewidth-30,
                 xmin=0, xmax=1, ymin=0.75, ymax=5.25,
                 ytick={1, 3, 5},
                 xticklabels={},
                 ylabel=\(\Delta\),
                 cycle list/Reds-3, cycle list/Purples-3,
                 legend style={at={(0.5, 1.05)}, anchor=south, legend columns=2, font=\scriptsize}]
        \addplot+[forget plot, name path=2A, very thick, no markers, index of colormap=2 of Purples-3] table[y expr=\thisrowno{2}] {data/hubbard/continua.dat};
        \addplot+[forget plot, name path=2B, very thick, no markers, index of colormap=2 of Purples-3] table[y expr=\thisrowno{1}] {data/hubbard/continua.dat};
        \addplot+[forget plot, index of colormap=1 of Purples-3, opacity=0.5] fill between[of=2A and 2B];
        \addplot+[forget plot, only marks, mark size=0.8pt, index of colormap=1 of Reds-3] table {data/hubbard/data-c.dat};
        \addplot+[only marks, mark size=0.8pt, index of colormap=1 of Reds-3] table[x expr=1-\thisrowno{0}] {data/hubbard/data-c.dat};
        \addplot+[only marks, mark size=0.8pt, index of colormap=2 of Reds-3] table {data/hubbard/data-c2.dat};
        \legend{Energy minimization~, Variance minimization};
        \node[anchor=north west] at (rel axis cs: 0, 1) {(a)};
        \node[index of colormap=2 of Reds-3] at (0.2, 3) {Chargons};
        \node[index of colormap=2 of Purples-3, align=center] at (0.75, 3) {Charge–spin–spin\\continuum};
    \end{axis}
    \begin{axis}[width=0.4\linewidth, height=0.28\linewidth, yshift=0.57\linewidth-35, xshift=25,
                 xmin=0, xmax=1, ymin=0, ymax=1.25,
                 xtick={0, 0.5, 1}, ytick={0, 1},
                 every tick label/.append style={font=\footnotesize},
                 cycle list/Blues-3]
        \fill[white, opacity=0.5] (rel axis cs: 0, 0) -- (rel axis cs: 0, 1) -- (rel axis cs: 1, 1) -- (rel axis cs: 1, 0) -- cycle;
        \addplot+[only marks, mark size=0.5pt, index of colormap=2 of Blues-3] table[x expr=0.5-\thisrowno{0}] {data/hubbard/data-s.dat};
        \addplot+[only marks, mark size=0.5pt, index of colormap=2 of Blues-3] table[x expr=0.5+\thisrowno{0}] {data/hubbard/data-s.dat};
        \node[index of colormap=2 of Blues-3] at (0.5, 0.3) {\footnotesize Spinons};
    \end{axis}
    \begin{axis}[width=\linewidth, height=0.3\linewidth,
                 xmin=0, xmax=1, ymin=-0.01, ymax=0.07,
                 ytick={0.00, 0.03, 0.06},
                 xlabel=\(k/\pi\),
                 ylabel=\(\sigma^2_\Delta\),
                 every axis x label/.style={at={(ticklabel* cs:0.5)}, anchor=north, yshift=-5pt},
                 cycle list/Reds-3, cycle list/Blues-3, cycle list/Purples-3]
        \addplot+[only marks, mark size=0.8pt, index of colormap=1 of Reds-3] table[y expr=\thisrowno{2}] {data/hubbard/data-c.dat};
        \addplot+[only marks, mark size=0.8pt, index of colormap=1 of Reds-3] table[x expr=1-\thisrowno{0}, y expr=\thisrowno{2}] {data/hubbard/data-c.dat};
        \addplot+[only marks, mark size=0.8pt, index of colormap=2 of Reds-3] table[y expr=\thisrowno{2}] {data/hubbard/data-c2.dat};
        \node[anchor=north west] at (rel axis cs: 0, 1) {(b)};
    \end{axis}
\end{tikzpicture}

%% file: figures/hubbard-wave-packets.tex
\begin{tikzpicture}
    \pgfplotsset{every tick label/.append style={/pgf/number format/fixed}}
    \begin{axis}[width=\linewidth, height=0.4\linewidth, yshift=0.8\linewidth-70,
                 xmin=-50, xmax=150, ymin=-0.02, ymax=0.10,
                 xticklabels={},
                 cycle list/Reds-3, cycle list/Blues-3,
                 legend style={at={(0.5, 1.05)}, anchor=south, legend columns=1, font=\scriptsize}]
        \addplot+[very thick, no markers, index of colormap=2 of Reds-3] table[y expr=\thisrowno{1}-1] {data/hubbard/data.t0.0.dat};
        \addplot+[very thick, no markers, index of colormap=2 of Blues-3] table[x expr=\thisrowno{0}-0.5, y expr=\thisrowno{3}] {data/hubbard/data.t0.0.dat};
        \legend{{Excess charge \(\hat{n}_i - 1\)}, {Spin (two-site average) \((\hat{S}^z_{i-1/2}+\hat{S}^z_{i+1/2})/2\)}};
        \draw[->, index of colormap=1 of Reds-3] (axis cs: 15, 0.025) -- (axis cs: 40, 0.025);
        \draw[->, index of colormap=1 of Blues-3] (axis cs: 90, 0.025) -- (axis cs: 80, 0.025);
        \node[anchor=north west] at (rel axis cs: 0, 1) {(a) \(\tau = 0\)};
    \end{axis}
    \begin{axis}[width=\linewidth, height=0.4\linewidth, yshift=0.4\linewidth-35,
                 xmin=-50, xmax=150, ymin=-0.02, ymax=0.10,
                 xticklabels={},
                 cycle list/Reds-3, cycle list/Blues-3]
        \addplot+[very thick, no markers, index of colormap=2 of Reds-3] table[y expr=\thisrowno{1}-1] {data/hubbard/data.t30.0.dat};
        \addplot+[very thick, no markers, index of colormap=2 of Blues-3] table[x expr=\thisrowno{0}-0.5, y expr=\thisrowno{3}] {data/hubbard/data.t30.0.dat};
        \node[anchor=north west] at (rel axis cs: 0, 1) {(b) \(\tau = 30\)};
    \end{axis}
    \begin{axis}[width=\linewidth, height=0.4\linewidth,
                 xmin=-50, xmax=150, ymin=-0.02, ymax=0.10,
                 xlabel=Site index \(i\),
                 cycle list/Reds-3, cycle list/Blues-3]
        \addplot+[very thick, no markers, index of colormap=2 of Reds-3] table[y expr=\thisrowno{1}-1] {data/hubbard/data.t75.0.dat};
        \addplot+[very thick, no markers, index of colormap=2 of Blues-3] table[x expr=\thisrowno{0}-0.5, y expr=\thisrowno{3}] {data/hubbard/data.t75.0.dat};
        \node[anchor=north west] at (rel axis cs: 0, 1) {(c) \(\tau = 75\)};
    \end{axis}
\end{tikzpicture}

%% file: figures/hubbard-wave-packets-full.tex
\begin{tikzpicture}
    \pgfplotsset{every tick label/.append style={/pgf/number format/fixed}}
    \begin{axis}[width=0.925\linewidth, height=0.6\linewidth, yshift=0.6\linewidth-35, axis on top,
                 ylabel=Time \(\tau\),
                 ytick={0, 25, 50, 75},
                 xticklabels={},
                 xmin=-50, xmax=150, ymin=0, ymax=75]
        \addplot graphics[xmin=-50, xmax=150, ymin=0, ymax=75] {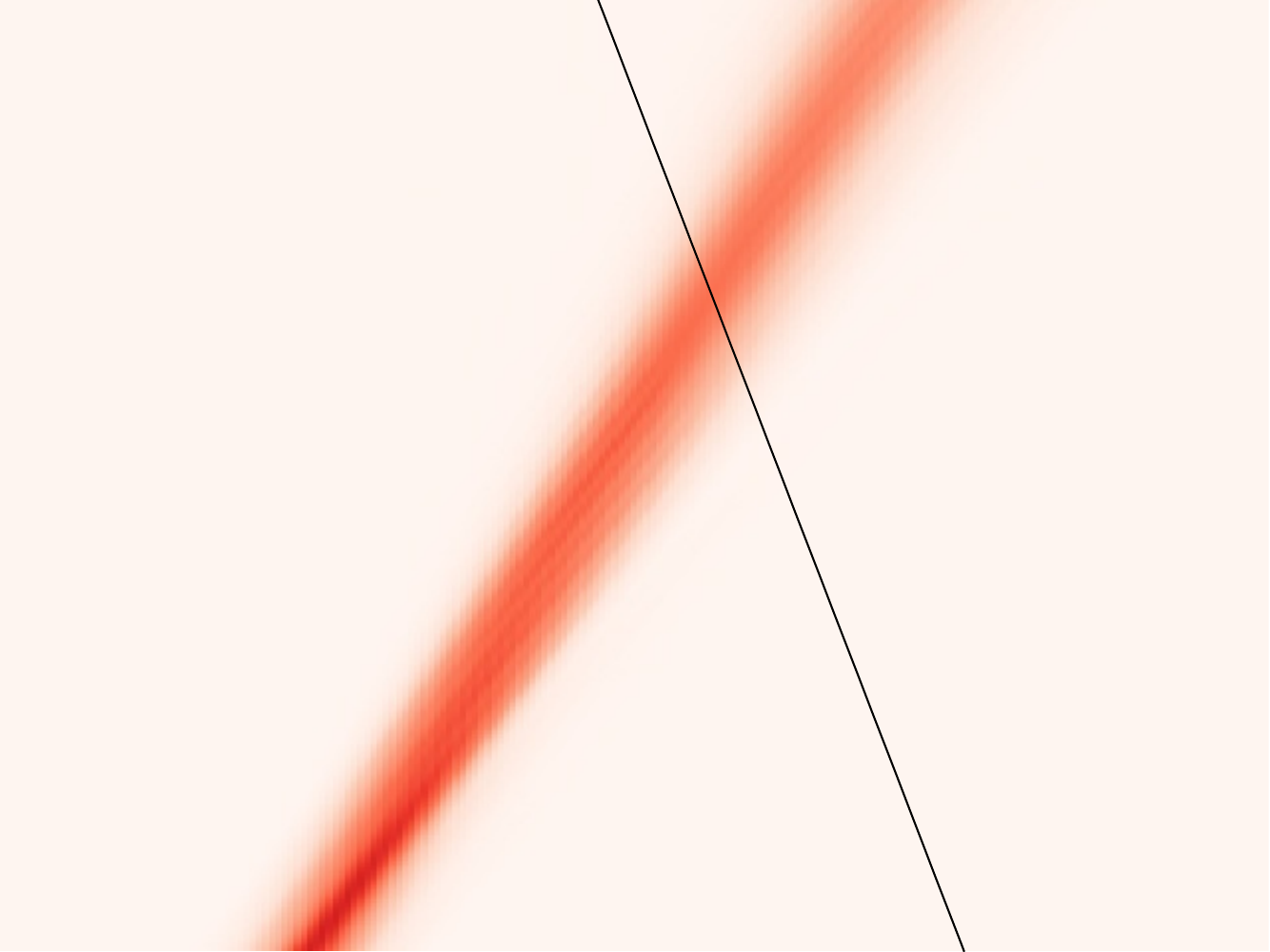};
        \node[anchor=north west, fill=white!50, fill opacity=0.5, text opacity=1, align=left] at (rel axis cs: 0.0, 1.0) {(a) \footnotesize Excess charge\\\phantom{(a) }\(\hat{n}_i - 1\)};
    \end{axis}
    \begin{axis}[width=0.225\linewidth, height=0.6\linewidth, yshift=0.6\linewidth-35, xshift=0.775\linewidth, axis on top,
                 xtick=\empty,
                 ylabel near ticks, yticklabel pos=right,
                 xmin=0, xmax=1, ymin=0, ymax=0.1]
        \addplot graphics[xmin=0, xmax=1, ymin=0, ymax=0.1] {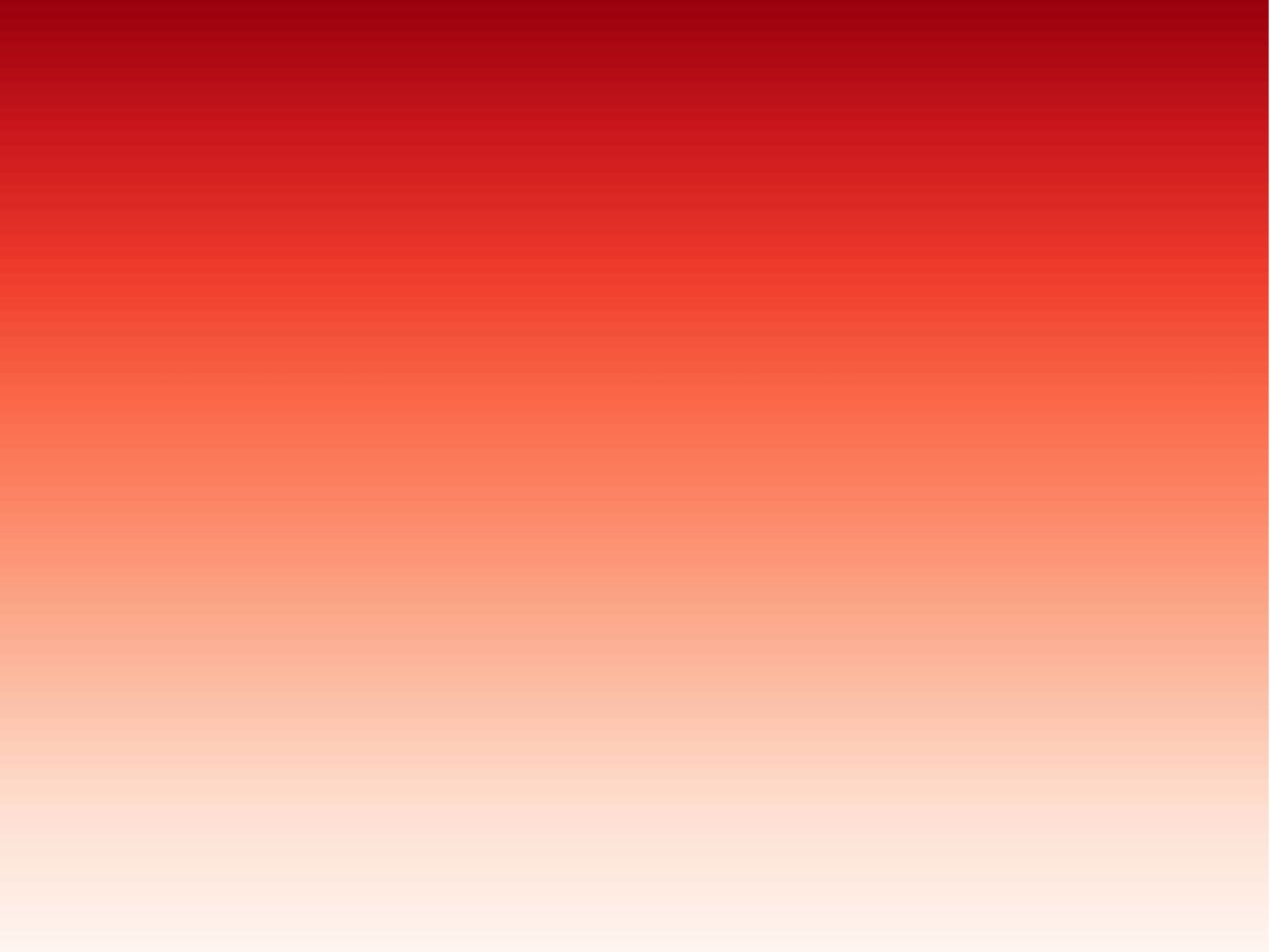};
    \end{axis}
    \begin{axis}[width=0.925\linewidth, height=0.6\linewidth, axis on top,
                 xlabel=Site index \(i\), ylabel=Time \(\tau\),
                 ytick={0, 25, 50, 75},
                 xmin=-50, xmax=150, ymin=0, ymax=75]
        \addplot graphics[xmin=-50, xmax=150, ymin=0, ymax=75] {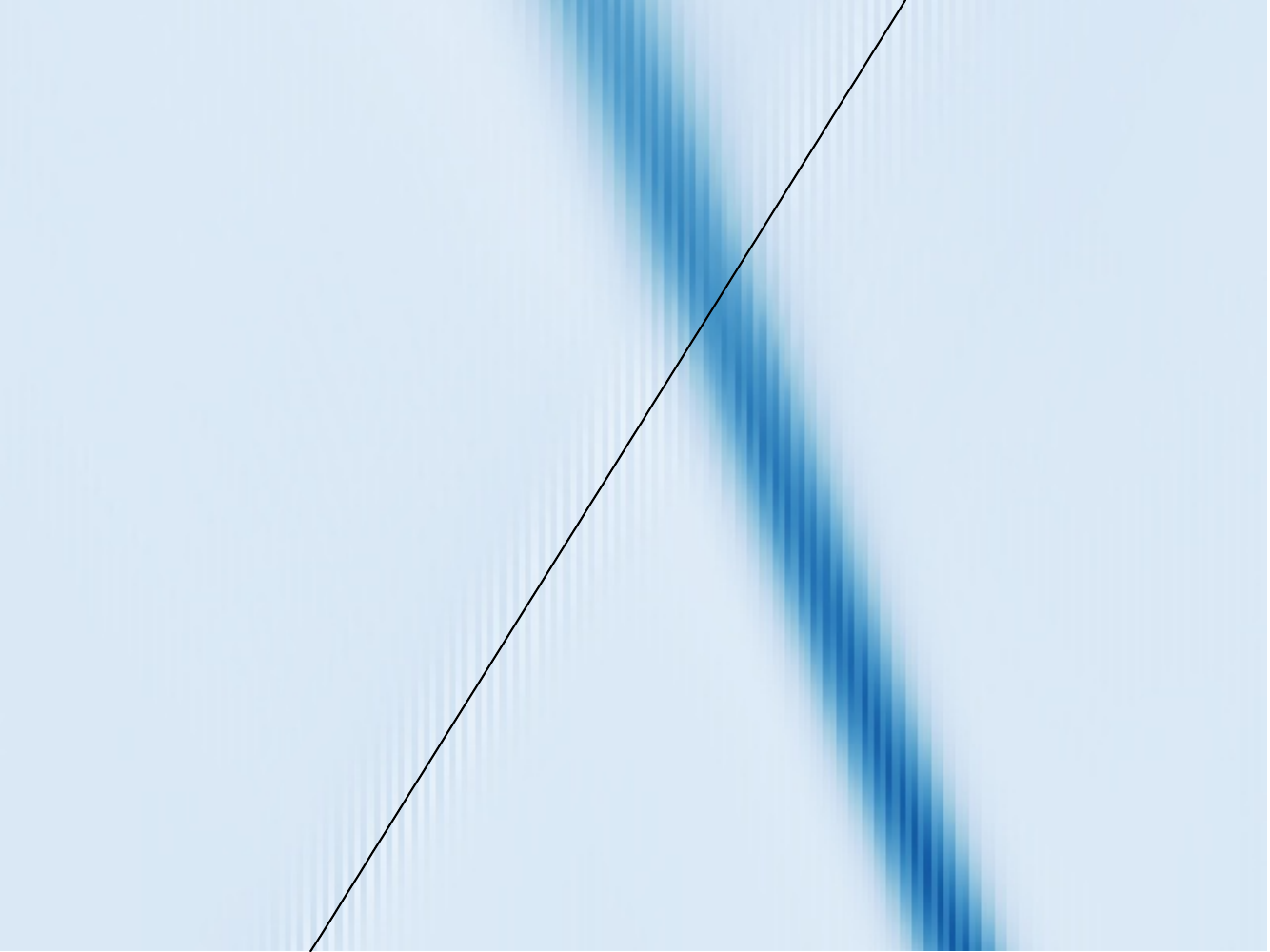};
        \node[anchor=north west, fill=white, fill opacity=0.5, text opacity=1, align=left] at (rel axis cs: 0.0, 1.0) {(b) \footnotesize Spin (two-site average)\\\phantom{(b) }\((\hat{S}^z_{i-1/2}+\hat{S}^z_{i+1/2})/2\)};
    \end{axis}
    \begin{axis}[width=0.225\linewidth, height=0.6\linewidth, xshift=0.775\linewidth, axis on top,
                 xtick=\empty, scaled ticks=false,
                 ytick={-0.01, 0, 0.01, 0.02, 0.03, 0.04, 0.05},
                 ylabel near ticks, yticklabel pos=right,
                 xmin=0, xmax=1, ymin=-0.01, ymax=0.05]
        \addplot graphics[xmin=0, xmax=1, ymin=-0.01, ymax=0.05] {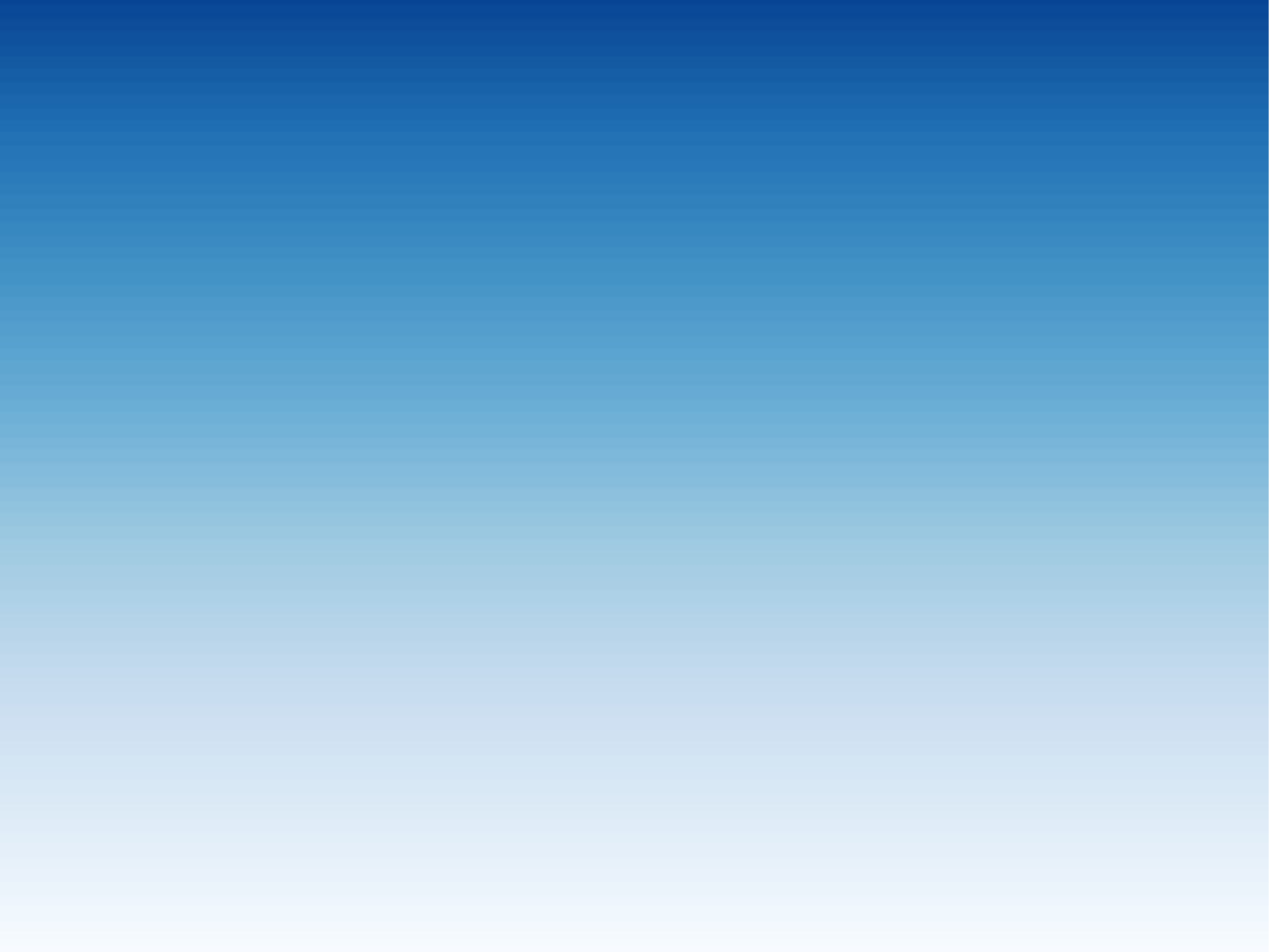};
    \end{axis}
\end{tikzpicture}

%% file: figures/hubbard-wave-packets-domain-wall.tex
\begin{tikzpicture}
    \pgfplotsset{every tick label/.append style={/pgf/number format/fixed}}
    \begin{axis}[width=\linewidth, height=0.4\linewidth,
                 xmin=-50, xmax=150, ymin=-0.075, ymax=0.10,
                 xlabel=Site index \(i\),
                 cycle list/Reds-3, cycle list/Blues-3,
                 legend style={at={(0.5, 1.05)}, anchor=south, legend columns=2, font=\scriptsize}]
        \addplot+[very thick, no markers, index of colormap=2 of Reds-3] table[y expr=\thisrowno{1}-1] {data/hubbard/data.t0.0.dat};
        \addplot+[thick, no markers, index of colormap=2 of Blues-3] table[x expr=\thisrowno{0}, y expr=\thisrowno{2}] {data/hubbard/data.t0.0.dat};
        \legend{{Excess charge \(\hat{n}_i - 1\)}, {Spin \(\hat{S}^z_i\)}};
    \end{axis}
\end{tikzpicture}